\documentclass[12pt]{article}
\usepackage{amsmath,amssymb,amsfonts,epsfig,cite,setspace,bigstrut,longtable,array,url}
\usepackage[paper=letterpaper,margin=1in]{geometry}
\usepackage{graphicx}
\usepackage{hyperref}
\usepackage{caption}
\usepackage{subcaption}
\usepackage[titles]{tocloft}
\usepackage{float}

\newcommand{\nn}{\nonumber}

\newcommand{\comment}[1]{}

\newcommand{\IP}{\mathbb{P}}

\newcommand{\IR}{\mathbb{R}}

\newcommand{\IZ}{\mathbb{Z}}

\newcommand{\vev}[1]{\left<#1\right>}

\renewcommand{\thefootnote}{\fnsymbol{footnote}}

\newcommand{\captionN}[1]{\caption{{\sf {\small #1}}}}

\numberwithin{equation}{section}

\newcommand\be{\begin{equation}}
\newcommand\ee{\end{equation}}
\newcommand\bea{\begin{eqnarray}}
\newcommand\eea{\end{eqnarray}}
\newcommand\eref[1]{\eqref{#1}}

\begin{document}
\thispagestyle{empty}
\renewcommand{\thefootnote}{\fnsymbol{footnote}}

${}$

\vspace{2 cm}

\begin{center}

\textbf{\LARGE Patterns in Calabi--Yau Distributions}

\vspace{1cm}

{\large
Yang-Hui He$^{a,}$\,\footnote[1]{\texttt{hey@maths.ox.ac.uk}},
Vishnu Jejjala$^{b,}$\,\footnote[2]{\texttt{vishnu@neo.phys.wits.ac.za}},
Luca Pontiggia$^{b,}$\,\footnote[3]{\texttt{lucatpontiggia@gmail.com}}}

\vspace{1cm}
$^{a}$\textit{
School of Physics, NanKai University, Tianjin, 300071, P.R.~China and \\
Department of Mathematics, City University, London, EC1V 0HB, UK and\\
Merton College, University of Oxford, OX1 4JD, UK}\\

\vspace{0.5cm}
$^{b}$\textit{NITheP, School of Physics, and Mandelstam Institute for Theoretical Physics,\\
University of the Witwatersrand, Johannesburg, WITS 2050, South Africa}\\

\end{center}

\vspace{1.5cm}

\begin{abstract}
  We explore the distribution of topological numbers in Calabi--Yau manifolds, using the Kreuzer--Skarke dataset of hypersurfaces in toric varieties as a testing ground.
  While the Hodge numbers are well-known to exhibit mirror symmetry, patterns in frequencies of combination thereof exhibit striking new patterns.
  We find pseudo-Voigt and Planckian distributions with high confidence and exact fit for many substructures.
  The patterns indicate typicality within the landscape of Calabi--Yau manifolds of various dimension.
\end{abstract}

\newpage
\renewcommand{\thefootnote}{\arabic{footnote}}
\setcounter{footnote}{0}

\tableofcontents

\section{Introduction}
A Calabi--Yau $n$-fold is a K\"ahler manifold of $n$ complex dimensions with a trivial canonical bundle.
In superstring theory, it serves as a compactification manifold wherein a ten dimensional theory at high energies reduces to an effective theory in four spacetime dimensions.
In particular, global $SU(n)$ holonomy ensures that $2^{1-n}$ of the original supersymmetry is preserved.
Thus, confronted by the vacuum selection problem, Calabi--Yau compactifications present an avenue for Standard Model building especially in the context of the heterotic string~\cite{chsw,cicy}.
Indeed, the basis of the landscape is to consider flux compactifications on these geometries~\cite{hitchin,douglas}.

To facilitate this approach to a low-energy phenomenology derived from string theory, mathematicians and physicists have constructed large datasets of Calabi--Yau threefolds~\cite{Candelas:1989hd,BB,Avram:1997rs,Kreuzer:2000qv,Altman:2014bfa,Gray:2014kda,FourFold,Anderson:2015iia,Davies:2011fr,Candelas:2008wb,He:2013epn,Kreuzer:1995cd,Kreuzer:1998vb,Kreuzer:2000xy} as well as various refined analyses of properties thereof \cite{Taylor:2012dr,Taylor:2015ppa, Gao:2013pra,Blumenhagen:2011xn,Gray:2012jy,Candelas:2012uu,Braun:2010vc,Candelas:2007ac}.
By far the largest database was constructed in a \textit{tour de force} of algebraic geometry, combinatorics, physics, and computer algorithms by Kreuzer and Skarke based on the theorems of Batyrev and Borisov~\cite{BB,Avram:1997rs,Kreuzer:2000qv,Kreuzer:1995cd,Kreuzer:1998vb,Kreuzer:2000xy,Kreuzer:2002uu,Braun:2012vh}.
In short, these Calabi--Yau $n$-manifolds $X_n$ are realized as a smooth hypersurface embedded in a toric variety $A_{n+1}$ of complex dimension $n+1$; the Calabi--Yau condition simply translates to the requirement that the polytope defining $A_{n+1}$ be \textbf{reflexive}.
We will henceforth consider only such Calabi--Yau manifolds, of which there are a plethora.

Let us briefly recollect what all this means.
The (possibly singular) toric variety $A_{n+1}$ is specified by an integer polytope $\Delta$ in $\IR^{n+1}$, which is a collection of vertices (dimension $0$) each of which is an $(n+1)$-vector with integer entries and such that each pair of neighboring vertices defines an edge (dimension $1$), each pair of edges defines a face (dimension $2$), etc., all the way up to a facet (dimension $n$).
Alternatively, $\Delta$ can be defined by a set of integer linear inequalities, each of which slices a facet.
The polytope is then the convex body in $\IR^{n+1}$ enclosed by these facets.
We will always include the origin as being contained in $\Delta$.
Using the usual dot product $\vev{~,~}$ inherited from $\IR^{n+1}$, the dual polytope is defined by
\begin{equation}
  \Delta^\circ :=
  \left\{
  v \in \IR^{n+1} | \vev{m,v} \ge -1, \forall~m \in \Delta
  \right\} \ .
\end{equation}
The polytope $\Delta$ is \textit{reflexive} if all the vertices of $\Delta^\circ$ are integer vectors.
In this case, we can define the Calabi--Yau hypersurface $X_n$ explicitly as the polynomial equation
\begin{equation}
\sum\limits_{m \in \Delta} c_m \prod\limits_{r = 1}^k x_r ^{\vev{m,v_r}+1} = 0 \ ,
\end{equation}
where $v_{r=1,\ldots,k}$ are the vertices of $\Delta^\circ$ with $k$ being the number of vertices of $\Delta^\circ$ (or equivalently the number of facets of $\Delta$), $x_r$ are the coordinates of $A_{n+1}$, and $c_m$ are numerical coefficients parameterizing the complex structure of $X_n$.
Indeed, the reflexivity of $\Delta$ ensures that the exponents are integral whereby making the hypersurface polynomial as required.

The classification of these Calabi--Yau manifolds thus amounts to that of reflexive polytopes in various dimensions, and the intense computer work of Kreuzer and Skarke was to combinatorially find such polytopes.
For $n=1$, there are $16$ such polytopes in $\IR^2$, and we have Calabi--Yau onefolds, or elliptic curves.
For $n=2$, there are $4319$ such polytopes in $\IR^3$, and we have Calabi--Yau twofolds, or K3 surfaces.
For $n=3$, there are $473,800,776$ such polytopes (which was a formidable computer task!), and we have the Calabi--Yau threefolds.
This sequence
\begin{equation}
\{ 1, 16, 4319, 473800776, \ldots \}
\end{equation}
of remarkable growth rate, can be found in the Online Encyclopedia of Integer Sequences~\cite{oeis}.
The numbers in higher dimension are still not known, nor has there been an asymptotic analysis of their growth.
It should be emphasized that generically a reflexive polytope corresponds to a \textit{singular} toric variety even though the hypersurface is chosen (by generic coefficients $c_m$) to miss the singularities and hence ensuring the smoothness of the Calabi--Yau $X_{n}$.
For example, of the some half-billion reflexive polytopes in $\IR^4$, only $136$ $A_{4}$ are in fact smooth~\cite{He:2009wi}.
As we desingularize the toric variety by various star-triangulations of $\Delta$, we are led to potentially \textit{inequivalent} Calabi--Yau manifolds.
In principle, the \textit{same} Calabi--Yau geometry can arise from different reflexive polytopes or triangulations of a given reflexive polytope.
Whereas K3 is essentially unique, we do not know how many Calabi--Yau threefolds there are.
A systematic study to classify the desingularizations, to compute the necessary topological data, and to build an interactive online database~\cite{Altman:2014bfa} is under way.
The moral is that there are almost certainly far more than half a billion Calabi--Yau threefolds!

\begin{figure}[t]
	\begin{center}	
	(a)
		\includegraphics[scale=0.9]{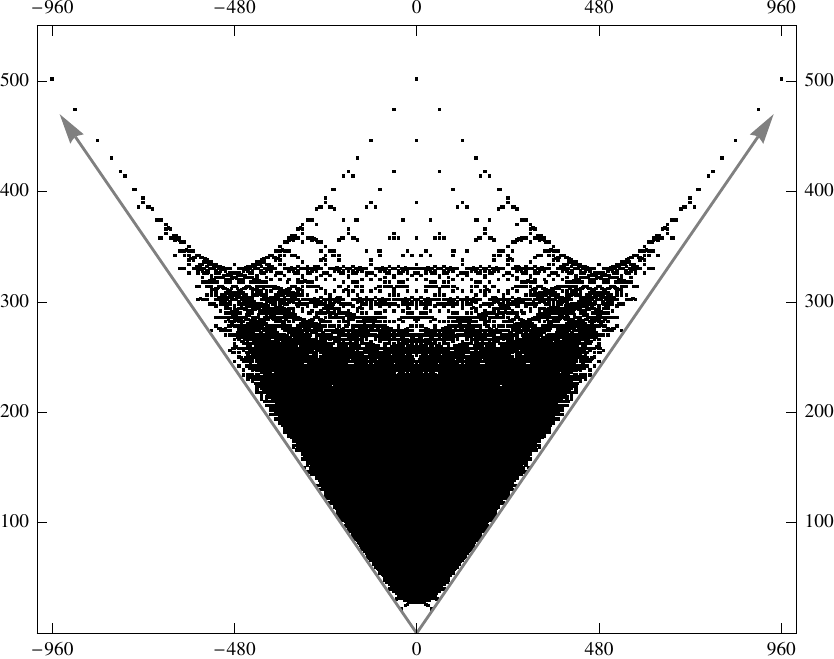}
	(b)
		\includegraphics[scale=0.5]{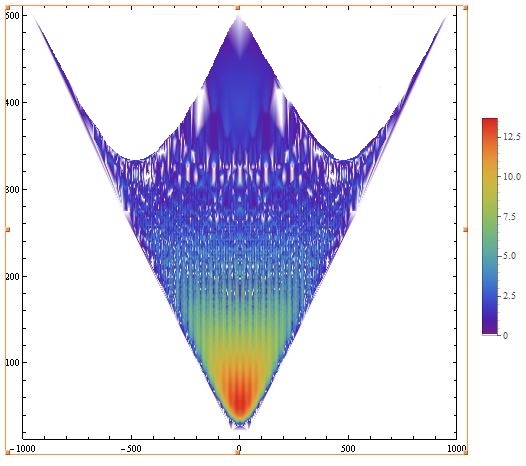}
	\end{center}
	\captionN{(a) The cumulative plot of $\chi=2(h^{1,1}-h^{1,2})$ on the abscissa versus $h^{1,1}+h^{1,2}$ on the ordinate for Calabi--Yau threefolds as hypersurfaces in toric fourfolds; (b) marking also the natural logarithm of the multiplicity of the Hodge pair with a color grading.}  	
	\label{f:hodge3}
\end{figure}

Luckily, the Hodge numbers depend only on the polytope and not on the choice of desingularization. (The intersection numbers, however, do depend on the choice.)
For Calabi--Yau threefolds, the pair of Hodge numbers $(h^{1,1},h^{1,2})$ is a famous quantity.
Indeed, the plot in Part (a) of Figure~\ref{f:hodge3} has become iconic.
Here, the sum $h^{1,1}+h^{1,2}$ is plotted against the Euler number $\chi=2(h^{1,1}-h^{1,2})$, and the left-right symmetry supplies ``experimental evidence'' for \textit{mirror symmetry}.
There is enormous redundancy in this data: of the some half a billion reflexive polytopes, there are only $30,108$ distinct pairs of Hodge numbers and the pair $(27,27)$ dominates the multiplicity, totaling almost one million. In Part (b) of Figure~\ref{f:hodge3} we have attempted to visualize the distribution of the multiplicity by having a color density plot of the logarithm of the number over each Hodge pair.

Understanding this multiplicity forms the inspiration for the present work.
While there have been analyses on the \textit{shape} of the funnel-like plot~\cite{Candelas:2007ac,Candelas:2012uu,Taylor:2012dr}, there has not been much work on its \textit{density}, \textit{i.e.}, the distribution of the multiplicity of Hodge data for the Calabi--Yau manifolds of various dimension.
Of course, fundamentally, this is entirely due to the combinatorics of reflexive polytopes and might in principle be analytically determined.
However, given the complexity of the problem it is expedient to analyze the available data which have been compiled over the years, observe intriguing patterns, and draw statistical inferences before turning to analytic treatments.
This is what we achieve in this work.

The organization of the paper is as follows.
We perform a detailed analysis on the structure and behavior of the threefold data in Section~\ref{ThreeAna}.
This is motivated by looking for an exact function describing the relationship of the distribution of the Hodge pairs $(h^{1,1},h^{1,2})$ with frequency.

In Section~\ref{Hdiff}, we study the distribution of $(h^{1,1}-h^{1,2},f)$. We find that this distribution is composed of a family of curves, for which each curve can be described using a modified pseudo-Voigt model. Although an approximation, the model is able to describe the general trend of the data, as well as some additional fine structure within each individual data point. Performing an analysis on the parameter relationships shows that three out of the five parameters can be expressed as a single variable, but conclude that additional modifications need to be introduced in the model to overcome certain shortfalls.

Subsequently, Section~\ref{Hsum} performs an analysis on the structure of $(h^{1,1}+h^{1,2},f)$. Similarly, this distribution is composed of a family of curves for which each curve can be described using a Planckian profile. Combining the regression analysis for each curve within the distribution, we construct a single function able to approximately model the entire distribution of $(h^{1,1}+h^{1,2},f)$ with only two variables. Section~\ref{Euler} uses the model developed in Section~\ref{Hdiff} to describe the distribution of the Euler number $\chi$.

Section~\ref{GOF} is dedicated to the description of model validation in our context, as the usual statistical tests are inadequate. Section~\ref{IfP} discusses possible implications to physics by referencing recent advancements in F theory and further investigations of structures within the Kreuzer--Skarke database. In Section~\ref{Picard} and Section~\ref{FourFolds}, we perform primary analyses of Calabi--Yau twofolds (Picard number and multiplicity) and Calabi--Yau fourfolds. Due to the lack of a complete data set, we are unable to provide a thorough analysis of the fourfolds as with threefolds. Finally the Appendix presents many supplementary plots and figures for the various sections.
We conclude with a summary and outlook in Section~\ref{s:conc}.

\section{Calabi--Yau Threefolds}\label{ThreeAna}
As advertised in the Introduction, we will begin with the analysis of threefolds and identify patterns within this rich distribution of Hodge numbers and their frequency as plotted in Figure~\ref{f:hodge3}.
It turns out striking patterns do exist, pointing to a definite structure within the threefold data, which consists of the triple
$
(h^{1,1}, h^{1,2},f) \ ,
$
where $f$ is the number of reflexive polytopes in the Kreuzer--Skarke database with the given Hodge pair.
Here, $h^{1,1}$ and $h^{1,2}$ respectively count the K\"ahler and complex structure moduli of the Calabi--Yau obtained from the reflexive polytope.
More precisely \cite{batyrev}, we have that
\begin{align}
  \nn
  h^{1,1}(X) &= \ell(\Delta^*) - \sum\limits_{{\rm codim} \theta^* = 1} \ell^*(\theta^*) +
  \sum\limits_{{\rm codim} \theta^* = 2} \ell^*(\theta^*) \ell^*(\theta)
  - 5 ;
  \\
  h^{1,2}(X) &= \ell(\Delta) - \sum\limits_{{\rm codim} \theta = 1} \ell^*(\theta) +
  \sum\limits_{{\rm codim} \theta = 2} \ell^*(\theta) \ell^*(\theta^*)
  - 5\ .
  \label{hodgeDelta}
\end{align}
In the above, $\Delta$ is the defining polytope for the Calabi--Yau threefold $X$ and $\Delta^*$ is its dual. Moreover, $\theta$ and $\theta^*$ are the faces of specified codimension of these polytopes respectively; $\ell(~)$ is the number of integer points of the polytope while $\ell^*(~)$ is the number of interior integer points.
Indeed, our analysis of the distribution of Hodge numbers ultimately reduces to counting these integer points.

\begin{figure}[H]
	\begin{center}	
		\includegraphics[scale=0.21]{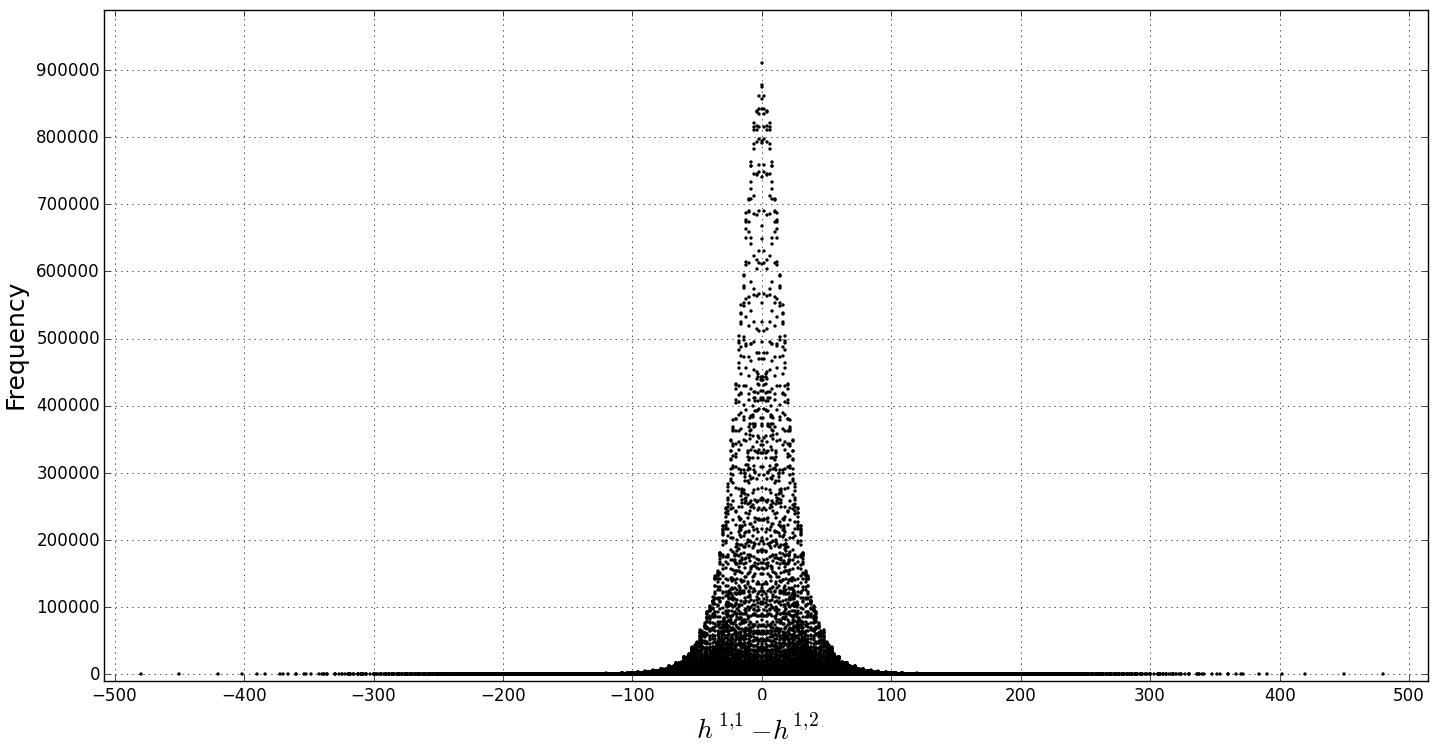}
		\includegraphics[scale=0.21]{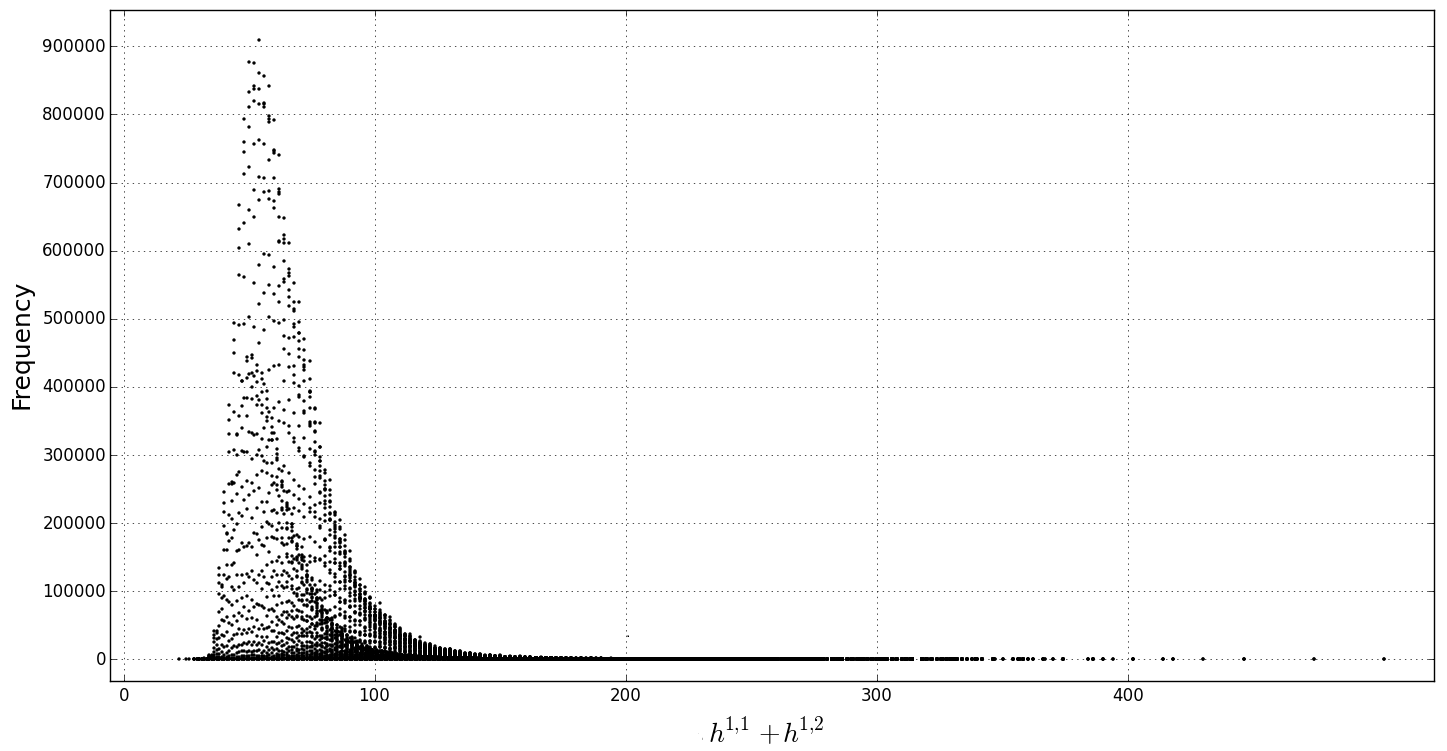}
	\end{center}
\captionN{
  (a) Frequency $f$ plotted against $\frac12 \chi = h^{1,1} - h^{1,2}$; 	
  (b) Frequency $f$ plotted against the sum of Hodge numbers $h^{1,1} + h^{1,2}$.}  	
\label{Fig:Main}
\end{figure}

\begin{figure}[h!]
	\begin{center}	
(a)		\includegraphics[scale=0.352]{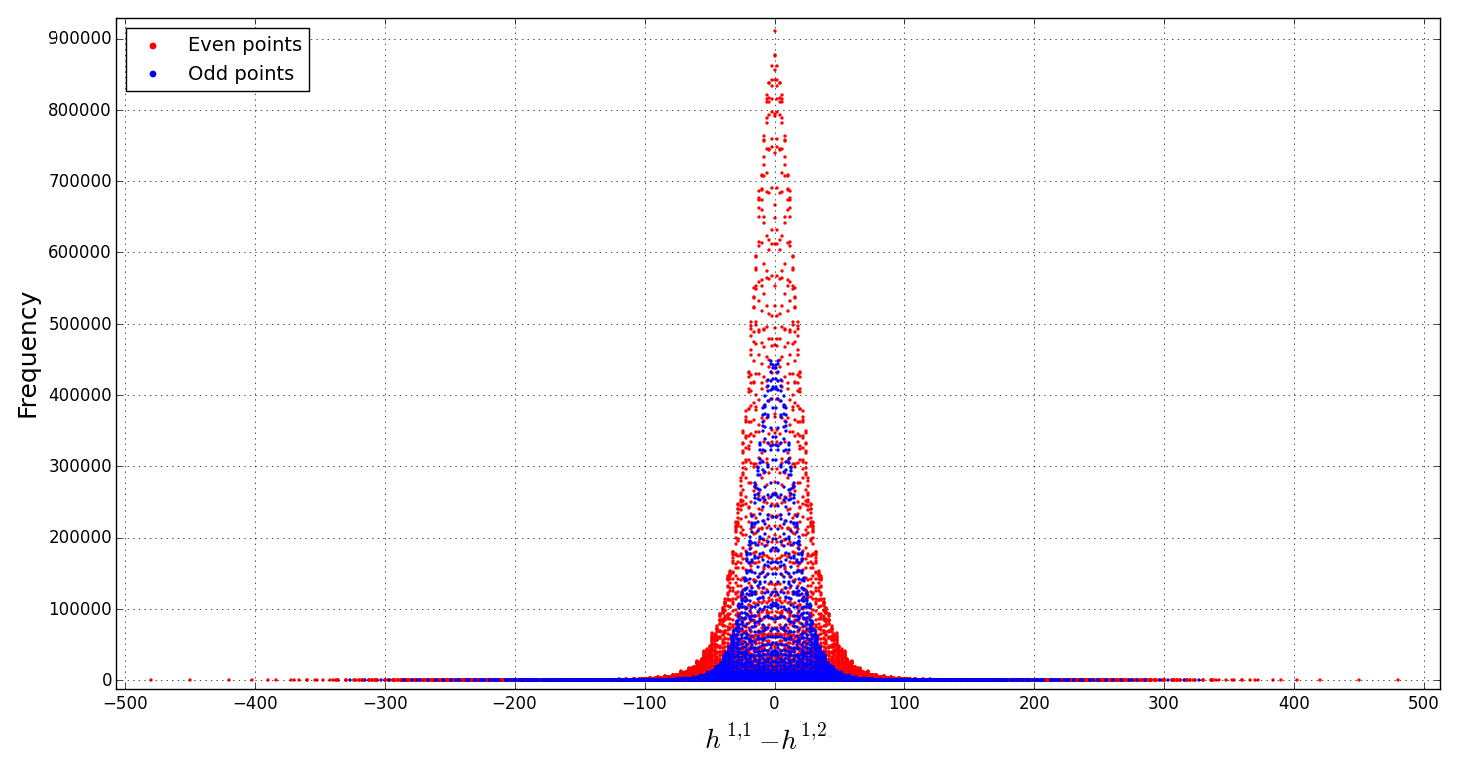}
	\end{center}
	\label{Fig:MainEvenOddDiff}
	\begin{center}	
(b)		\includegraphics[scale=0.35]{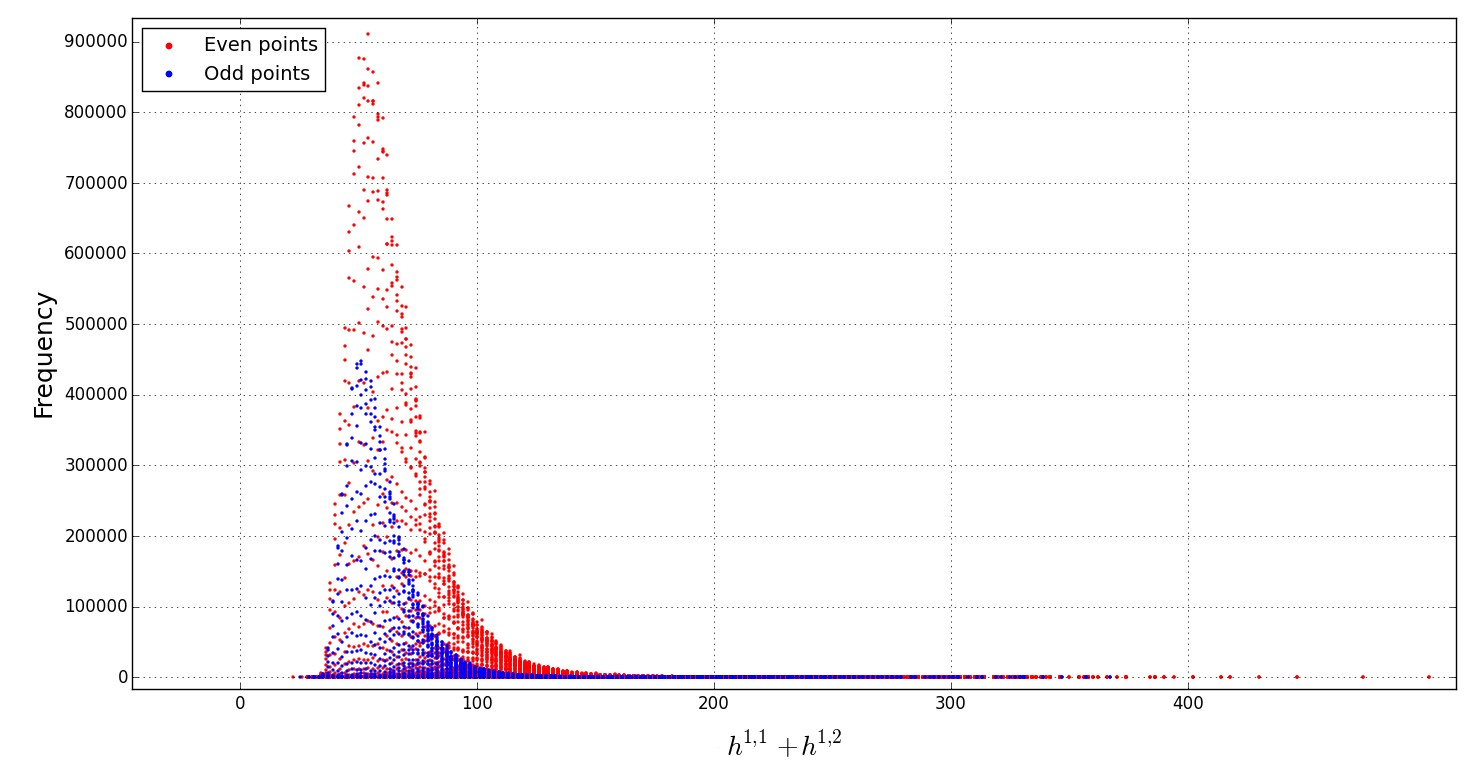}
	\end{center}
	\captionN{
          (a) The $h^{1,1} - h^{1,2}$ distribution for threefolds, highlighting the two sub-distributions, where red and blue data points correspond to even and odd values of $h^{1,1} - h^{1,2}$, respectively;
          (b) The same, but for  $h^{1,1} + h^{1,2}$.}  	
	\label{Fig:MainEvenOdd}
\end{figure}

To facilitate the analysis, we plot $(h^{1,1} - h^{1,2},f)$ and $(h^{1,1} + h^{1,2},f)$ as shown in (a) and (b) of Figure~\ref{Fig:Main}, respectively.
Recall that the Euler number $\chi = 2(h^{1,1}-h^{1,2})$.
We will use the difference $h^{1,1}-h^{1,2}$ rather than the Euler number.
In the simplest heterotic constructions, $|h^{1,1}-h^{1,2}|$ corresponds to the index of the Dirac operator and gives the number of generations of particles in the low-energy spectrum~\cite{chsw}.
By inspection, these plots already exhibit two patterns.
Firstly, in both the $h^{1,1} - h^{1,2}$ and $h^{1,1} + h^{1,2}$ plots, there appears to be an inner distribution contained within the outer distribution.
We find that these inner and outer distributions are related to the parity of $h^{1,1} \pm h^{1,2}$.
Figure \ref{Fig:MainEvenOdd} elucidates this point by having the odd and even values in different colors.
Though this parity structure may be a result of the Kreuzer--Skarke algorithm, its consistent appearance means we need to treat the distributions of even and odd distinctly for now.

The second evident structure which can been seen by inspection, is that the outer edge of the distribution of $h^{1,1} - h^{1,2}$ (Figure~\ref{Fig:MainEvenOdd}(a)) appears to follow a normal like curve, whereas the edge of $h^{1,1} + h^{1,2}$ (Figure~\ref{Fig:MainEvenOdd}(b)) follows a Planck like curve.
It is through the analysis of these distributions that we deduce their characteristic behavior and underlying structure.
In the main body of this paper, we outline the results and analysis of only the even distributions for $h^{1,1} - h^{1,2}$ and $h^{1,1} + h^{1,2}$, except where it is important to present both.
It turns out that any structure and patterns which are found in the even distributions for $h^{1,1} - h^{1,2}$ and $h^{1,1} + h^{1,2}$ are found identically in the odd distribution (see Appendix for various plots).

\subsection{Analysis of $h^{1,1} - h^{1,2}$}\label{Hdiff}
Before we can present the results, it is important to explain some notation.
When working with the distribution of $h^{1,1} - h^{1,2}$, we find that it is composed of many curves, whose individual structure is the same as the ``edge'' or boundary of the distribution mentioned earlier.
As a consequence of this, we refer to $h^{1,1} - h^{1,2}$ as being composed of a ``family of curves.'' Each curve is then classified by its \textbf{$r$-value}, where $r =h^{1,1} + h^{1,2}$. It is important to be clear that in this analysis, although $h^{1,1} - h^{1,2}$ is just half the Euler number, we are not summing over all the possible values of $h^{1,1} + h^{1,2}$. We are keeping these values distinct: hence, the $r$-curves we obtain. Later on in Section~\ref{Euler} we sum over all possible values of  $h^{1,1} + h^{1,2}$ to get two plots representing the full Euler number distribution.

\begin{figure}[H]
  \begin{center}
    (a)
    \includegraphics[scale=0.3]{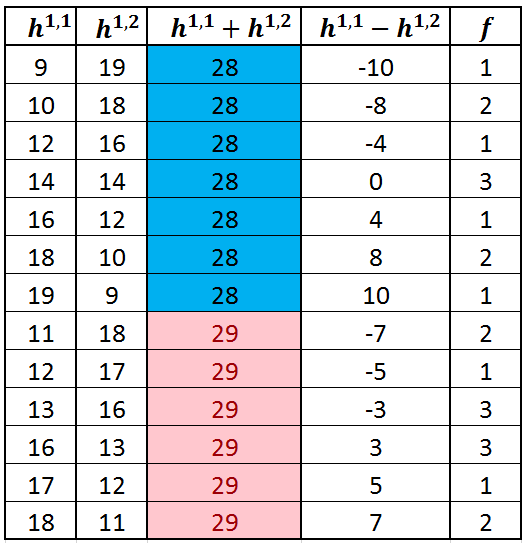}
     \ \ \ \ \ \ \ \ 
    (b) \hspace{-0.1in}
    \includegraphics[scale=0.25]{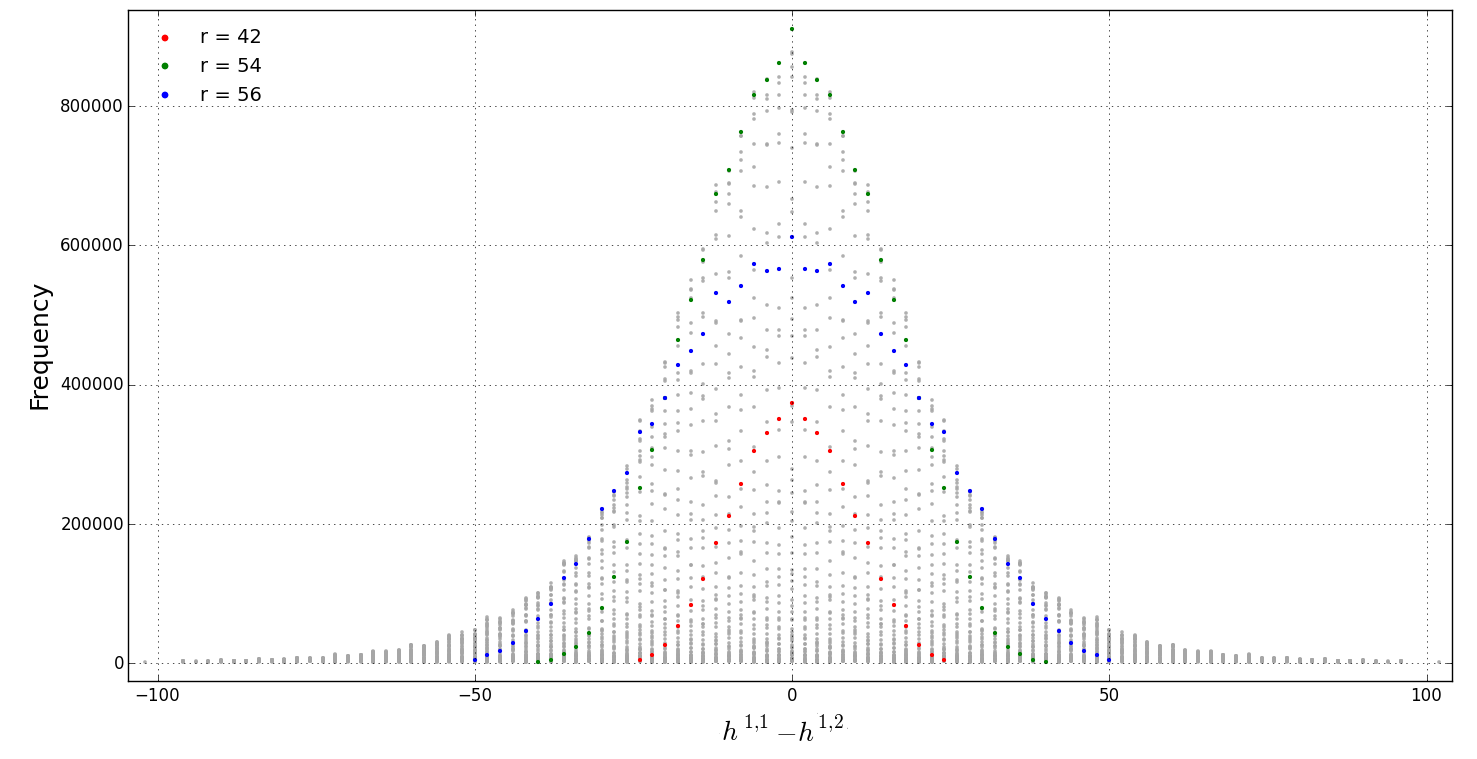}
  \end{center}
  \captionN{
(a) Example of repeated values of the sum $h^{1,1}+ h^{1,2}$ being 28 and 29; 
    (b) Three highlighted curves ($r = 42,54,66$) within the even $h^{1,1} - h^{1,2}$ distribution. The transparent grey data dots are all the data plots for the distribution. Refer to Figure~\protect\ref{Fig:HdiffExampleOdd} for the corresponding odd plot.}
  \label{Fig:HdiffExample}
\end{figure}

Consider the example in Figure~\ref{Fig:HdiffExample}(a).
By ordering the data in terms of $h^{1,1} + h^{1,2}$, one can classify data sets within  $h^{1,1} - h^{1,2}$ by an $r$-value.
Holding $r$ fixed, we can plot the frequency $f$ versus the difference $h^{1,1} - h^{1,2}$.
We call each value of $r$ a curve, which we can overlay on the same plot.
In this example, we tabulate data for curves identified by $r=28$ and $r = 29$.
As a further illustration, we show explicitly the curves of the even distribution within $h^{1,1} - h^{1,2}$ for $r = 42,54,66$ in Figure~\ref{Fig:HdiffExample}(b).
By mirror symmetry, the curve is symmetric about the vertical axis, where $h^{1,1}-h^{1,2} = 0$.

We can now perform a regression analysis for each individual curve, in the quest of obtaining a function describing the distribution.
In the analysis, we indeed find an approximate function predicting the fine structure of the data.
We operate with one caveat: we ignore data points which have a frequency lower than $2000$.
At large $r$, the data, whose frequency is below $2000$, begins to deviate from our model.
The reason for such deviations, comes down to the fact that our model, though remarkably accurate, is still an approximation.
We suspect that with further modifications, such deviations can be accounted for and that consequently, it may be possible to find an exact function to map the frequency distribution of $h^{1,1} - h^{1,2}$.
Such statements also apply to the distribution of $h^{1,1} + h^{1,2}$.

\subsubsection{A Pseudo-Voigt Fit}
Due to the normally-distributed, peak-like nature of these curves, we performed a regression analysis using the following models:
  Gaussian; 
  Cauchy (Lorenztian);
  Pearson7;
  Breit--Wigner;
  Voigt; and
  Pseudo-Voigt.
In the Appendix~\ref{AppenModels}, we perform a side by side comparison.
It turns out that both the Voigt model (\ref{fig:VoigtEO}) as well as the Pseudo-Voigt model (\ref{fig:PVEO}) give excellent fits.

We focus on the \textbf{pseudo-Voigt model} as it gives the best fits.
This is a linear combination of a Gaussian and Lorentzian (Cauchy) distribution:
\begin{equation}\label{eqn:PV}
f(x,A,\mu,\sigma,\alpha) = (1-\alpha)\frac{A}{\sigma\sqrt{2\pi}}e^{\frac{-(x-\mu)^2}{2\sigma^2}} + \alpha\frac{A}{\pi}\left[\frac{\sigma^2}{(x-\mu)^2+\sigma^2}\right] ~,
\end{equation}
with amplitude ($A$), center ($\mu$), Gaussian width ($\sigma$), and fractional parameter alpha ($\alpha$).
However, we can modify the above distribution slightly so that the amplitude $A$ of the distribution has an oscillating component
\begin{align}\label{eqn:modified}
A(x,A_0,a,b)= A_0 + a\cos(2\pi b\cdot x) ~,
\end{align}
where $A_0$ is the original amplitude of a particular curve described by the Pseudo-Voigt distribution, $a$ is the amplitude of oscillations, and $b$ represents the period.
By doing a regression analysis one curve at a time using this modified Pseudo-Voigt model, we are almost able to replicate not just the basic structure of each curve, but even the individual behavior of each data point in the entire distribution.
(See Appendix~\ref{Appenfirst} for a comparative plot of the all the regression curves using the standard, unmodified, Pseudo-Voigt model.)

\begin{figure}[H]
    \centering
    \begin{subfigure}[h]{\textwidth}
        \includegraphics[width=\textwidth]{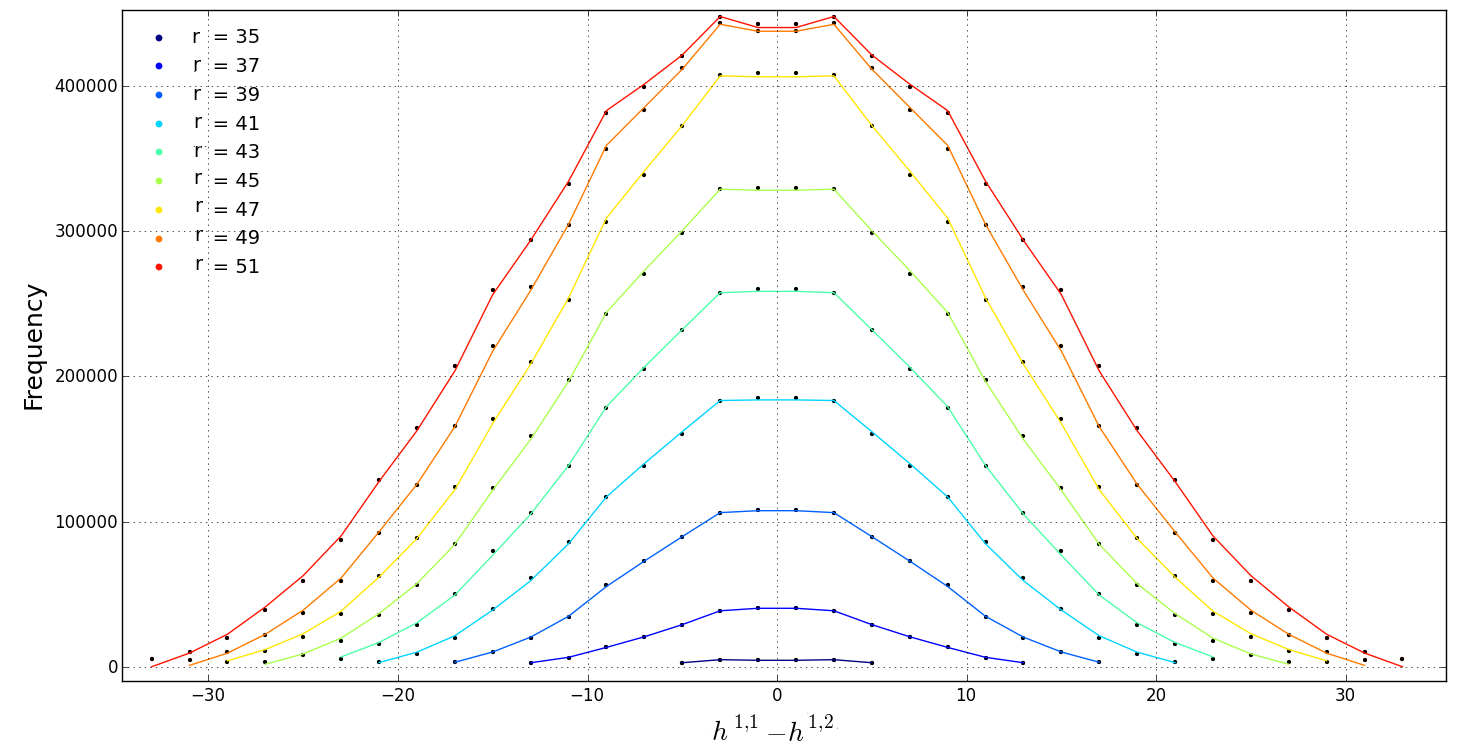}
         \caption{Regression lines for all odd $r$ valued curves, with $r\in[35,51]$.}
        \label{Fig:OddHdiff1}
    \end{subfigure}

    \begin{subfigure}[h]{\textwidth}
        \includegraphics[width=\textwidth]{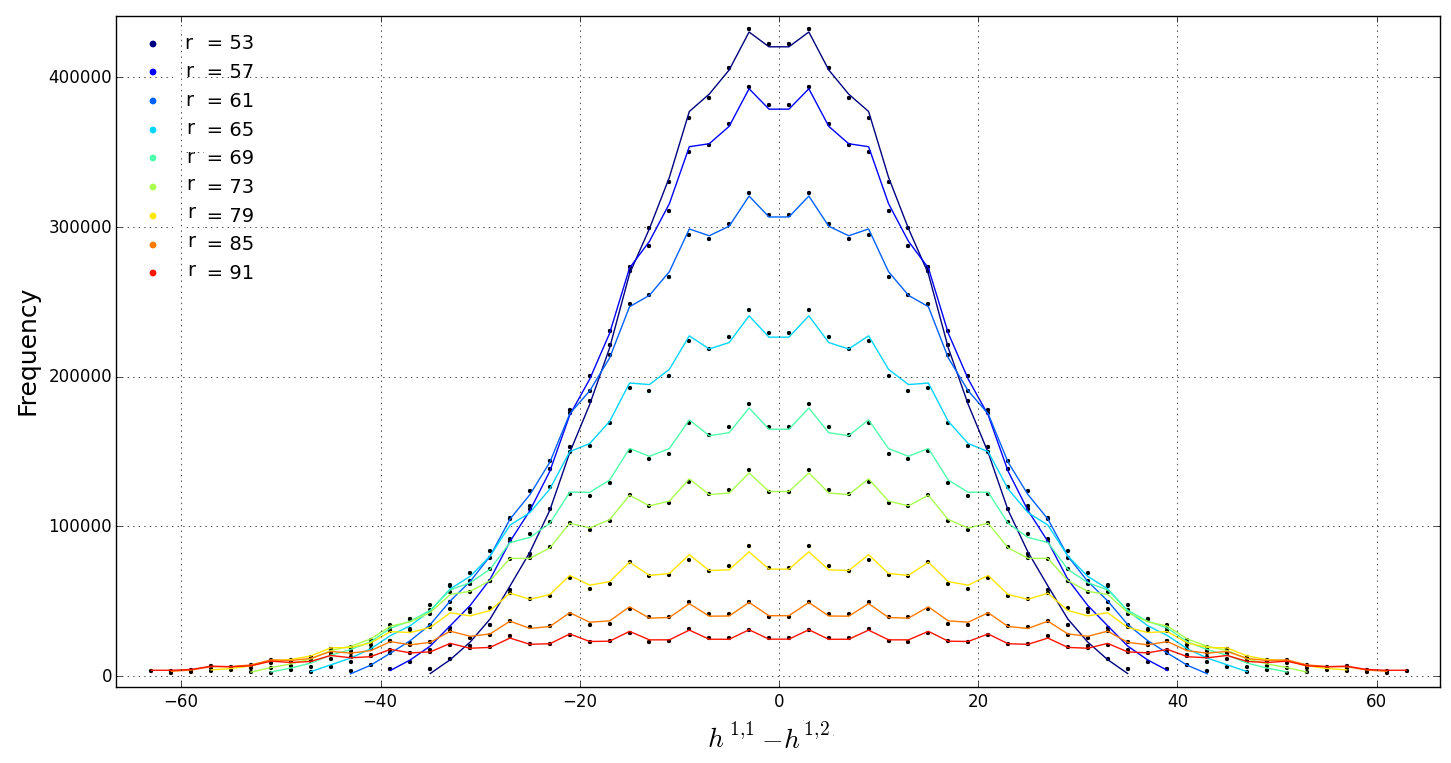}
         \caption{Regression lines for few select odd $r$ values, with $r>51$.}
        \label{Fig:OddHdiff2}
    \end{subfigure}
    
    \caption{Plots of frequency against $h^{1,1}-h^{1,2}$ for various odd values of $r$. Each line represent a modified pseudo-Voigt profile based on the regression analysis for each curve. See \protect\ref{fig:AllPVEven2} for a plot of all even curves.}
    \label{Fig:OddHdiffs}
\end{figure}

\begin{figure}[H]
    \centering
    \begin{subfigure}[h]{\textwidth}
        \includegraphics[width=\textwidth]{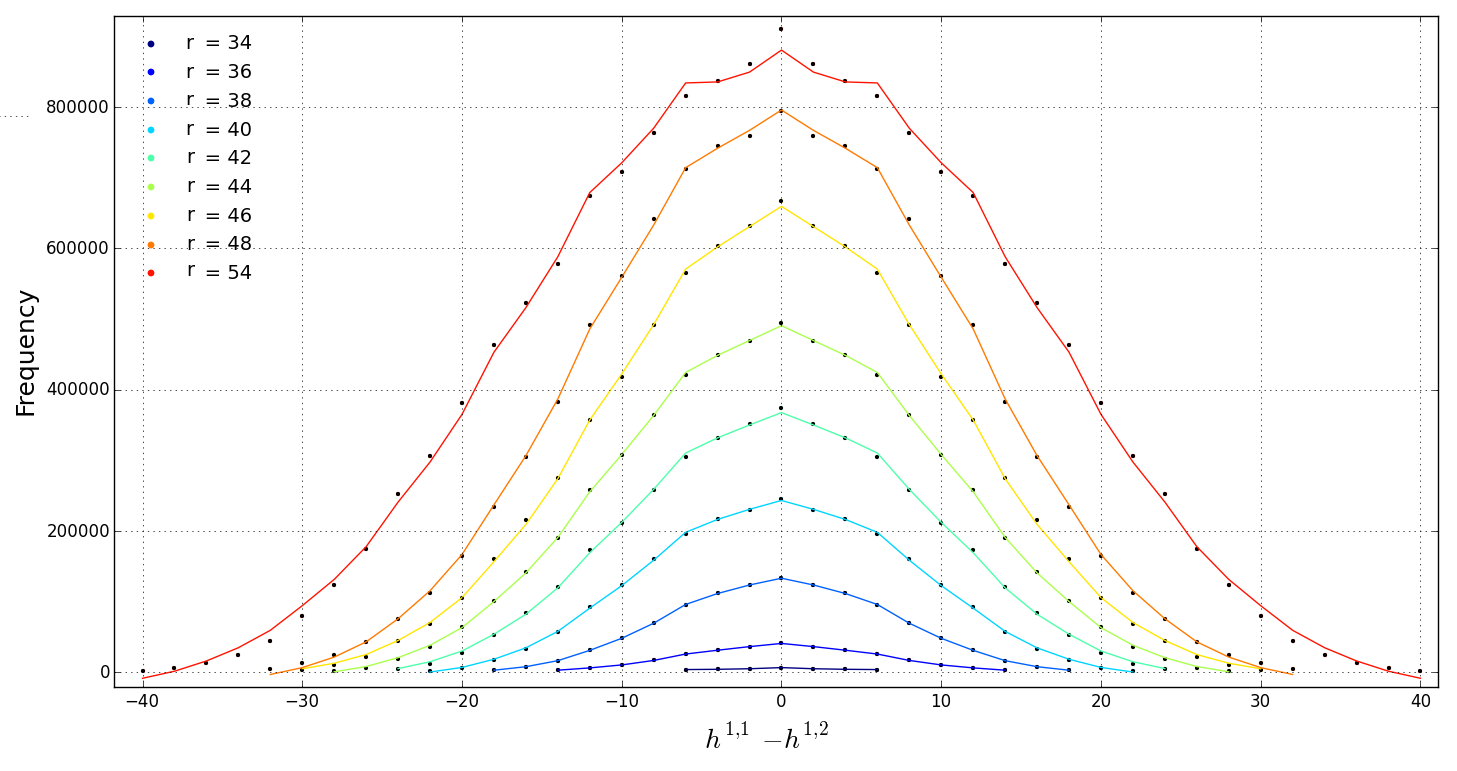}
        \caption{Regression lines for few select even $r$ values, with $r\leq 54$.}
        \label{Fig:OddHdiff1}
    \end{subfigure}

    \begin{subfigure}[h]{\textwidth}
        \includegraphics[width=\textwidth]{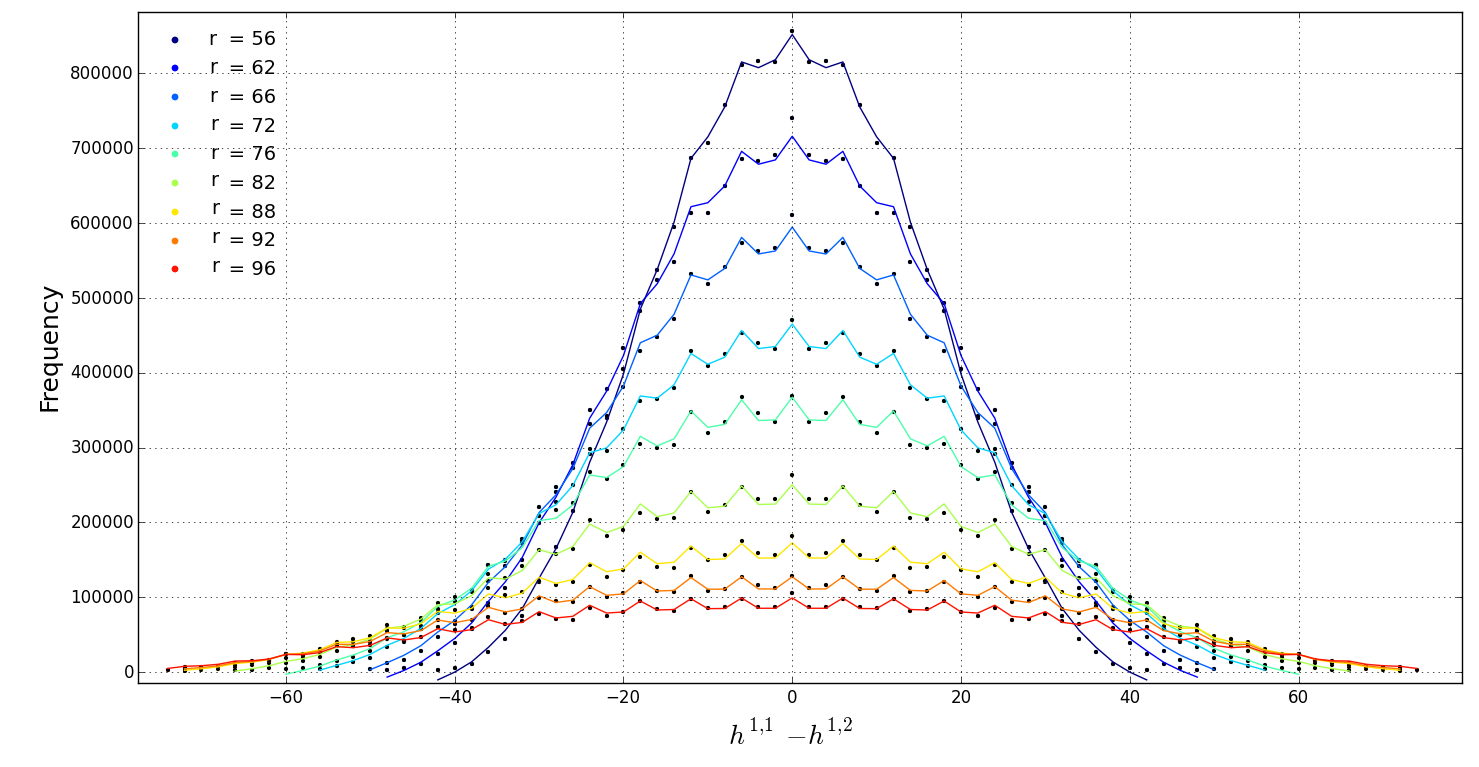}
        \caption{Regression lines for few select even $r$ values, with $r>54$.}
        \label{Fig:OddHdiff2}
    \end{subfigure}
    
    \caption{Plots of frequency against $h^{1,1}-h^{1,2}$ for various even values  of $r$. Each line represent a modified pseudo-Voigt profile based on the regression analysis for each curve.  See \protect\ref{fig:AllPVOdd2} for a plot of all odd curves.}
    \label{Fig:EvenHdiffs}
\end{figure}

We plot the frequency against $h^{1,1}-h^{1,2}$ for various values of $r$ (odd and even).
Figures~\ref{Fig:OddHdiffs} and \ref{Fig:EvenHdiffs} are striking in their accuracy.

As these figures illustrate, each curve follows a pseudo-Voigt profile, however the individual data points seem to ``jump'' up and down, as if oscillating.
It is this behavior of the data points which can be accounted for by the modified pseudo-Voigt model.
To do the regression analysis, we used Python \textit{lmfit} with a custom model which is just the modified pseudo-Voigt model.
The parameters that were fitted are $(A_0,a,b,\sigma,\alpha)$.
Due to mirror symmetry, $\mu = 0$.
In Appendix~\ref{ADiffPValues}, one can find a table with the value of every parameter for every curve as well as their reduced $\chi^2$ values.

\begin{figure}[ht]
	\begin{center}	
		\includegraphics[scale=0.45]{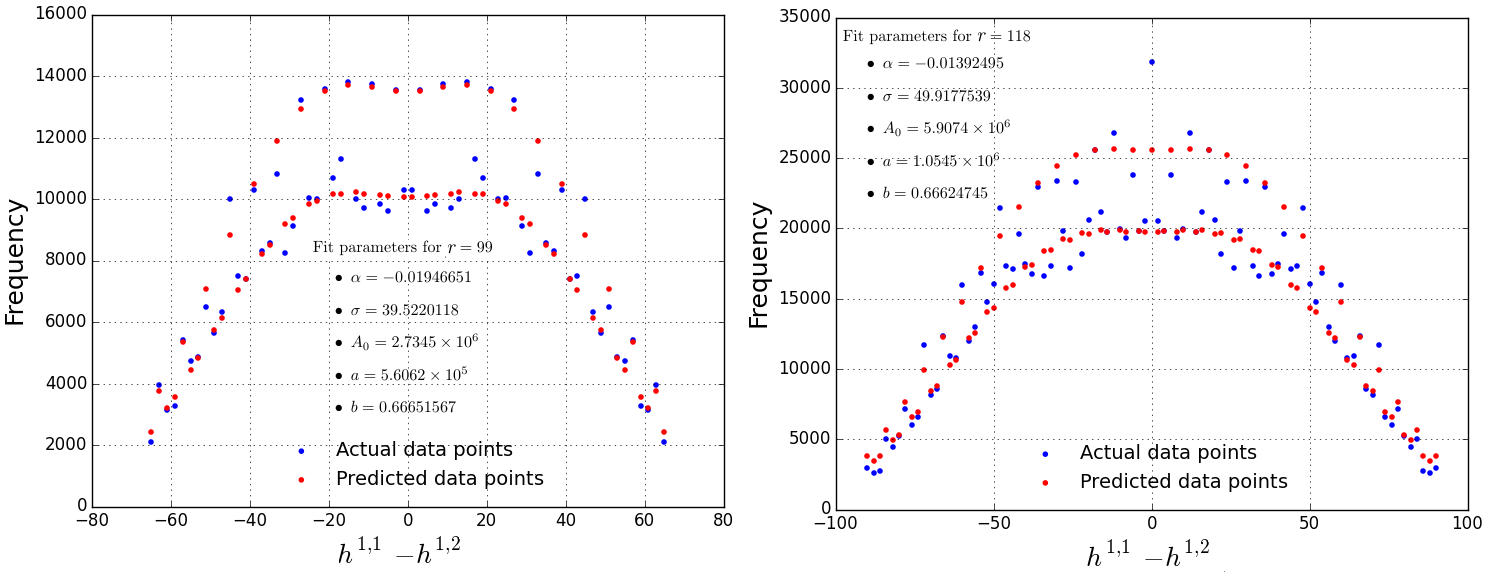}
	\end{center}
	\captionN{These two plots serve two purposes. The first is to show how the modeled data should really look by using data points (red points) instead of the (perhaps misleading) lines (refer to Comment~1 below).
	The second purpose is to illustrate that as $r$ becomes large (left plot has $r = 99$, right plot has $r = 118$), the actual data points deviate more and more from the modeled data, implying that there is a missing function in the modified Pseudo-Voigt model which would allow one to describe the data at much lower frequencies.}  	
	\label{Fig:LargePEvenOdd}
\end{figure}

A few comments explicate the regression lines and the behavior of the distributions.
\begin{enumerate}
\item[$1.$]
When we refer to the model as being an ``excellent fit,'' it is principally a statement made by inspection of the curves and the data.
If one inspects the reduced $\chi^2$ values (Figure~\ref{fig:ParamTableEvenOddHdiff}), the numbers are  large, which statistically does not refer to a good fit.
This is misleading however.
Firstly, we need to consider that the number of parameters used in the model is five.
This allows for a larger $\chi^2_R$ value.
Secondly, the distribution is based on a discrete set of data.
When doing a regression analysis using the modified Pseudo-Voigt model, one obtains an equation which describes a continuous curve. Lastly, the frequency values span over several orders of magnitude. The tiniest deviation from a parametric model --- in this case, the modified pseudo-Voigt profile --- will be detected in cases where there is such a huge sample size. Typically the predicted model gives data points which are in the range of 0.02\% to 3\% accuracy from the actual data point. The tail behavior of the model is less accurate however, here the predicted values can be off from between 60\% and 80\%. For cases with a very poor fit, the last data point (large value of $h^{1,1}-h^{1,2}$) can have an error of up to 300\% --- this is another example of the model being less accurate at lower frequency. When one is dealing with such sample sizes, even a 1\% error can give a difference of up to a couple of thousand. This difference summed over all the data points for a particular curve result in a large $\chi^2_R$ value. Due to the discussion in Section~\ref{GOF} we from now on ignore the $\chi^2_R$ as a test for model validation. Instead we opt for probability plots --- which can also be seen in Section~\ref{GOF}.

\item[$2.$]
One obtains a continuous model to describe the discrete data, in reality, we should not be plotting fitted curves, but rather fitted data points --- as can be seen  in Figure~\ref{Fig:LargePEvenOdd}. It is just illustratively more clear to display the curves. One could in principal work out what the discrete approximation is to our continuous model. 

\begin{figure}[H]
	\begin{center}	
		\includegraphics[scale=0.40]{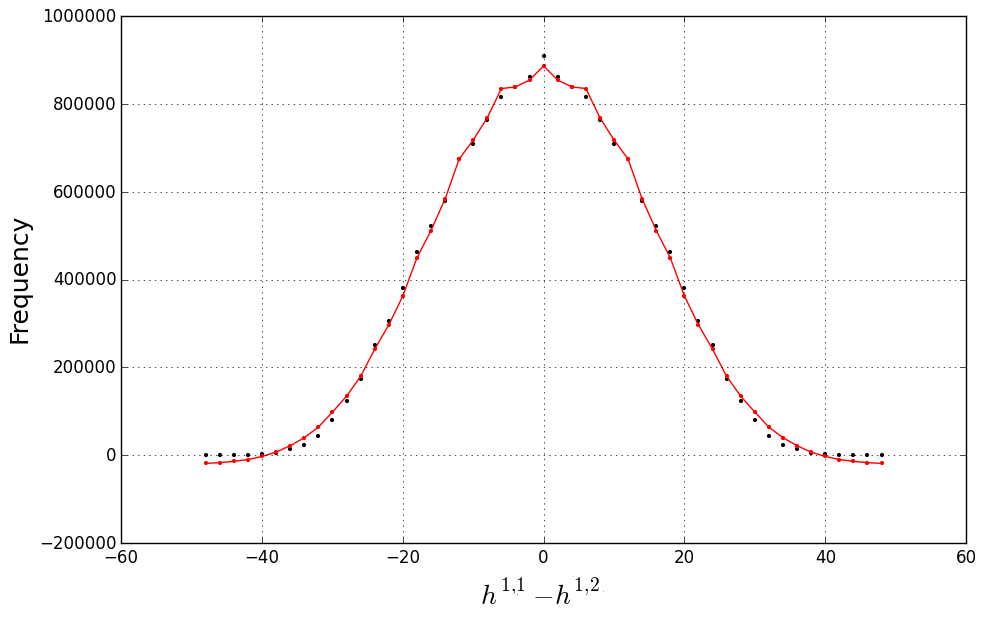}
	\end{center}
	\captionN{By considering the entire frequency range, the model is not able to adequately describe the tail behavior. The model goes into the negative frequency range instead of tapering off to 0. }
	\label{Fig:NegativeFit}
\end{figure}
\item[$3.$]
Although the modified pseudo-Voigt distribution does a good job to model the behavior of the data, one still needs to address the problems experienced with our model at low frequency. A problem which is hidden, by virtue of our cut-off frequency, is that the tail of our models predicts negative values, Figure \ref{Fig:NegativeFit}. There is a possibility that by having different variances $\sigma_g,\sigma_c$ for the mixing of the two distributions (Gaussian, Cauchy), one could adjust the tail behavior. Introducing more and more parameters however does not always resolve the problem, as it is possible to over-fit the data. Yes, the model may be more accurate, but one loses physical significance. In a situation like ours, where one does not have any physical backing for choice in models, this line between fitting and over fitting is not so clear.

\item[$4.$]
The odd distribution's behavior is more regular.
In comparison to the even distribution, as one increases in $r$ value, the behavior of the individual data points remain somewhat constant relative to the fitted curve.
The even distribution becomes more and more irregular as one increases the $r$ value.
This suggests that there is an added parameter which seems as if it should be function of $r$.
By regular and irregular we are referring to how well the data point is described by the model.

\item[$5.$]
Both distributions become very irregular as the value of $r$ becomes large ($r>100$ and $r>120$ for odd and even distributions respectively --- see Figure~\ref{Fig:LargePEvenOdd}).
A large $r$ value refers to curves which have a relatively low frequency.
Again this suggests that the Pseudo-Voigt model needs to some how have some function of $r$ which ``distorts'' the behavior of the curves as $r$ increases (by the looks of how the real data deviates from the modeled one, it seems that the missing functions is also oscillating in nature).
\end{enumerate}

There exist, however, certain cases where the model is exact.
In other words predicted values are the same as the actual values.
This happens when one adjusts the frequency cutoff for each $r$ curve individually.
That is to say, we only examine data points with at least $f_0$ reflexive polytopes with a given value of $r$ and $h^{1,1}-h^{1,2}$.
If there are fewer than $f_0$ cases, the data is ignored.

\begin{figure}[H]
	\begin{center}	
		\includegraphics[scale=0.42]{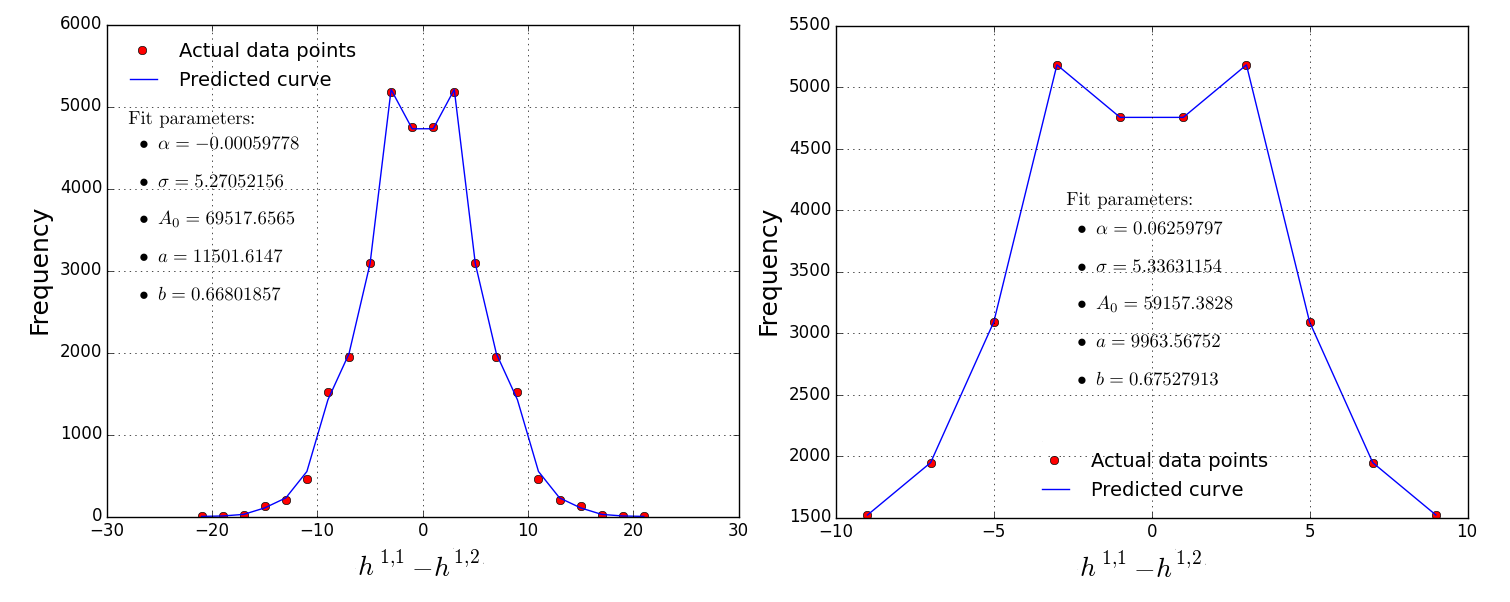}
	\end{center}
	\captionN{Left plot shows the modeled line according to the modified pseudo-Voigt distribution with no cutoff frequency. We obtain a good fit to the data. The right plot has a cutoff frequency of $460$, which is equivalent to a percentage cut off of $9.68\%$ (calculated relative to the peak frequency for that $r$-curve). This curve is exact.}  	
	\label{fig:ComparisonExact1}
\end{figure}

This trend persists for all values of $r$, however what becomes apparent is that it's not the percentage cutoff frequency that determines whether or not one gets an exact fit, but rather, the number of data points that remains after the percentage cut of has been effected. Figure \ref{fig:PerfectFits} gives a table of how many data points remain after an appropriate cut off percentage has been chosen to achieve a perfect fit. From this table we see that for even curves, one almost always requires 7 data points to achieve a perfect fit; for the odd curves, the number of data points is 10. The reason for this constant number throughout all the curves is that the centers of all the distributions for the various curves are all similar. As soon as one includes a larger number of data points we cannot achieve exact fits, and the model becomes approximate. At very low $r$ values the number of data points remaining after cutoff are not too different to the total number of points. As $r$ increase, the total number of points increase --- the fact that we can achieve exact fits becomes less meaningful. The other models --- even when including an oscillatory component were unable to give exact fits. 

The model is thus much more accurate at low $r$ values, and as $r$ increases the actual data deviates more and more from the fit.
This reinforces the statements from the comments that the pseudo-Voigt model can be modified further with some function $g(x,r)$ such that it will greatly improve the accuracy of the fit, and perhaps even become exact.

\begin{figure}[H]
  \begin{center}	
    (a) \includegraphics[scale=0.35]{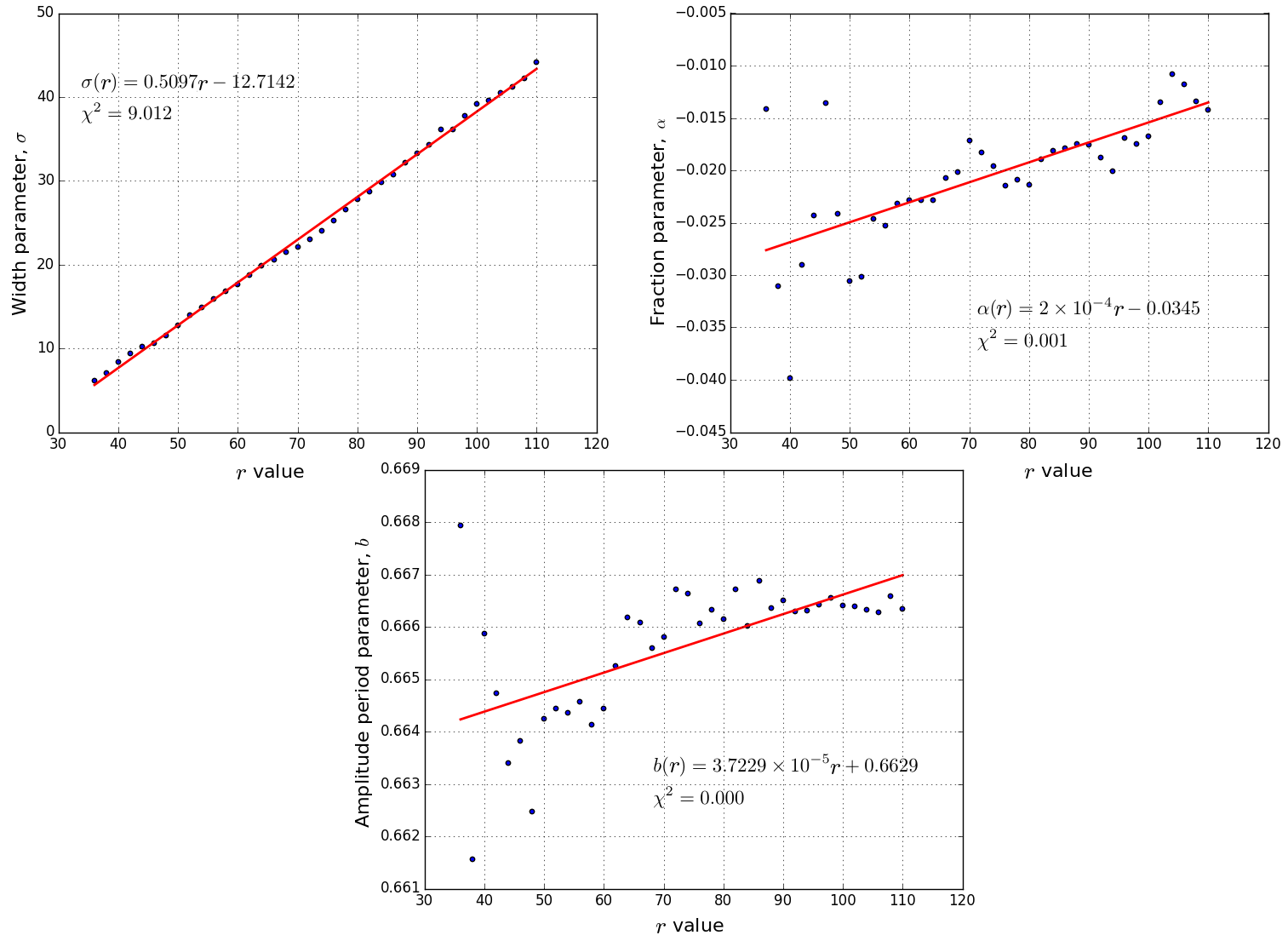}
  \end{center}
  \begin{center}
    (b) \includegraphics[scale=0.40]{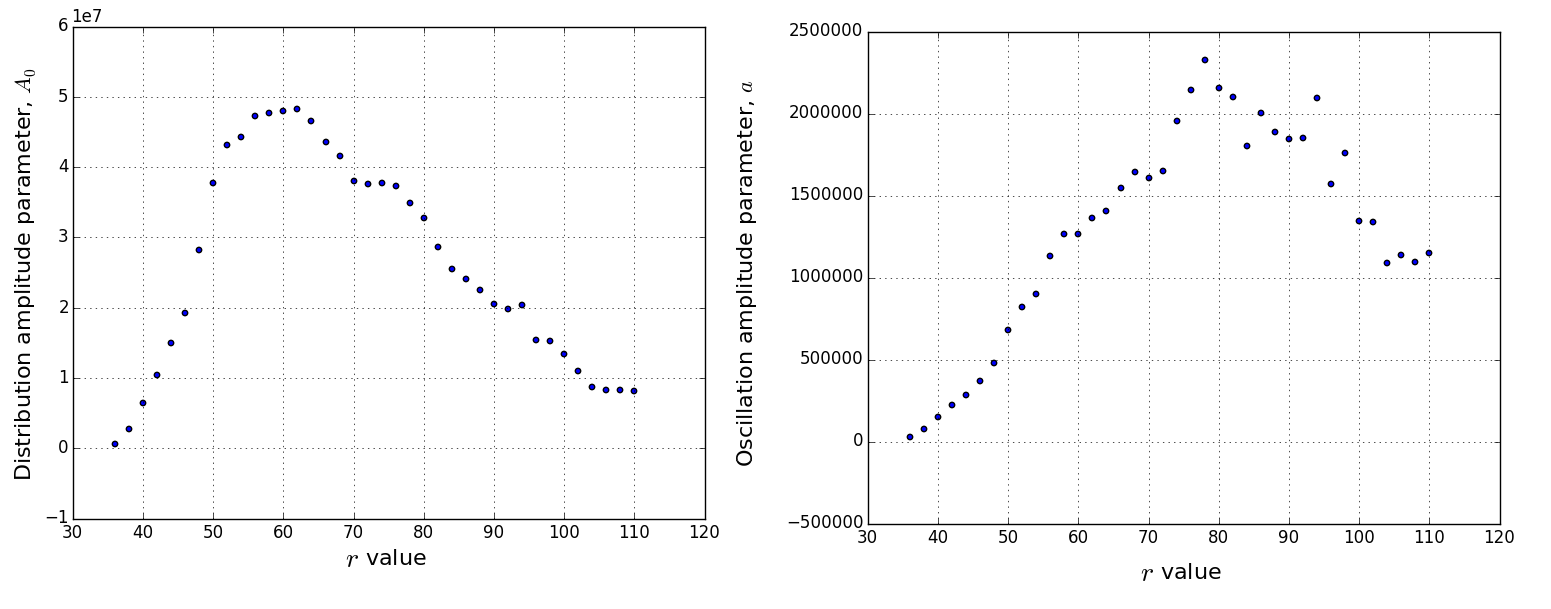}
  \end{center}
  \captionN{
    For the even distribution of $h^{1,1} - h^{1,2}$.
    (a)The width parameter $\sigma$ has a linear relationship with $r$ such that $\sigma(r) = 0.5097r - 12.7142$. The amplitude period parameter, $b$, also has a linear relationship, however, since $r$ is at most order $3$ in magnitude, we can regard it as a constant such that $b(r) = 0.6629 \sim 2/3$. The same goes for the fraction parameter,$\alpha$; we can regard it as a constant such that $\alpha(r) = -0.0345$. For odd parameter fit statistics see Figure~\protect\ref{fig:OddHdiffParam1};
    (b) Plots of $A_0$ vs $r$ (left) and  $a$ vs $r$ (right). Both exhibit a similar pattern, however it is difficult to discern any nice relationships. For odd parameter plots see Figure~\protect\ref{fig:OddHdiffParam2}.
    \label{Fig:EvenHdiffParamAll}}
\end{figure}

After the above analysis, we return to our goal of finding a single function describing the distributions.
It is clear from the above that the function has to be a function of at least two variable, $f=f(x,r)$.
We thus continue the analysis by plotting all the parameters vs $r$, in search for any relationships.
We find that three parameters $\sigma$,$b$ and $\alpha$ can be expressed in terms of $r$, the other parameters, while they show trends, do not give a precise relationship with $r$.
For the even distribution of $h^{1,1} - h^{1,2}$, the $r$ values range from $36$ to $110$, whereas for the odd distribution (see Figures~\ref{fig:OddHdiffParam1}, \ref{fig:OddHdiffParam2}) the $r$ values range from $37 $ to $99$.
By looking at Figure~\ref{Fig:EvenHdiffParamAll}(a), it turns out that:
\begin{align}
\alpha(r) = c_{\alpha} \ , \quad
b(r) = c_b \ , \quad
\sigma(r) = c_{\sigma_1}r + c_{\sigma_2}
\ . 
\end{align}
Our model of $h^{1,1} - h^{1,2}$ now looks as follows:
\begin{align}\label{eqn:PV}
f(x,r,A_0,a) = (1-c_{\alpha})&\frac{A_0(r) + a(r)\cos(2\pi c_b\cdot x)}{\sqrt{2\pi}(c_{\sigma_1}r + c_{\sigma_2})}e^{\frac{-(x)^2}{2(c_{\sigma_1}r + c_{\sigma_2})^2}} + \nonumber\\
&c_{\alpha}\frac{A_0(r) + a(r)\cos(2\pi c_b\cdot x)}{\pi}\left[\frac{(c_{\sigma_1}r + c_{\sigma_2})^2}{x^2+(c_{\sigma_1}r + c_{\sigma_2})^2}\right],
\end{align}
where $A_0(r)$ and $a(r)$ are two unknown functions yet to be determined (see Figure~\ref{Fig:EvenHdiffParamAll}(b) for relationship plots).
For replicating the plots as precisely as possible, one would need to keep the parameters, as they are, up to their $17$ decimal values, without excluding terms as we have done.
If one wants to reproduce the data from the model, one has to use the exact expressions.
Making an approximation from an already approximate model leads to large errors.

The first plot in Figure~\ref{Fig:EvenHdiffParamAll}.(a) in particular evinces a sinusoidal fluctuation about the mean.
This again indicates the possibility of refining the plots by adding an extra function.

\subsection{Analysis of $h^{1,1} + h^{1,2}$}\label{Hsum}

\begin{figure}[h]
	\begin{center}	
		\includegraphics[scale=0.43]{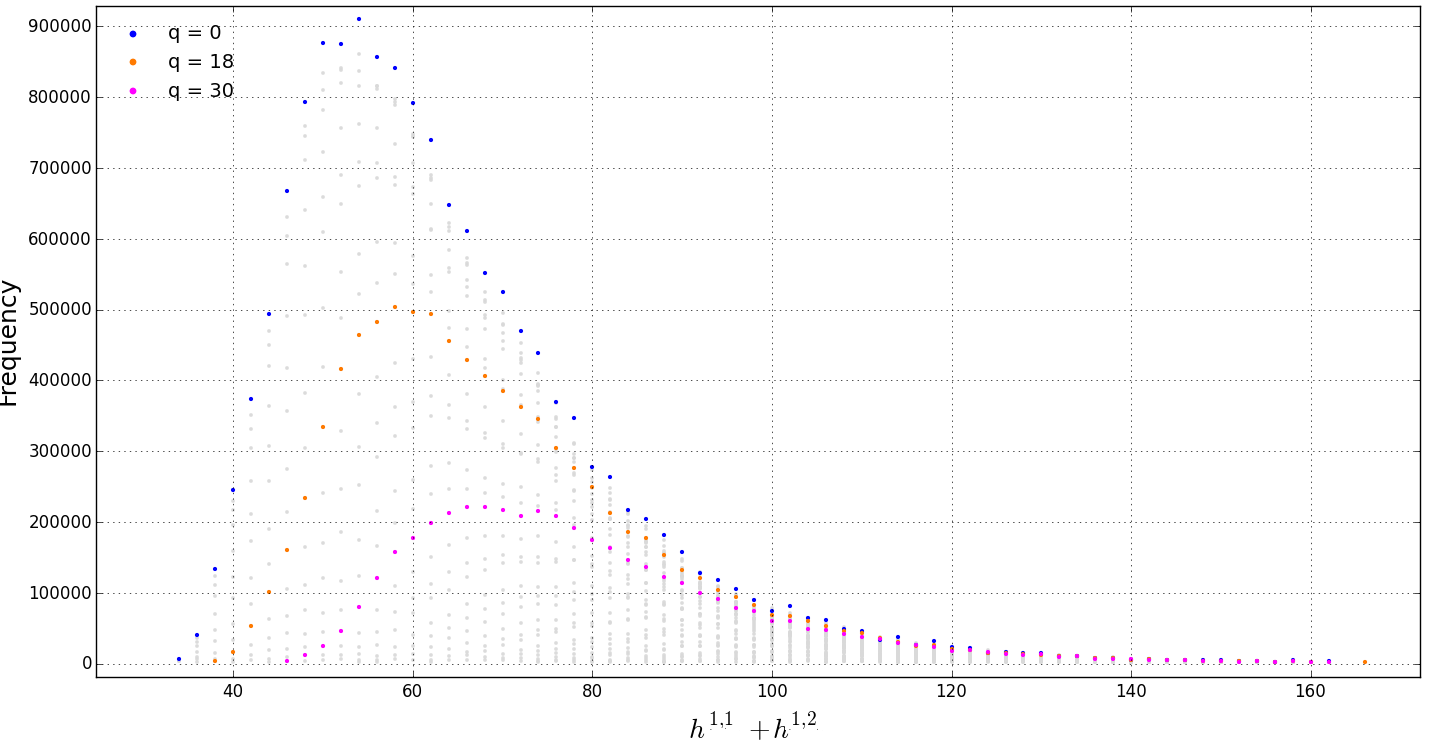}
	\end{center}
	\captionN{Three curves ($q = 0,18,30$) within the even $h^{1,1} + h^{1,2}$ distribution. The transparent grey data dots are all the data plots for the distribution. Refer to Figure~\protect\ref{fig:ExampleOdd1} to see the same example for the classification of odd curves within the odd distribution.}
	\label{fig:CurveExample}
\end{figure}

We begin by classifying the curves within the $h^{1,1} + h^{1,2}$ distribution (Figure~\ref{Fig:Main}) in an analogous way to how it was explained before.
This time, we order the data by $h^{1,1} - h^{1,2}$ such that a single curve within $h^{1,1} + h^{1,2}$ can be identified by its $q$-value, where $q = h^{1,1} - h^{1,2}$.
Due to mirror symmetry, the curve for $q=-a$ is the same curve as $q=a$, thus within our two-dimensional plots will only have $q>0$.
In continuation to the analysis on $h^{1,1} - h^{1,2}$, we use a cutoff frequency of $2000$ and only present results from the even distribution within $h^{1,1} + h^{1,2}$, unless stated otherwise.
As an example, illustrating the classification of curves within $h^{1,1} + h^{1,2}$, consider the curves $q = 0,18,30$ in Figure~\ref{fig:CurveExample}.

\subsubsection{A Planckian Fit}
Each curve within the $h^{1,1} + h^{1,2}$ distribution behaves the same.
Just like in the $h^{1,1} - h^{1,2}$ distribution, we do a regression analysis for each curve within the distribution independently, in the quest to describe the entire $h^{1,1} + h^{1,2}$ with a single function.
The model we chose to describe $h^{1,1} + h^{1,2}$ is the simplest possible Planckian model
\begin{align}\label{eqn:PlanckMod}
f(x,A,n,b) = \frac{A}{x^n}\frac{1}{e^{b/(x-22)}-1}
\end{align}
The parameter names in the fit results are the amplitude $A$, the power $n$, and some real constant $b$.
The shift in $x$-axis is so that the distribution begins at $0$ as the smallest $h^{1,1} + h^{1,2}$ above the cutoff is $22$.
The choice of a Planckian model in the above form is greatly motivated by the blackbody distribution $f(T,\lambda)$.
The $q$ curves within $h^{1,1} + h^{1,2}$ appear to behave in a manner analogous to the curves of constant $T$ within the blackbody distribution.
This is an initial trial.
Later, we will discover additional structure in the distribution by trying to mimic the blackbody distribution exactly.
It turns out that the general behavior of the distribution is modeled very well, \textit{cf.}\ Figure~\ref{Fig:CurvesEven1}.

Consider the maximum of each of the curves.
As indicated in Figure~\ref{Fig:CurvesEven1}, we can fit the maxima to a curve as indicated using the data plotted for the given values of $q$.
From the above analysis, the $h^{1,1} + h^{1,2}$ distribution behaves analogously to a blackbody spectrum --- except for one small subtlety.
It is in this subtlety that the added structure within $h^{1,1} + h^{1,2}$ is observed. 

\begin{figure}[H]
    \centering
    \begin{subfigure}[h]{0.9\textwidth}
        \includegraphics[width=\textwidth]{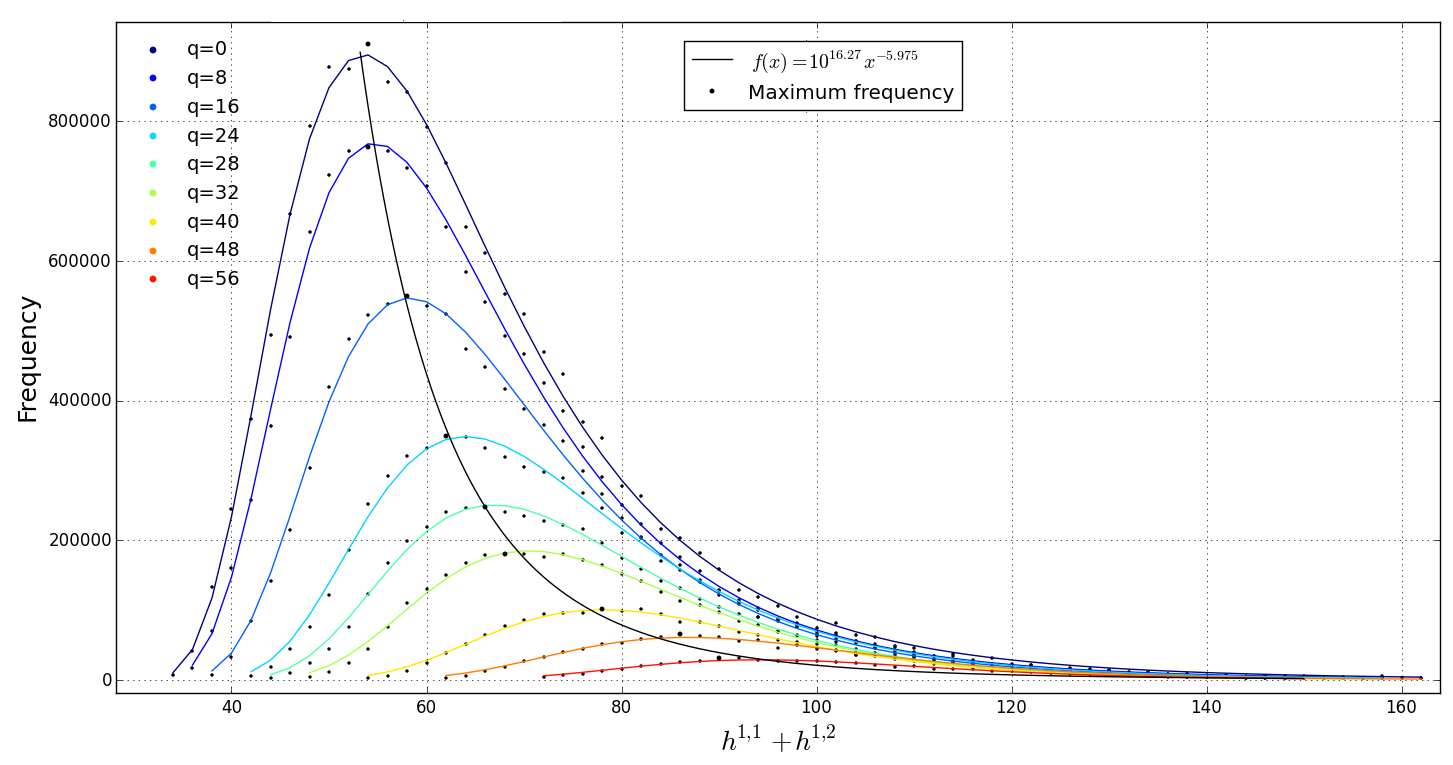}
        \captionN{Lines of best fit from a regression analysis for a few select curves. The black data points represent the maximum frequency for that particular $q$-curve. the Black line is a line of best fit to describe the points of maximum frequency --- this is analogous to a blackbody spectrum. See Figure~\protect\ref{fig:SeveralOddActual} for the curves within the odd distribution. }
        \label{Fig:CurvesEven1}
    \end{subfigure}
\end{figure}
\begin{figure}[H]
\ContinuedFloat 
  \centering 
    \begin{subfigure}[h]{0.9\textwidth}
        \includegraphics[width=\textwidth]{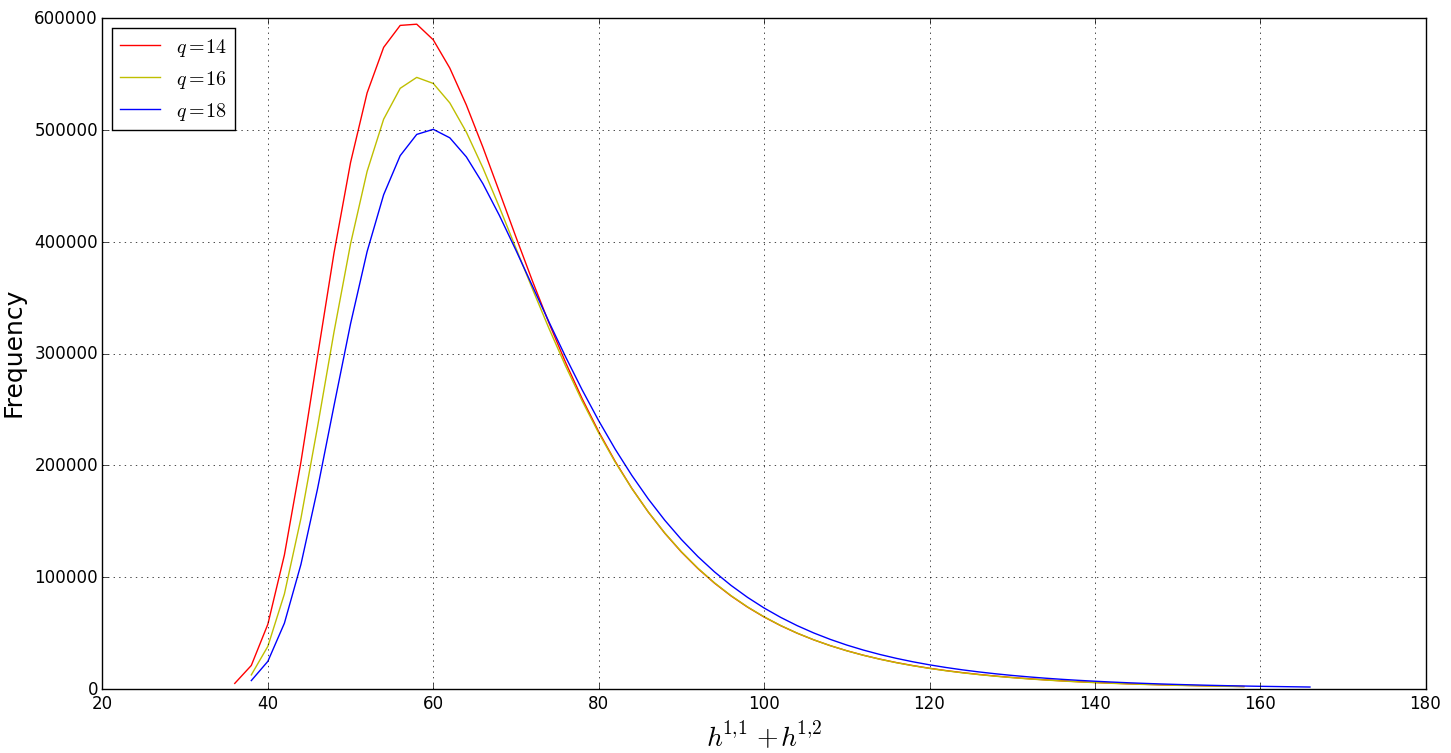}
        \captionN{The curves segregate into three classes determined by the value of the even integer modulo $6$. A similar pattern occurs in the odd distribution; see Figure~\protect\ref{fig:OverLapOdd}.}
        \label{Fig:Overlap1}
    \end{subfigure}
    
    \captionN{In the attempt to describe the data analogously to a blackbody distribution (a), we discover some subtle structure (b).}
    \label{Fig:BlackEven}
\end{figure}

Just as was seen in Figure~\ref{Fig:Main}, $h^{1,1} + h^{1,2}$ appears to split up into two smaller distributions based on the parity of $h^{1,1} + h^{1,2}$.
One can then further break up both the even and odd distributions into three further sets.
The manner we observed this added fine structure is again motivated by a blackbody spectrum.
In a true blackbody distribution, the curves of constant $T$ never overlap.
However, if you consider the lines of best fit only, when looking at our distribution one sees an overlap of certain curves.
For example, observe the following plot of curves which clearly cross in Figure~\ref{Fig:Overlap1}.

It turns out that this overlapping occurs consistently to the point where one can classify the curves (defined by their $q$ value) into residue classes $q_n$ distinguished by $n\ \mathrm{mod}\ 6$.
On the left hand side of the $h^{1,1}+h^{1,2}$ axis, the curves are ordered with red (residue class $q_2$) above yellow (residue class $q_4$) above blue (residue class $q_0$), whereas on the right hand side of the axis, the order is reversed.
Similar behavior is observed in the odd distribution of $h^{1,1} + h^{1,2}$ with the curves in the residue classes $q_1$, $q_3$, and $q_5$ (see Figure~\protect\ref{fig:OverLapOdd}).

\begin{figure}[H]
    \centering
    \begin{subfigure}[h]{0.9\textwidth}
        \includegraphics[scale=0.4]{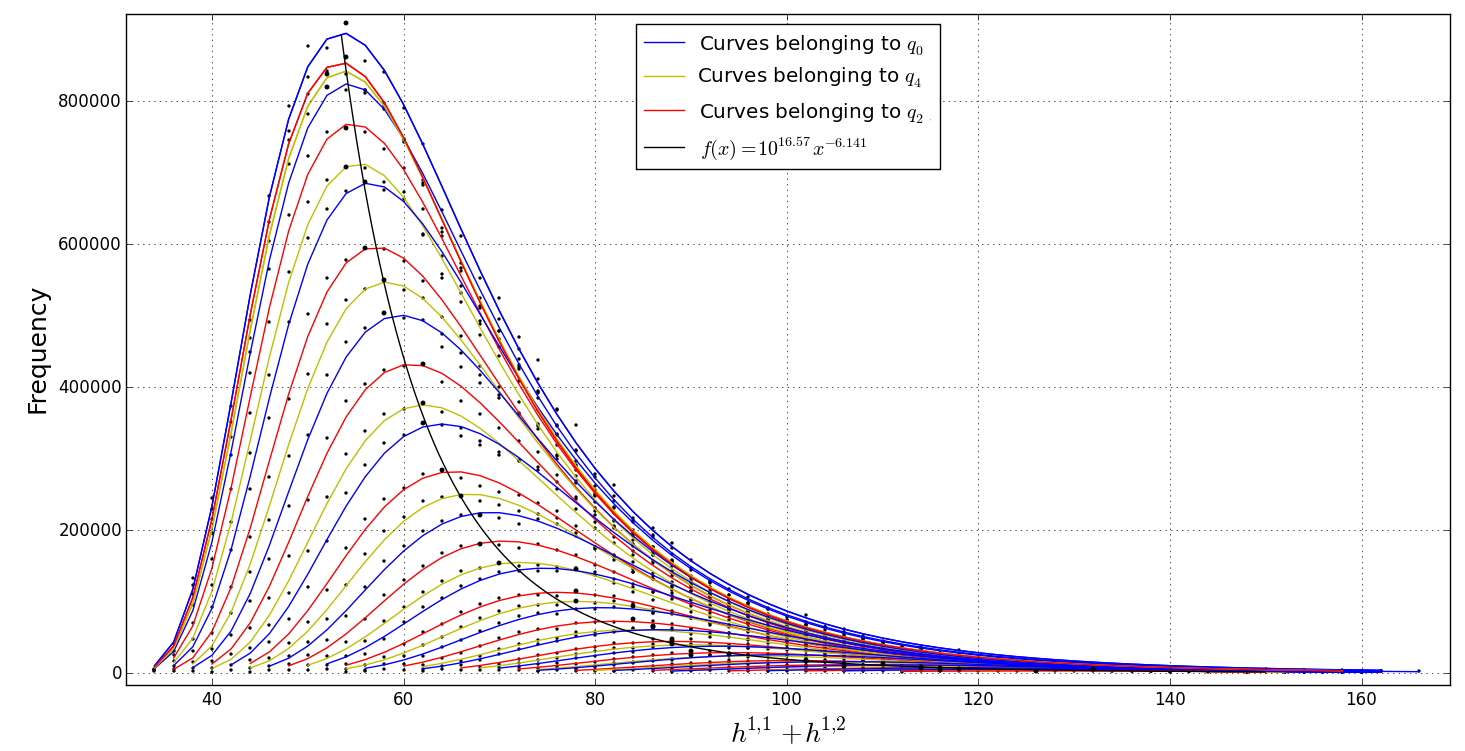}
        \captionN{All the curves color coded according to what residue class their curves $q_n$ belong to. }
        \label{Fig:OddHdiff1}
    \end{subfigure}

    \begin{subfigure}[h]{0.4\textwidth}
        \includegraphics[width=\textwidth]{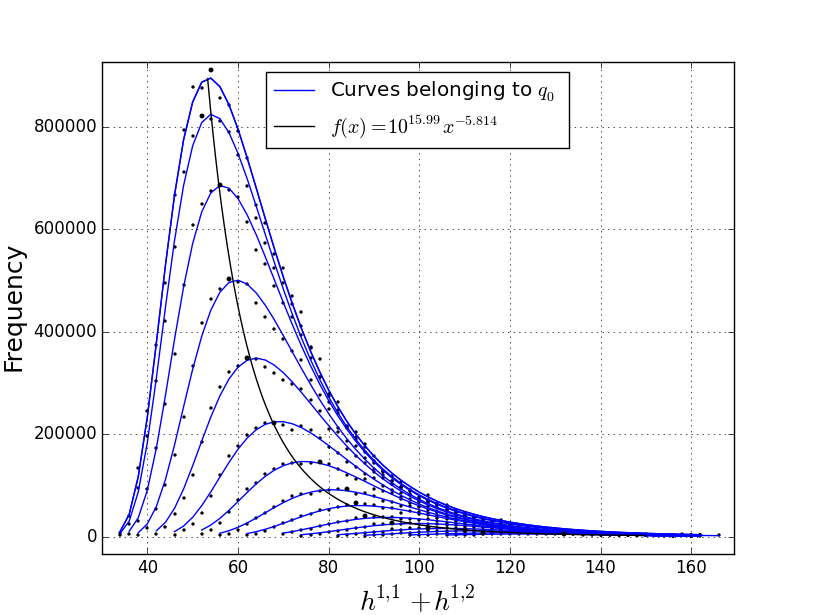}
        \captionN{Family of curves all belonging to $q_0$.}
        \label{Fig:Curvesq_0}
    \end{subfigure}
    ~
     \begin{subfigure}[h]{0.4\textwidth}
        \includegraphics[width=\textwidth]{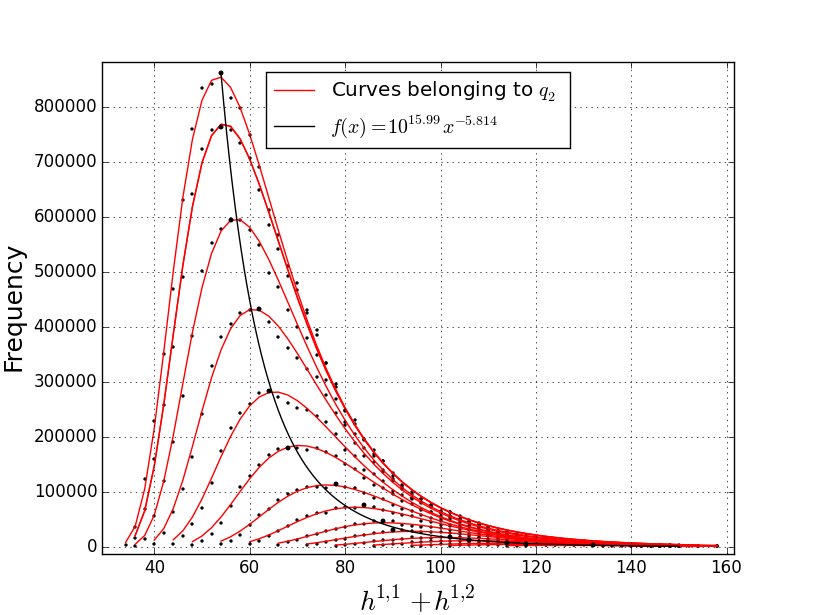}
        \captionN{Family of curves all belonging to $q_2$.}
        \label{Fig:Curvesq_2}
    \end{subfigure}

     \begin{subfigure}[h]{0.4\textwidth}
        \includegraphics[width=\textwidth]{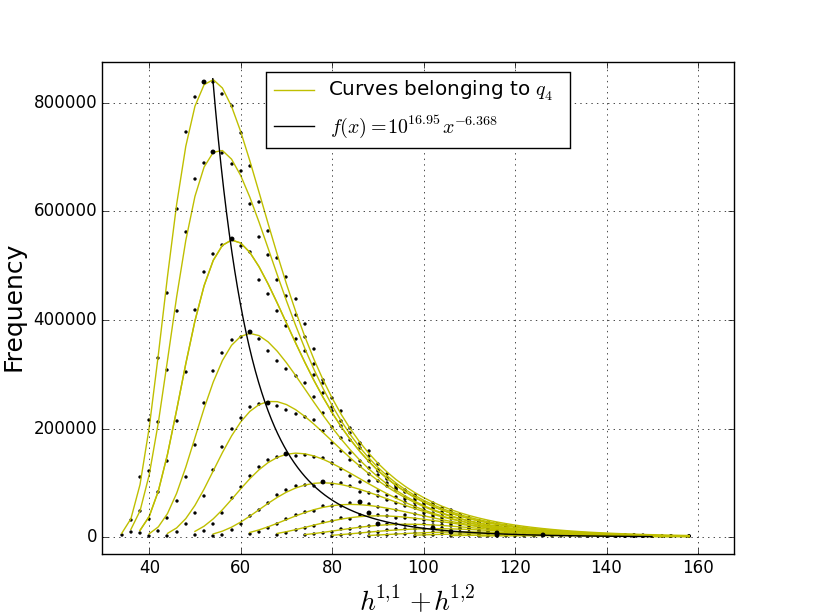}
        \captionN{Family of curves all belonging to $q_4	$.}
        \label{Fig:Curvesq_2}
    \end{subfigure}

    \caption{We illustrate the added structure for even $h^{1,1}+h^{1,2}$ data, by displaying how the regression curves can be divided into residue classes. For the list of odd curves, refer to Figure~\protect\ref{Fig:AllOddResidue}.}
    \label{Fig:AllEvenResidue}
\end{figure}

The clusters of curves constitute an entire set of mod $6$ residue classes.
These classes now define a set of curves which belong to  very ``nice'' distributions that behave exactly like a blackbody distribution.\footnote{
Of course $h^{1,1} + h^{1,2}$ is not continuous.
It is discrete.
However, the structure of the best fit curve to the data points appears very similar to that of a continuous blackbody distribution.}
Compare, for example, a plot of the all the curves for even distribution of $h^{1,1} + h^{1,2}$, separated into their residue classes, Figure~\ref{Fig:AllEvenResidue}

%Overlap2B.png

As a first approximation we have successfully modeled the general trend of the data.
There is, however, a fine structure to the individual data points that we would like to model.
Introducing an oscillating term in the amplitude, as seen in the analysis of $h^{1,1} - h^{1,2}$, unfortunately did not seem to improve the fits.\\
Again, it appears that the least number of variables our functions can have is two, $f=f(x,q)$.
This function will be slightly different in the values of coefficients, depending on which residue class one is modeling.\\
Just as for $h^{1,1} - h^{1,2}$, we wish to express the parameters for the $h^{1,1} + h^{1,2}$ model \eref{eqn:PlanckMod} in terms of $q$. We therefore write $A=A(q)$, $b=b(q)$, $n=n(q)$ and seek to find expressions for the coefficients.

While the $x$-axis of $h^{1,1} + h^{1,2}$ has only positive $q$ values --- due to the fact the data points will overlap --- when plotting them against the parameter values, we also have to consider the negative values of $q$.
We present the various relationships (see Figure ~\ref{fig:AllOddHsumParam} for the plots for the odd distribution of $h^{1,1} + h^{1,2}$ analogous to Figure ~\ref{fig:AllEvenHsumParam}).

\begin{figure}[H]
    \centering
    \begin{subfigure}[h]{\textwidth}
        \includegraphics[width=\textwidth]{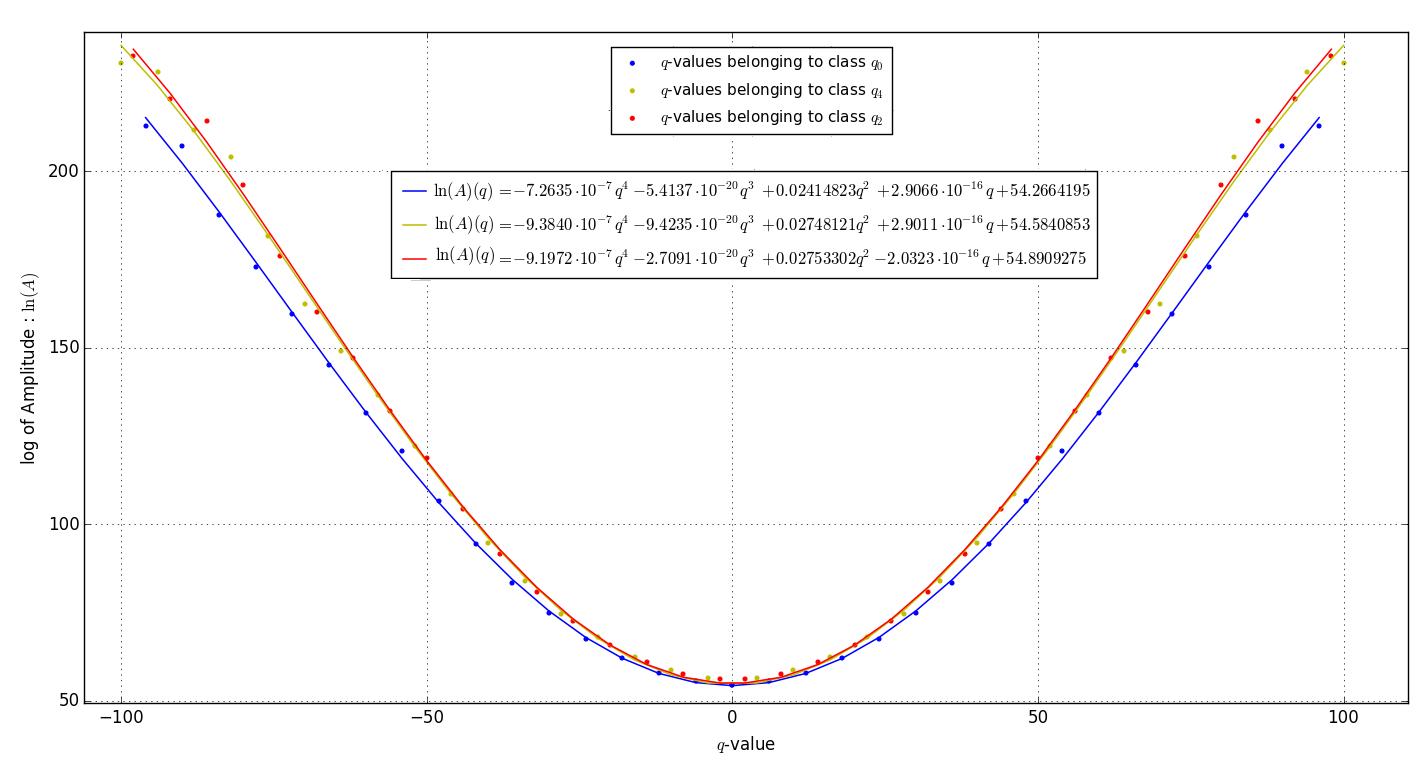}
        \captionN{Plotting the $q$- value parameter vs the $\log(A)$ parameter.}
        \label{fig:EParamLogA}
    \end{subfigure}
\end{figure}

\begin{figure}[H]
 \centering
 \ContinuedFloat
    \begin{subfigure}[h]{\textwidth}
        \includegraphics[width=\textwidth]{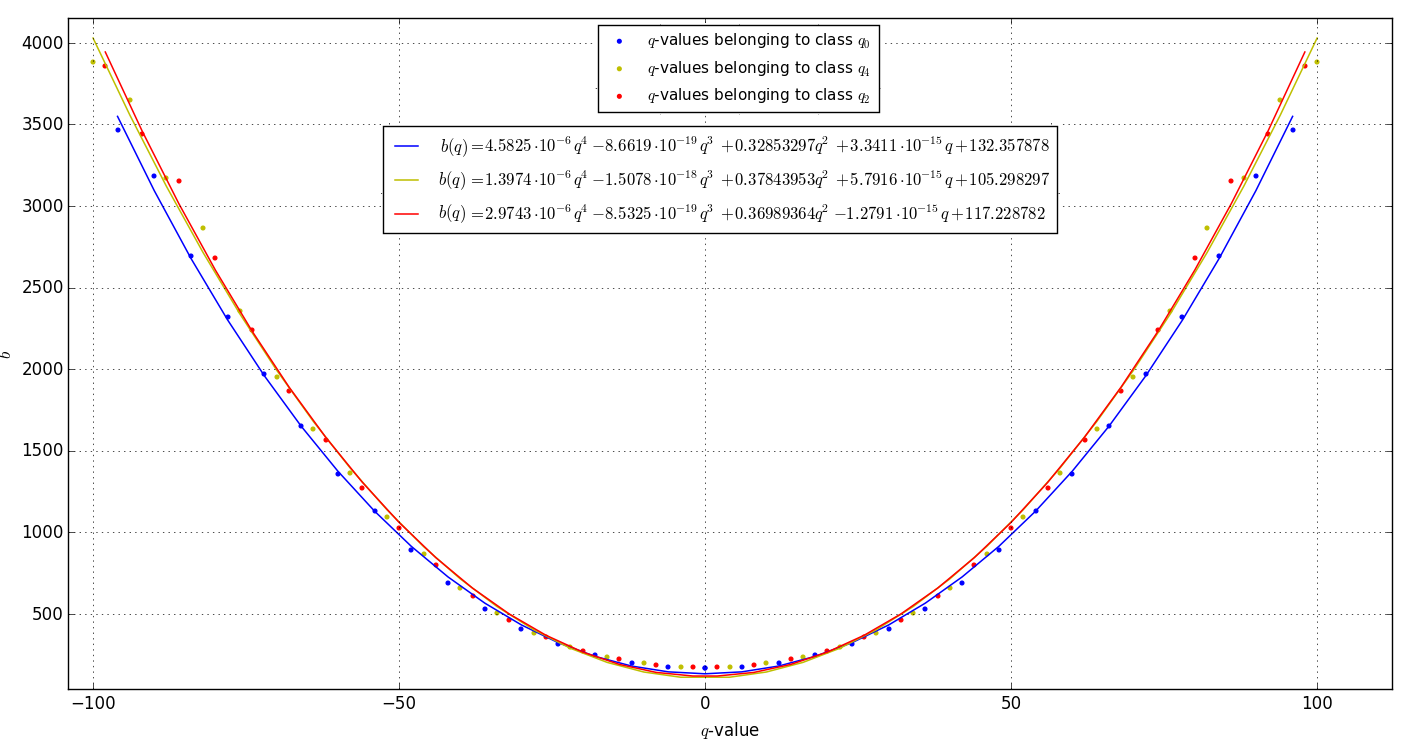}
        \captionN{Plotting the $q$- value parameter vs the $b$ parameter.}
        \label{fig:EParamB}
    \end{subfigure}

    \begin{subfigure}[t]{\textwidth}
        \includegraphics[width=\textwidth]{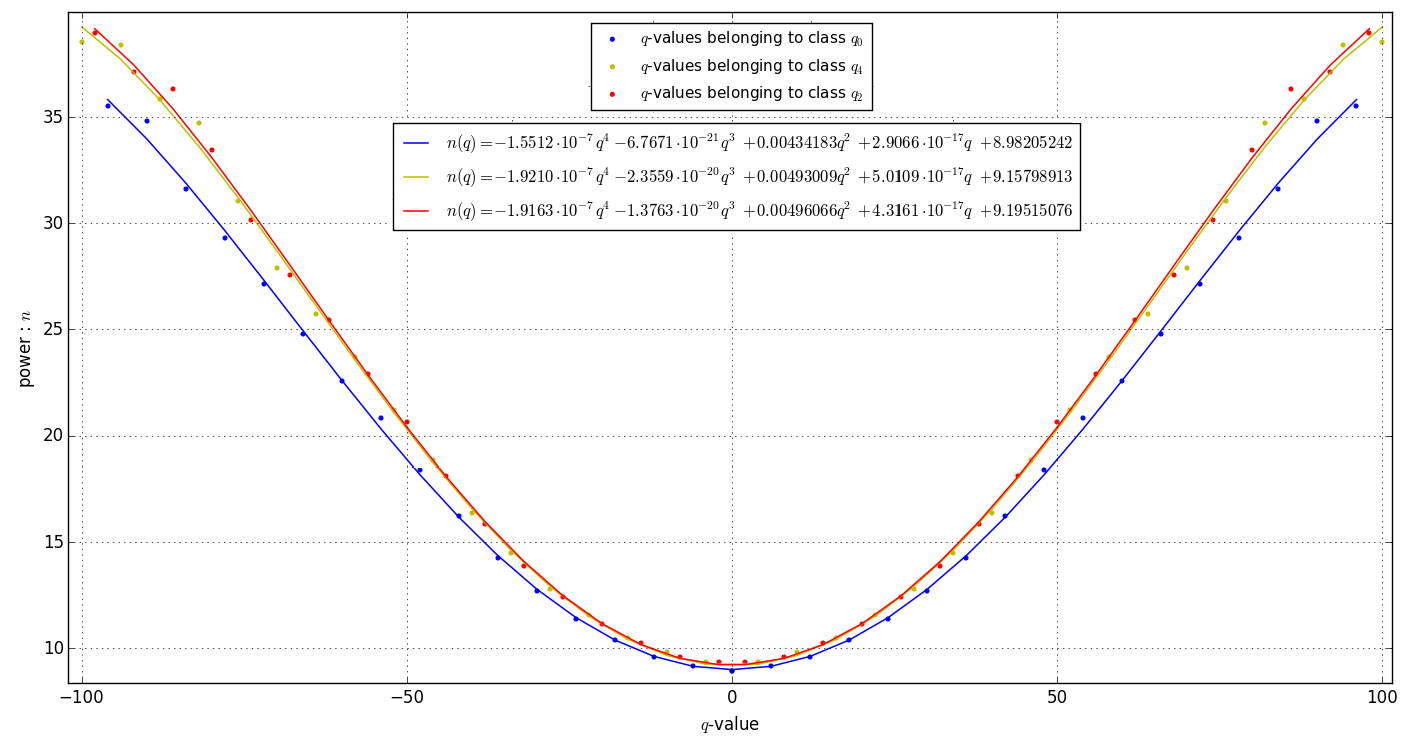}
        \captionN{Plotting the $q$- value parameter vs the power $n$ parameter.}
        \label{fig:EParamN}
    \end{subfigure}
    \captionN{The parameter plots are color coded according to what residue class their $q$ value belong to.}\label{fig:AllEvenHsumParam}
\end{figure}

Each distribution has an equation with different parameter values. However, the fact that we can express all the parameters in terms of $q$ means we are able to get a generalized formula to describe the entire $h^{1,1} + h^{1,2}$ distribution --- as long as the frequency is above $2000$. For succinctness we use the following notation for the coefficients
\begin{align}
A_{k,i}, \quad 
n_{k,i}, \quad
b_{k,i} \ ,
\end{align}
where the subscript $k=0,1,2,3,4,5$ refers to residue class $q_k$, and $i = 0,1,2,3,4$ refers to the coefficient of the $i^{th}$ power of $q$. Thus, we have:
\begin{align}
A_k(q) = \exp(\sum_{i=0}^{4}A_{k,i}q^i) ~,\quad
n_k(q) = \sum_{i=0}^{4}n_{k,i}q^i ~, \quad
b_k(q) = \sum_{i=0}^{4}b_{k,i}q^i ~,
\end{align}
where the matrix of coefficient values for $A_{k,i},n_{k,i}$ and $b_{k,i}$ can be found in Appendix \ref{AHsumTables}.\footnote{Perhaps it is important to state explicitly --- due to potential confusion --- that the coefficients $A_{k,i}$ refers to the natural logarithm of the amplitude values while $A_k$ is the actual amplitude seen in the model.}
Our function \eref{eqn:PlanckMod} now is able to approximately describe the entire $h^{1,1} + h^{1,2}$ distribution:
\begin{align}
f_k(x,q) = \frac{e^{\sum_{i=0}^{4}A_{k,i}q^i}}{x^{\sum_{i=0}^{4}n_{k,i}q^i}}\frac{1}{\left(e^{\frac{\sum_{i=0}^{4}b_{k,i}q^i}{(x-22)}}-1\right)},
\end{align}
Of course there are certain constraints on the values of $q$.
For a given $k$, $q$ has to be an integer which falls within the residue class $q_k$.
For even values of $k$, $x = 2m$, and for odd $k$, $x = 2m+1$.
We have $m>12$.

A few comments about the analysis on the $h^{1,1} + h^{1,2}$ distribution are in order.
\begin{itemize}
    \item[1.] The Planckian model used in \eref{eqn:PlanckMod} could be modified in some manner such that there is some oscillating behavior in the amplitude.
    Any kind of oscillatory term we introduce, only has a mild effect on the model's behavior.
    As the $q$ values exceed $100$, the model is not able to describe the data very well.\\
    \item[2.] Assuming one adds an oscillatory component to the model, the module used in python to do the regression analysis called \textit{lmfit} is sensitive to the initial conditions set by the user.
    Since the model is a custom model, it is difficult to find the correct initial conditions such that the best fit line oscillates close to every point (as with $h^{1,1} - h^{1,2}$).%
 \item[3.] It is possible that the model used does not have the features required to describe the oscillatory ``up and down" behavior of the data points.
 The Planckian model was chosen in that the $h^{1,1} + h^{1,2}$ distribution resembled a blackbody distribution.%
 \item[4.] In choosing a polynomial model for Figures \ref{fig:EParamLogA},\ref{fig:EParamB},\ref{fig:EParamN}, we picked the lowest order polynomial that gave the best fit. Choosing the order to be four for all the plots appeared to be convenient. However, it is apparent that the parameter relationship plot in Figure \ref{fig:EParamB} would be  better described by a polynomial of order 6. One could use an order 6 polynomial for all the other relationships plots too, but doing so might not have any physical significance. One can achieve an arbitrarily good fit the larger the order of the polynomial used, but that does not necessarily mean the chosen model is the correct model.
\end{itemize}

\begin{figure}[t]
	\begin{center}	
		\includegraphics[scale=0.45]{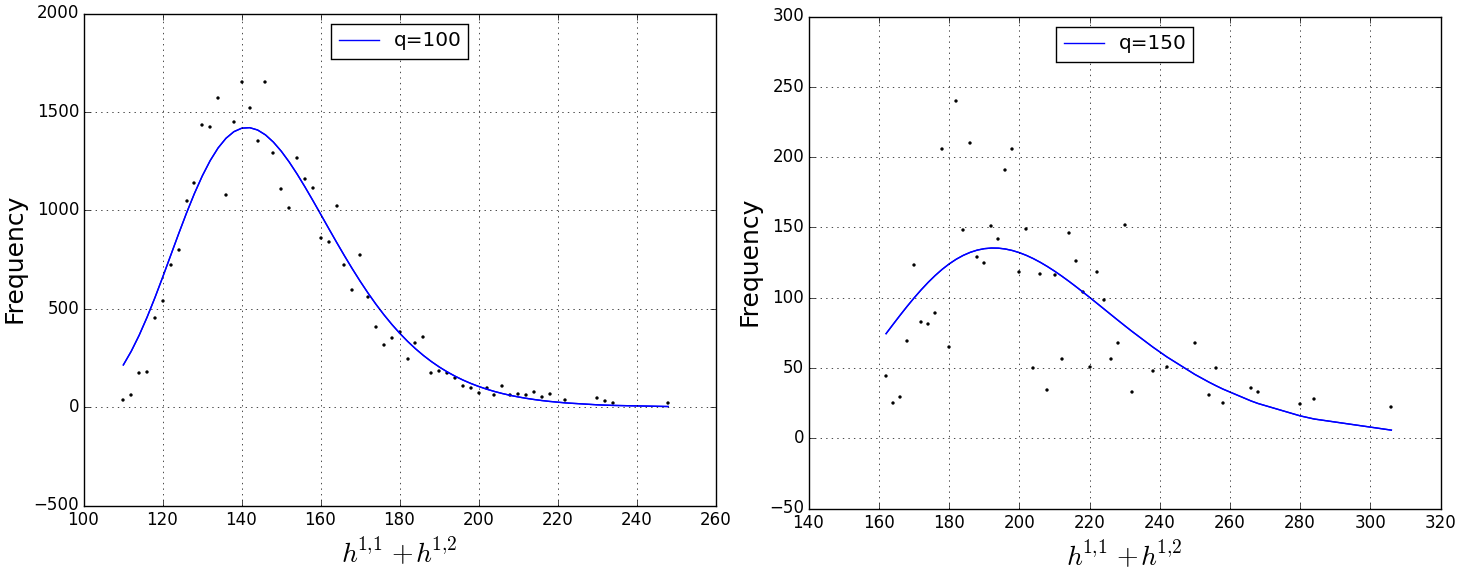}
	\end{center}
	\captionN{Left figure is the fitted model(blue line) for a $q$ value of 100 and right has a $q$ value of 150. As the $q$-value increases, the scattering of the data points within $h^{1,1} + h^{1,2}$ increases to the point where the model works no longer. For an example of how the model begins to break down at large $q$, see Figure~\protect\ref{fig:ExampleLargeOdd}.} 
		\label{fig:Highq1}
\end{figure}
\subsection{The Distribution of the Euler Number}\label{Euler}
The Euler number for Calabi--Yau threefolds is
\begin{align}
\chi = 2(h^{1,1}-h^{1,2}) ~.
\end{align}
As mentioned previously, we are summing over all the various $r$-curves to obtained the full-Euler number distribution. A plot of $\chi$ versus frequency yields the pseudo-Voigt distribution.
In particular, we can model the behavior of the distribution almost perfectly using the modified pseudo-Voigt curve \eqref{eqn:PV} and \eqref{eqn:modified}, which is repeated here for convenience:
\begin{equation}\label{eqn:PV}
f(x,A,\sigma,\alpha) = (1-\alpha)\frac{A}{\sigma\sqrt{2\pi}}e^{\frac{-(x)^2}{2\sigma^2}} + \alpha\frac{A}{\pi}\left[\frac{\sigma^2}{x^2+\sigma^2}\right] ~,
\end{equation}
where
\begin{align}\label{eqn:modified}
A(x,A_0,a,b)= A_0 + a\cos(2\pi b\cdot x) ~.
\end{align}
The results of the regression analysis for the Euler number distribution is presented in Figure~\ref{fig:EulerBoth}.

\begin{figure}[H]
    \centering
    \begin{subfigure}[h]{0.9\textwidth}
        \includegraphics[width=\textwidth]{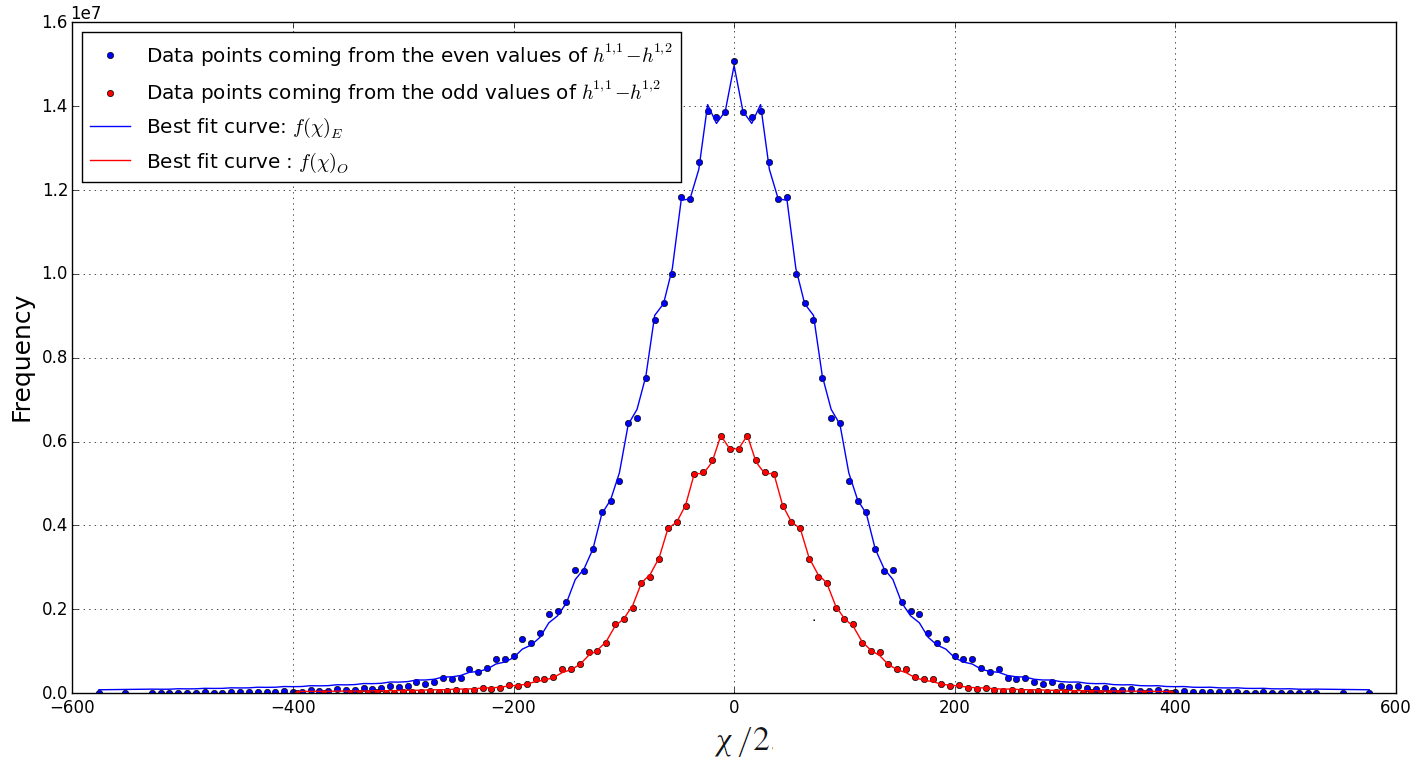}
        \captionN{The distribution of Euler numbers fitted to a modified pseudo-Voigt curve. The blue curve $f(\chi)_E$ represents even values of $\chi/2$. The red curve $f(\chi)_O$ represents odd values.}
        \label{fig:EulerBoth}
    \end{subfigure}

    \begin{subfigure}[h]{0.9\textwidth}
        \includegraphics[width=\textwidth]{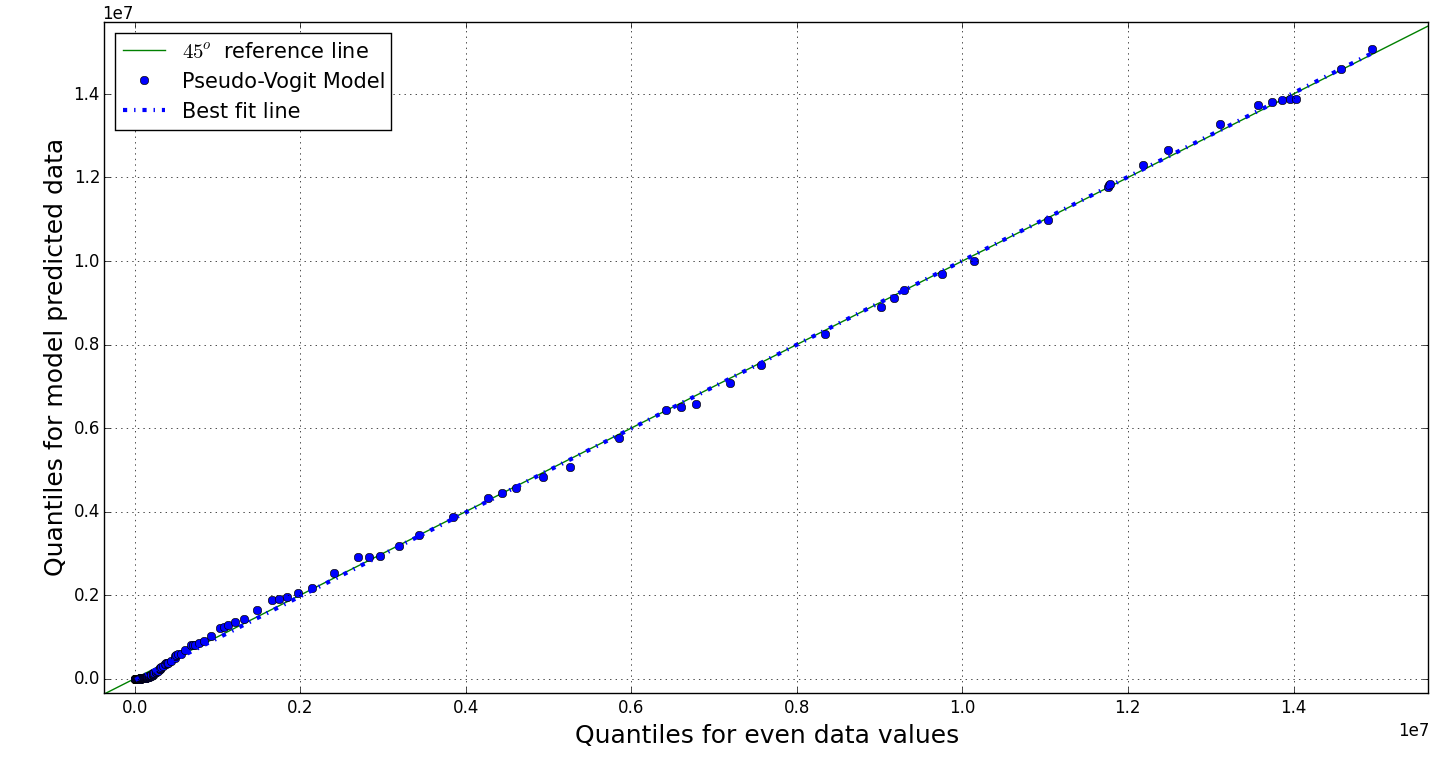}
        \captionN{Probability plot for the even values of $\chi/2$. The model fits the data with $R^2 = 0.99944$.}
        \label{fig:qqEven}
    \end{subfigure}
\end{figure}

\begin{figure}[H]
  \ContinuedFloat 
  \centering 
    \begin{subfigure}[h]{0.9\textwidth}
        \includegraphics[width=\textwidth]{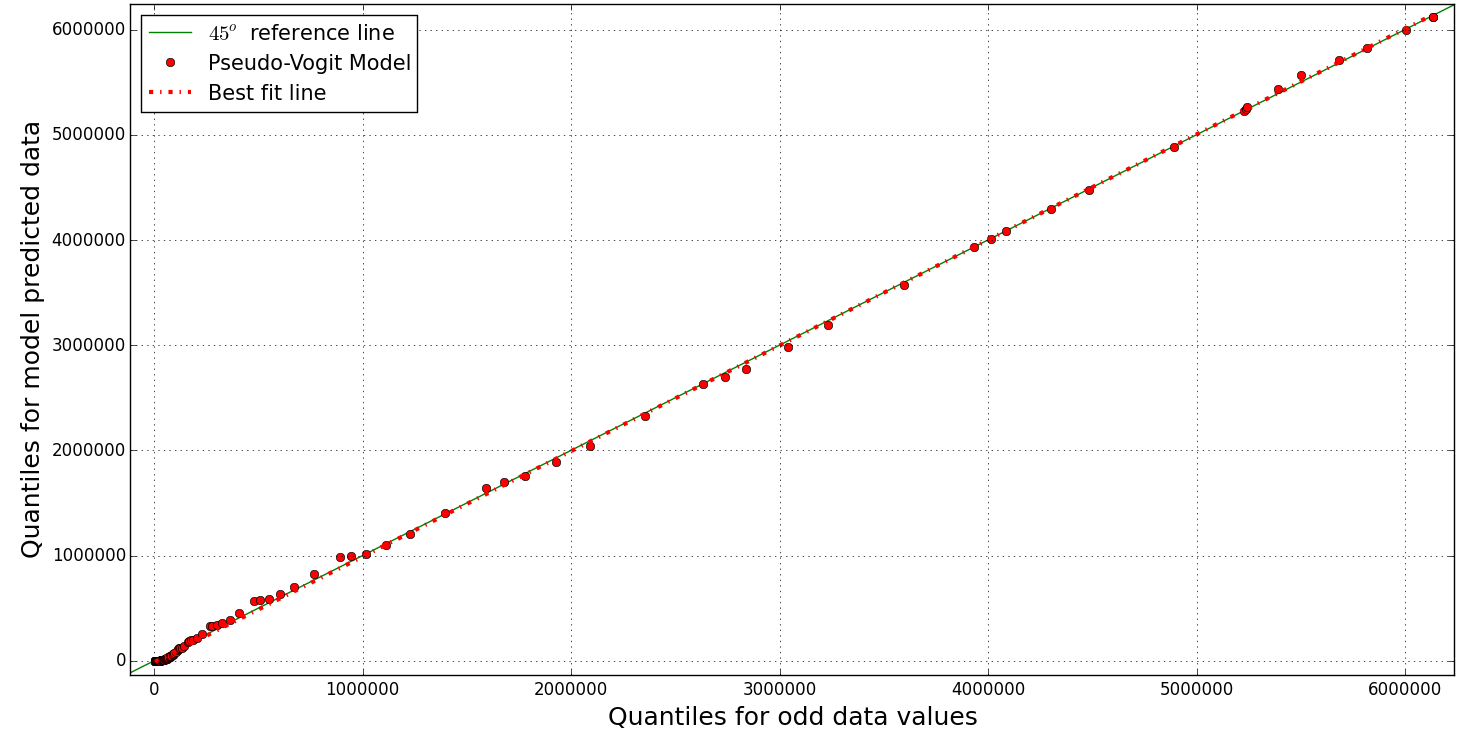}
        \captionN{Probability plot for the odd values of $\chi/2$. The model fits the data with $R^2 = 0.99965$.}
        \label{fig:qqOdd}
    \end{subfigure}
    \captionN{Various plots illustrating the actual fit of the modified pseudo-Voigt model. We can tell we have a good fit by looking at the probability plots for the quantiles of the standard pseudo-Voigt distribution vs quantiles for the actual data. The. The $R^2$ values in (b) and (c) are given relative to the line $y=x$.}\label{fig:qqOdd}
\end{figure}

The fitted parameter values for $f(\chi)_E$ corresponding to even values of $h^{1,1}-h^{1,2}$ are:
\begin{align}
  (A_0,\sigma,\alpha,b,a) = 
(1.9032\times 10^{9},
 75.8305889,
 0.00718459,
 0.58347826,
 8.7427\times 10^{7}) ~.
\end{align}
Likewise, the fitted parameter values for $f(\chi)_O$ corresponding to odd values of $h^{1,1}-h^{1,2}$ are:
\begin{align}
  (A_0,\sigma,\alpha,b,a) = 
  (7.6043\times 10^{8}, 64.9735680,0.00549425,0.83357720,3.6881\times 10^{7})
  \ .
\end{align}
Although $\chi$ is only even, the two curves originate from the fact that if you take $\chi/2$ you get even and odd values. The two curves arise from the parity of $\chi/2$ and are presented in Figures \ref{fig:EulerBoth}.

\subsection{Goodness-of-fit}\label{GOF}
A goodness-of-fit test is implemented as a means of testing how well a given model describes some given data. Typically the model validation process consists of only quoting a single statistically generated number like the $R^2,\chi^2$ or $p$ values. Based on the size of this number, one then makes inferences on how well the chosen model fits the observation. One needs to be careful however of misusing such indicators as an absolute measure for assessing goodness-of-fit. 

For a structural equation model (SEM) --- in our case, the modified pseudo-Voigt and Planckian models --- this assessment is not so straight forward as it would be for a simple regression analysis. To quantify the predictive power of an SEM, a single statistical test does not suffice - in fact, there is no single test. According to \cite{StatsTextBook}, the best one can do is assess three different aspects of what it means to have a good fit, these are: overall fit, comparative fits to a test model and model parsimony.\footnote{Parsimony refers to the ability of a model to give a certain degree of fit whilst having the least required number of predictor variables.} The only real test available is the chi-squared ($\chi^2$) test, when it comes to overall fit, this $\chi^2$ statistic is the most popular test. The $\chi^2$ test compares observed and predicted correlation matrices with each other, and so, statistical significance is evaluated based on the value of $\chi^2$. A large $\chi^2$ value signifies a considerable difference between the correlation matrices. A low value indicates there is little statistical difference between matrices. Since the $\chi^2$ test is between actual and predicted matrices only, when looking for overall fit, one searches for non-significant differences between the correlation matrices. Often, rather than presenting the $\chi^2$ or $\chi^2_R$ (the chi-squared value relative to the degrees of freedom for the model) value, a $p$ value is given instead. The $p$ value, in a way, informs us whether one should reject a null hypothesis or not. A small $p$-value suggests that the differences in observed vs. predicted are too large to be consistent with the null-hypothesised model i.e. assuming the null-hypothesised model, the probability of observing what we did is relatively small, suggesting either an absolutely fluke experimental outcome or an incorrect model null-hypothesis. The $p$-values can be determined by a $p$-value calculator by inputting the $\chi^2_R$ value. There is no standard way of choosing a significance level for the $p$-value, but typically $p<0.05$ is considered statistically significant.

In general, statistical non-significance given by appropriate values of the $\chi^2$ fit statistics is adequate. However, one must be careful of drawing similar conclusions for structural equation modeling. The fit statistic makes a statement of the correlation matrices only, not about whether or not the correct model is identified. This is largely due to the sensitivity to sample size of the $\chi^2$ test. In our analysis, the sample size (number of reflexive polytopes) is enormous --- almost one billion! For large samples ($>200$) the $\chi^2$ test will give significant differences for any model used. This sensitivity to a sample size, together with an \textit{effect size} and \textit{alpha value}, is related to what one calls the power of a test - the probability of not incorrectly accepting a null hypothesis that is actually false.

Without worrying too much about what an effect size and alpha value is; for any alpha value, the greater the sample size, the greater the power of the statistical test. However, increasing the sample size beyond a certain amount, can result in the test having ``too much" power\footnote{Power is the probability that you do detect deviations from your null-hypothesised model, when the null-hypothesised model is, in fact, incorrect}. Perceived effects in very large sample sizes, will always become significant\footnote{Conversely is also true, for extremely small sample sizes, any effect which should be significant, becomes insignificant}. Observe how in tables \ref{fig:ParamTableEvenOddHdiff} and \ref{fig:ParamOddEvenValues} the $\chi^2_R$ values for all the different curves is extremely large, naively indicating that we have a horrible fit --- which would be an incorrect conclusion.

It is clear from the above discussion that we cannot use the $\chi^2$ or $p$ values in validating our choice in model. What is not so clear, is the additional subtlety in using purely statistical means to asses goodness-of-fit for our data. This subtlety lies at the heart of almost all statistical tests --- the construction of a null hypothesis. The term frequency, as used in the statistical sense, refers to the number of outcomes for a certain event. The measurement of this outcome will often have certain known or unknown factors affecting it. These tests check for the probability that the errors found are too significant to be solely do to random variations in the data. For example, assume that statistical tests give non-significant results. If the residuals are small enough to be considered random errors in the measurement of the frequency, we could say that the model is appropriate. If however, the residuals are too large or present additional structure, we could say the model is good, but not quite the correct one as the residual errors are not ``random enough". In our case, there is no notion of measured frequency and error in measurement of frequencies. Our frequencies are generated as a result of a combinatoric calculation. Statistical tests assume that the input is from measurement and observations (obeying some null-hypothesis), thus they are inherently constructed with this notion in mind. By inputting our data, the tests are trying to calculate something from a data set which does not obey the very assumption they use in their calculations. We are not exactly clear how much this affects statistical outcomes, but it is important to keep in mind.

How do we validate then, that our chosen models are a good fit, or that our model is the best one at describing the data? We implement graphical methods. The first graphical method is obviously through pure inspection --- this is not quite statistically quantifiable. There is a statistically based graphical method to asses goodness-of-fit called probability plots, Q-Q plots or P-P\footnote{A P-P plot is the plot of the cumulative distribution frequency of the one data set against the CDF of the other. P-P plots are not as useful as Q-Q plots, thus are seldom used.} plots. These plots were initially constructed to test the ``normality" of a data set when the sample size is too large too depend on the $\chi^2$ and $p$ values. In principle, a standard probability plot tells you the likelihood that the a sample's distribution of data obeys a normal distribution --- hence checking for normality. The answer to the question is not given by a statistical value, but rather by a graphical representation --- from which one can extract statistical numbers. If the plotted data on this probability plot is a straight line, then we can determine that the sample set is normally distributed. 

We can extend this concept further: we can take two different samples, and take a probability plot to determine if two data sets come from populations with a common distribution. Such a probability plot is referred to as a Q-Q (quantile-quantile) plot. Extending this concept one more time --- as for our use --- we will take the quantiles of our theoretical distribution (the modified pseudo-Voigt and Planckian profiles) as our ``first sample" and plot them against the quantiles of our data as our ``second sample", this will give us our probability plot.  In all the probability plots, it is the quantiles of the respective data sets which are plotted against each other. 

Quantiles are basically just a generalization of quartiles. For example, the $k^{th}$ percentile of a set of values divides them, such that the number of values which lie below is $k$\%, and the number of values which lie above is $(100-k)$\%. The $25$th percentile is the lower quartile or the $\frac{1}{4}$ quantile. Quantiles are the same as percentiles, but indexed by sample fractions rather than by sample percentages. Suppose that $p\in[0,1]$, the aim is to find the value that is the fraction $p$ of the way through the ordered data set. As an example, if $p=\frac{1}{2}=0.5$, we want to know what is the value that sits at $p=0.5$ of the way through i.e. half way. The value that sits there (this value may have to be interpolated) will be called the quantile for the fraction $p=0.5$. There are many different algorithms for generating the quantiles for a given data set, we use python to generate the quantiles in a manner similar to that discussed above. For an ordered data set, $x_{1}\leq x_{2}\leq x_{1}\ldots \leq x_{n-1}\leq x_{n}$, the most common way of calculating quantiles is to first compute the empirical distribution function:
\begin{align}
F(x) = \frac{1}{n}\sum_{i=1}^n = 1(x_i\leq x),~~~ x\in \mathcal{R},
\end{align}
and then define the quantile function to be the inverse of $F(x)$:
\begin{align}
F^{-1}(p) = min\lbrace x\in \mathcal{R}:F(x) \geq p,~p\in(0,1)\rbrace.
\end{align}
By generating the quantiles of some theoretical model and comparing them to the quantiles of a given data set of equal length, one can determine if the data set belongs to the same distribution as the data set belonging to the theoretical model --- \textit{i.e.}, does the data fit the model. If the quantiles are roughly equal the plots will all be more or less on a straight line. 

In probability plots :
\begin{enumerate}
\item The length of data set needs to be equal. For unequal lengths, one must perform an interpolation of data.
\item If two identical data sets were compared to one another, the points would lie exactly on a $45$ degree line. Thus, for two different data sets, the deviation from this reference line determines the likelihood that the sets belong to similar distributions. To quantify this likelihood, one can calculate the $R^2$-value of the data, relative to the $y=x$ reference line.
 
\item Q-Q plots are not only limited to determining similarity in data sets. By analyzing the deviations which occur, one can determine how the scale and location of the data is shifted - the data would follow some line $y=mx +c$, where $m,c$ would be the estimates of these shifts in scale and location. Also, from the distribution of points above or below the reference line, one can infer aspects of the tails and skewness in the data.
\end{enumerate}

\begin{figure}[H]
    \centering
    \begin{subfigure}[h]{0.9\textwidth}
        \includegraphics[width=\textwidth]{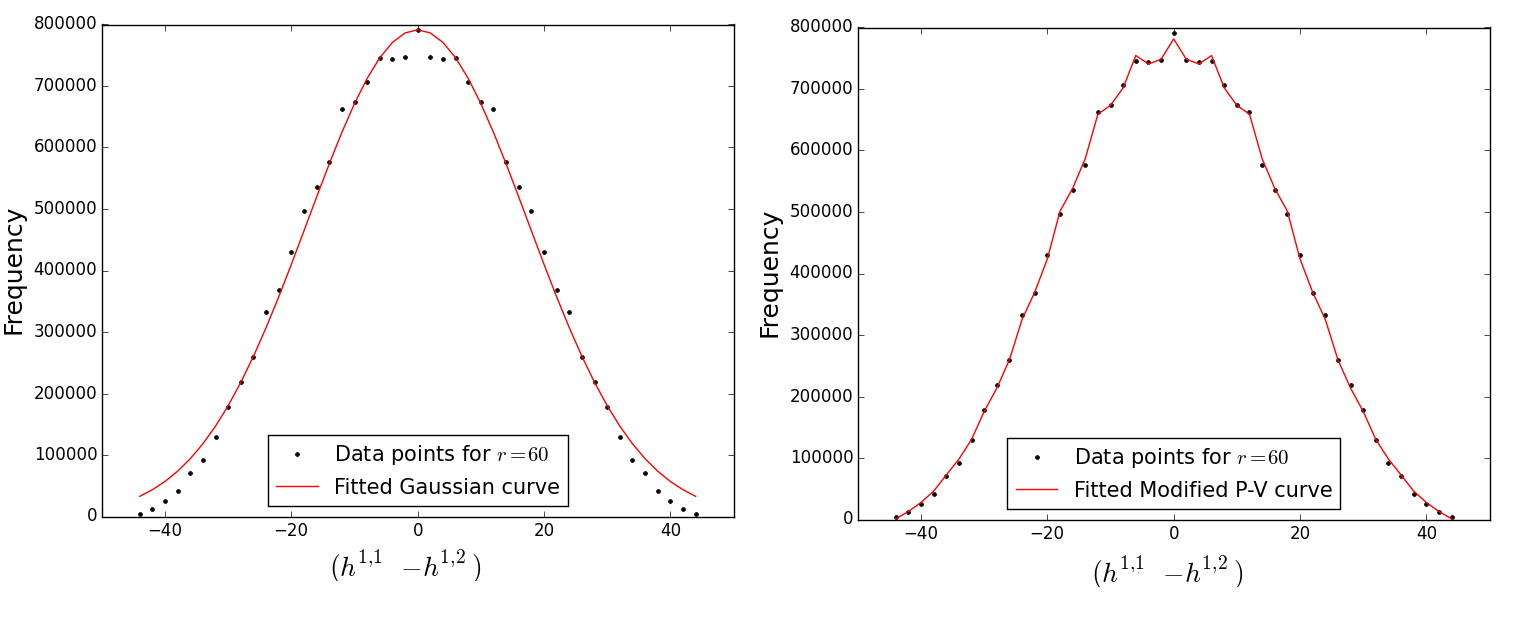}
        \captionN{Best fit curve for $r = 60$ based on the left: Gaussian model,  right: modified pseudo-Voigt model.}
        \label{fig:GGaussVSmpv}
    \end{subfigure}

    \begin{subfigure}[h]{0.9\textwidth}
        \includegraphics[width=\textwidth]{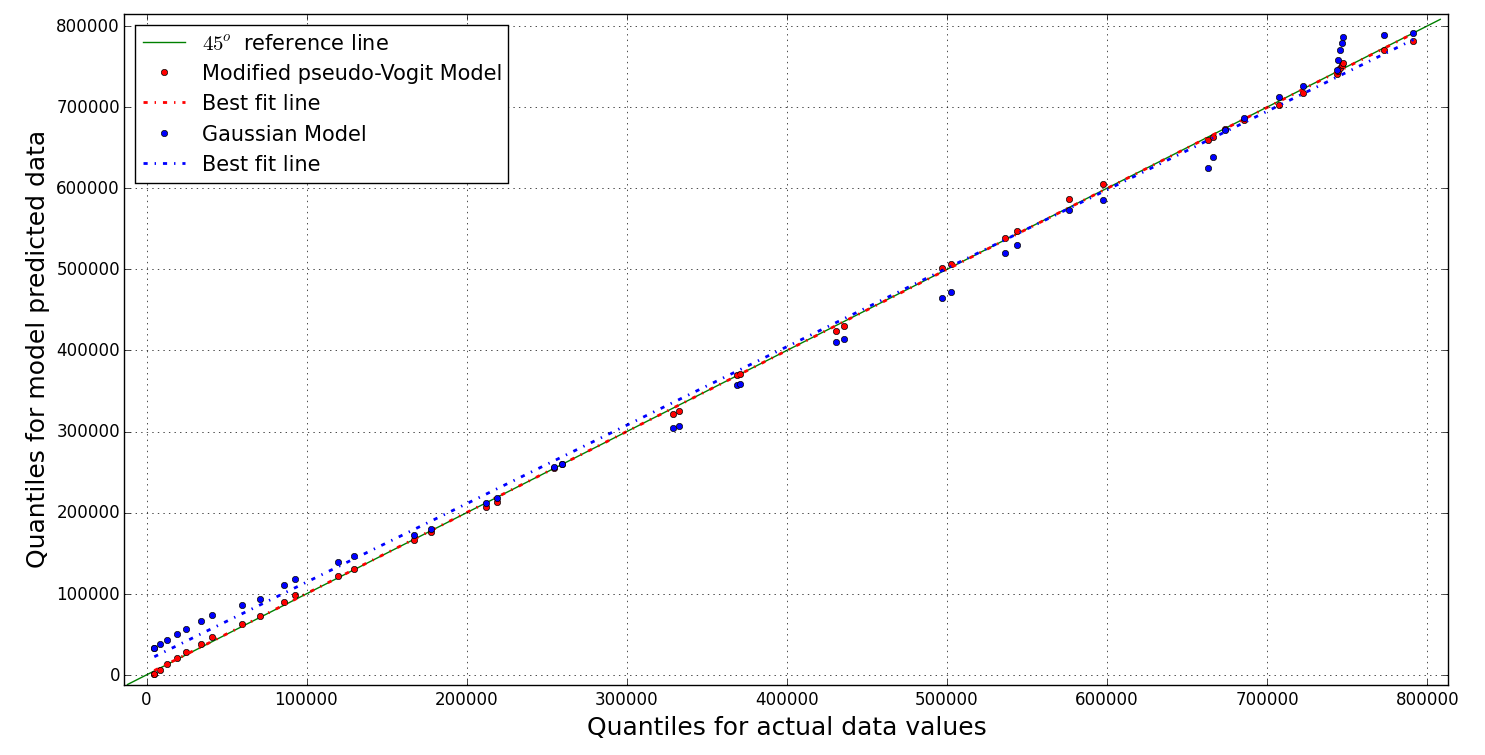}
        \captionN{Probability plot for Figure \protect\ref{fig:GGaussVSmpv}. The $x$-axis represents the quantiles for the actual data, the $y$-axis represents the theoretically predicted quantiles --- dependent on the model chosen (red: modified pseudo-Voigt model ($R^2 = 0.99974$); blue: Gaussian model ($R^2 = 0.99334$). The $R^2$ values are not relative to the best fit lines, but are relative to the $45^\circ$ reference line $y=x$. The closer  the $R^2$ value is to $1$, the more similar the predicted quantiles are to the actual ones, thus, the better the model describes the data. }
        \label{fig:GassAndMPVQQ}
    \end{subfigure}
    
    \captionN{Using probability plots, we are able to statistically see which model provides the better fit. We employ such graphical methods as standard goodness-of-fit tests such as the $\chi^2$ fail to give meaningful results.}
    \label{fig:qqOdd}
\end{figure}

Consider the following curves for the $h^{1,1}-h^{1,2}$ distribution with $r = 60 $ in Figures \ref{fig:GGaussVSmpv} and \ref{fig:GassAndMPVQQ}.

\begin{figure}[H]
	\begin{center}	
		\includegraphics[scale=0.45]{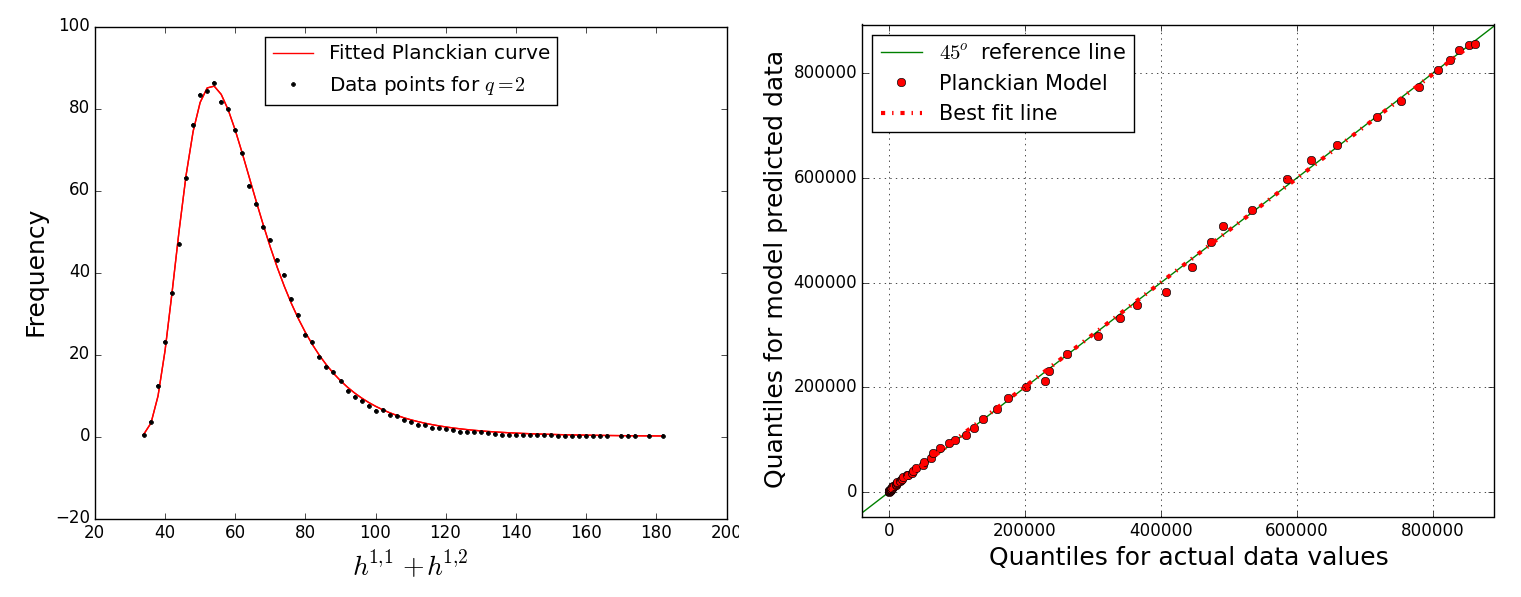}
	\end{center}
	\captionN{Left: best fit curve of $h^{1,1}-h^{1,2}$ distribution for curve $q=2$ based on the Planckian model. Right: probability plots of our fitted theoretical Planck model vs the $q = 2$, $h^{1,1}-h^{1,2}$ distribution.}  	
	\label{fig:PlanckANDqqPlanck}
\end{figure}

For the $h^{1,1}+h^{1,2}$ distribution we just plot the data of $q=2$ together with the corresponding probability plot in Figure \ref{fig:PlanckANDqqPlanck}.
In its current form, the probability plots do not allow us to calculate $p$-values of the various models. This due to the same issue encountered previously. If one however standardizes the data according to the Z-standardization:
\begin{align}
Z = \frac{X-\mu}{\sigma},
\end{align}
where $\mu$ and $\sigma$ are the mean and standard deviation, it is possible to calculate the $p$- values since the magnitude of each sample gets rescaled. The probability plot of all the models is displayed in the Appendix, with the relative $p$-values for each model --- Figure \ref{fig:Allqq51} and Figure \ref{fig:Allqq54}. What we see is that the modified pseudo-Voigt is statistically the model which provides the best fit.

\subsection{Implications for Physics}\label{IfP}

Calabi--Yau threefold compactifications of string theory have been the traditional approach to obtaining interesting phenomenological models.
The plethora of geometries and configurations, ranging from heterotic strings on Calabi--Yau threefolds endowed with stable bundles, to D-brane probes on local Calabi--Yau varieties, to F-theory compactification on elliptic fibrations, has over the years justified the landscape and inspired various statistical analyses of the space of vacua.

Of particular interest have been the investigation of further structures in the Kreuzer--Skarke database, including
identification of ``the tip'' where Hodge numbers are small
\cite{Candelas:2007ac,Candelas:2008wb,Candelas:2016fdy},
the top bounding curves where Hodge numbers are large
\cite{Johnson:2014xpa},
identifying elliptically fibered threefolds
\cite{Braun:2011ux,Taylor:2012dr,Taylor:2015ppa,Taylor:2015isa},
finding further fibrations such as K3-fibers
\cite{Candelas:2012uu,Anderson:2016cdu},
or a step-by-step construction of all possible smooth Calabi--Yau hypersurfaces from the
reflexive polytope data
\cite{Altman:2014bfa},
etc.
Now, it should be emphasized that each of the some 473 million reflexive polytopes admits, as an ambient toric variety, many\footnote{
  The actual numbers are not yet known, but even up to $h^{1,1}=7$, we already see from tens to thousands and with the number increasing potentially exponentially as we go up in Hodge number \cite{Altman:2014bfa}.
} so-called maximal projective crepant partial (MPCP) desingularization, each of which gives rise to a different Calabi--Yau threefold.
Therefore, the actually number of Calabi--Yau threefolds from the Kreuzer--Skarke database is many orders of magnitude larger than $10^{10}$.
While manifolds coming from the same reflexive polytope have different geometrical data such as triple intersection numbers, which in the standard embedding in heterotic compactification correspond to Yukawa couplings, they do share the same Hodge numbers because these, by virtue of \eqref{hodgeDelta}, depend only on the combinatorics of the polytope.
We need to wait for significant theoretical and/or computational advances to have the full data of the Hodge pairs in view of the Calabi--Yau manifolds themselves, which might give new statistics. It would be perhaps even more interesting if the statistic remain largely the same, thereby hinting at some universality in the distribution of such topological data.

In the context of the recent works on F-theory, it is an important fact the vast majority of the Kreuzer--Skarke threefolds are elliptic fibrations over some complex surface, and in fact birational to \cite{Braun:2011ux,Taylor:2015isa,Anderson:2016cdu} a Weierstrass model.
For example, some $10^6$ alone \cite{Braun:2011ux} come from elliptic fibrations over $\IP^2$.
Therefore the Kreuzer--Skarke dataset is directly relevant to F-theory.
In the more classical context of heterotic strings, the Hodge numbers dictate the number of (anti-)generations in the standard embedding.
In our above plots, the Euler number $\pm 6$ indicate the three generation models.
The generic paucity of $\chi = \pm 6$ manifolds led to the industry of non-standard embedding where extra vector bundle and Wilson line information is needed.
The advantage of F-theory models is that the compactification data comes only from the Calabi--Yau manifold.
In particular, the intersection theory of the cycles and fiber-degeneration structure determine the gauge group, anomaly cancellation, matter content, and Yukawa couplings.
Much of this can be extracted from the polytope data.

F-theory compactifications on threefolds, resulting in six dimensional gauge theories have been considered from the point of view of systematically classifying the base complex surfaces \cite{Taylor:2015isa} and the statistics have been performed therein.
Non-toric bases were considered and a number of Calabi--Yau threefolds beyond the Kreuzer--Skarke data were found.
It is remarkable that the overall distribution of Hodge numbers remains largely unchanged.
Indeed, in unpublished work of Kreuzer--Skarke, where they extended the hypersurface in toric fourfolds to double hypersurfaces in fivefolds, obtaining some $10^{10}$ more manifolds and the shape of Figure \ref{f:hodge3} persists.
All these point to the Kreuzer--Skarke data being a robust representative in the space of Calabi--Yau threefolds.
Our distribution subsequently seems a representative a sample, and we speculate that analyses of string vacua, in any context, should be thus weighted.
For example, in study of the ``typical'' number of generations in four dimensional heterotic compatification, or of charged matter in six dimensional F-theory compactification, one should superpose our pseudo-Voigt profile.

\section{Calabi--Yau Twofolds:  K3 Surfaces}\label{Picard}
\begin{figure}[H]
    \centering
    \begin{subfigure}[h]{0.75\textwidth}
        \includegraphics[width=\textwidth]{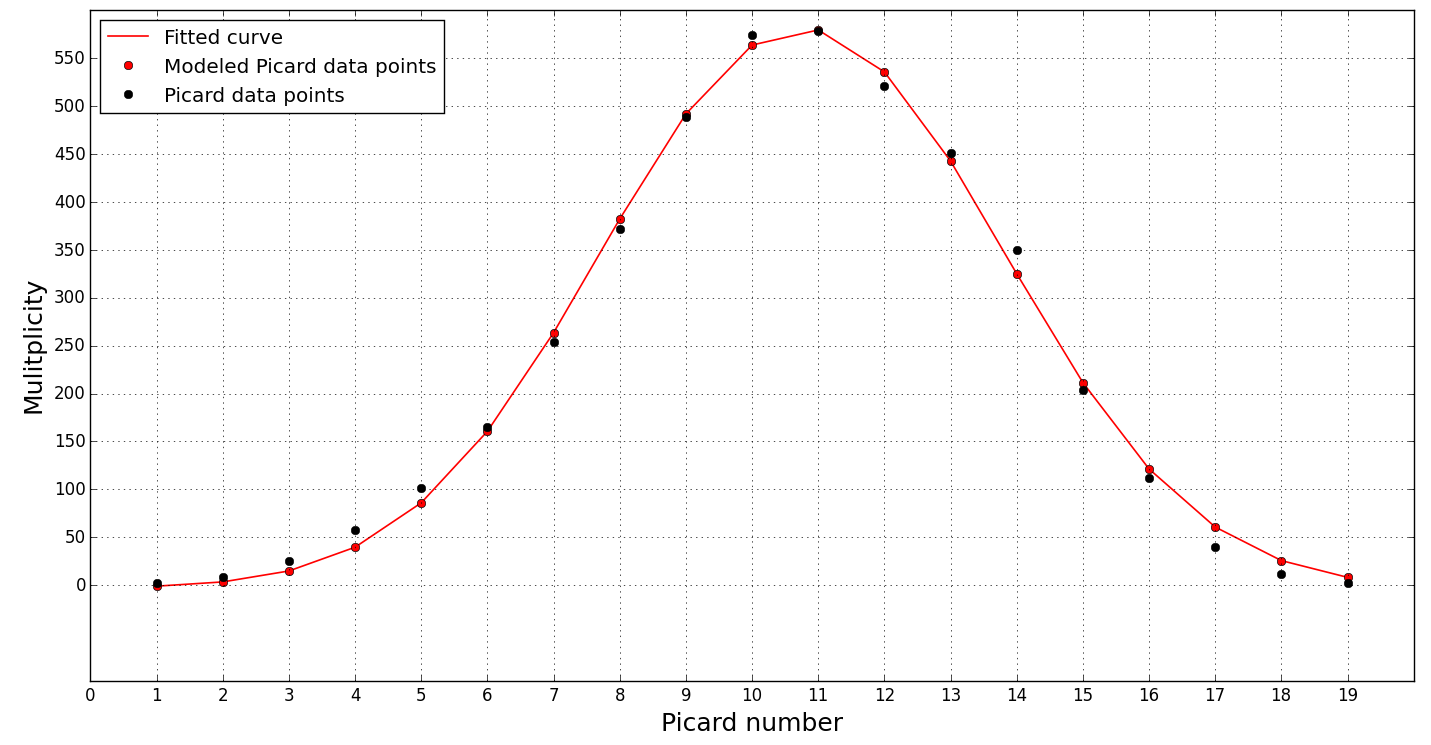}
        \captionN{For K3 surfaces, the multiplicity is plotted against Picard number with a pseudo-Voigt fit.}
        \label{fig:PicardFit1}
    \end{subfigure}

    \begin{subfigure}[h]{0.75\textwidth}
        \includegraphics[width=\textwidth]{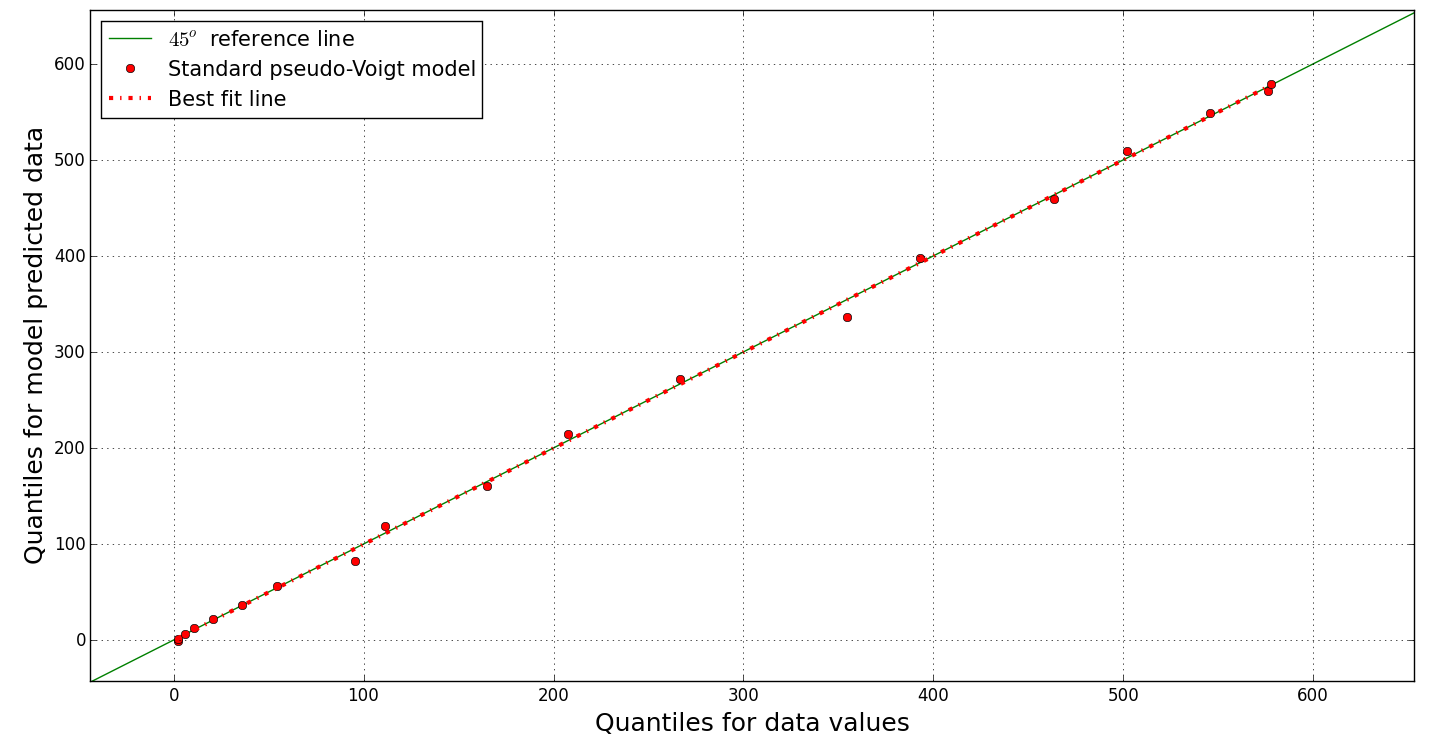}
        \captionN{Probability plot for the multiplicity quantiles vs the fitted standard pseudo-Voigt quantiles. The $R^2$ value is $0.99908$.}
        \label{fig:qqPicardFit}
    \end{subfigure}
    
    \captionN{Using probability plots, we are able to statistically see which model provides the better fit. We employ such graphical methods as standard goodness-of-fit tests, such as the $\chi^2$ test, fail to give meaningful results.}
    \label{fig:Picard}
\end{figure}

As noted in the Introduction, there are $4319$ data points, corresponding to hypersurfaces as Calabi--Yau twofolds, \textit{i.e.}, K3 surfaces, in reflexive three dimensional polytopes.
Being algebraic K3 surfaces, there is only one relevant topological invariant, the Hodge number, $h^{1,1} = 19$.
However, there is a further refined algebraic quantity for the K3 surface $X$, the rank of the Neron--Severi lattice
$H^{2}(X;\IZ) \cap H^{1,1}(X)$, which is the {\bf Picard Number} $\rho(X)$ and which enumerates the number of divisors on the surface up to algebraic equivalence.
The Picard numbers of the $4319$ K3 surfaces were computed in \cite{Kreuzer:1998vb}.
We present the distribution thereof in Figure~\ref{fig:PicardFit1}.

We only used the standard pseudo-Voigt profile as the modified one did not change the fit significantly. Here are the fit statistics for best fit curve: $(A,\mu,\sigma,\alpha) = (4517.45,10.76,2.97,-0.031)$, as shown in Figure \ref{fig:Picard}.

What is interesting about Figure \ref{fig:PicardFit1} is that the ``oscillations" of the actual data points above and below the modeled curve is very apparent, yet modifying the pseudo-Voigt profile is unable to give any significant improvement. This leads to two potential conclusions: (a) The pseudo-Voigt profile is not the best profile to use in combination with an oscillatory component; (b) The manner in which the oscillations occur is not so straight forward as introducing simple cosine function. An interesting exercise would be to superimpose a cosine function along the distribution, by rotating it as one traverses the profile. As long as the wavelength, amplitude and angle of rotation are all small enough, the continuously rotated cosine function should remain a function everywhere along the profile.

\section{Calabi--Yau Fourfolds}\label{FourFolds}
The analysis of the four fold data is performed in the same spirit as the threefold data.
We aim to look for patterns in the frequency plots.
Due to complex conjugation and Poincar\'e duality, the only topological invariants of fourfolds that vary are $h^{1,1}$, $h^{1,2}$, $h^{1,3}$, and $h^{2,2}$.
Three of these are independent~\cite{FourFold}:
\bea
h^{2,2} = 44+4h^{1,1}-2h^{1,2}+4h^{1,3} ~.
\eea

\begin{figure}[H]
	\begin{center}	
		\includegraphics[scale=0.45]{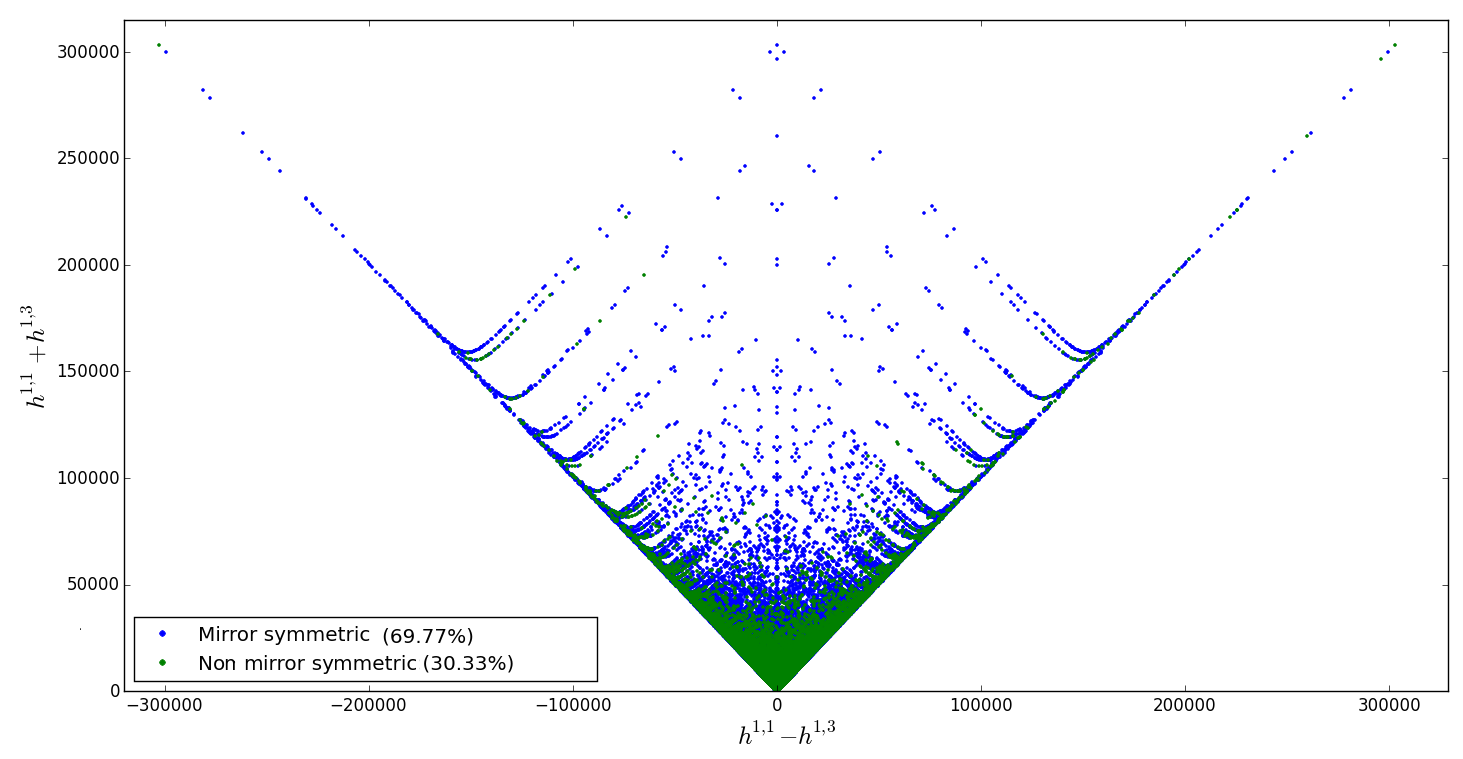}
	\end{center}
\captionN{The blue points correspond to manifolds with a mirror symmetric counterpart in the data set.} 	
\label{Fig:AllMirrorNoMirror}
\end{figure}

We compiled a database for the frequency of the triplets ($h^{1,1}, h^{1,2}, h^{1,3}$) to then obtain the following data structure 
\begin{align*}
(h^{1,1}, h^{1,2}, h^{1,3},f) ~.
\end{align*}
Since one expects mirror symmetry within the invariants ($h^{1,1}\pm h^{1,3}$) ~\cite{9812195}, a plot of $h^{1,1}-h^{1,3}$ against $h^{1,1}+h^{1,3}$ (Figure \ref{Fig:AllMirrorNoMirror}) should be symmetric about the line $h^{1,1}-h^{1,3} = 0$.

Doing a quick analysis of the data yields the following observations: only partial mirror symmetry is found.
For $69.77\%$ of data points, the point $(h^{1,1}-h^{1,3},h^{1,1}+h^{1,3})$ is accompanied by the point $(-h^{1,1}+h^{1,3},h^{1,1}+h^{1,3})$. Taking frequency into account, the percentage drops to $27.35\%$ --- see Figure \ref{Fig:AAllMirror1} in the Appendix. This is most likely due to an incomplete data base.\\
\\
For now, we have performed a primary analysis on the Euler distribution only.
The Euler number for fourfolds is~\cite{FourFold}:
\begin{align}
\chi = 6(8+h^{1,1} - h^{1,2} + h^{1,3}) ~.
\end{align}
Interestingly enough, the distinction between even and odd distributions persist in the fourfold data base.
For illustrative purposes, we show the distribution of $\chi/6$ against frequency.

\begin{figure}[H]
	\begin{center}	
		\includegraphics[scale=0.45]{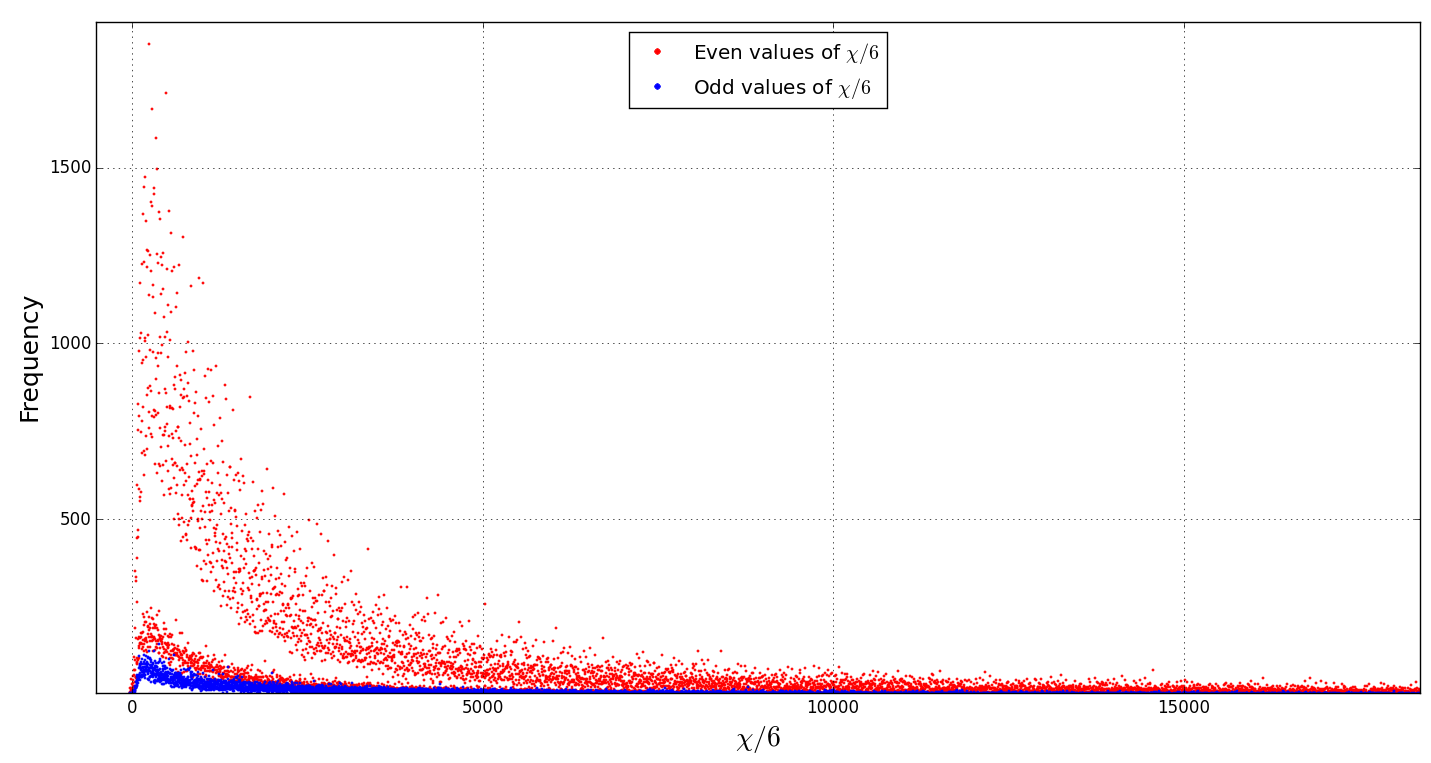}
	\end{center}
	\captionN{Frequency of Calabi--Yau fourfolds with a given Euler number.}  	
	\label{fig:MainEuler}
\end{figure}
It is not immediately clear what is the reason for the gap, presumably it could be a cluster of data points which is missing from the data base. Until one obtains the complete fourfold data base of Hodge numbers, one can't say much else. We also preset plots of the individual Hodge numbers $h^{i,j}$ vs. frequency.

\begin{figure}[H]
    \centering
    \begin{subfigure}[h]{0.45\textwidth}
        \includegraphics[width=\textwidth]{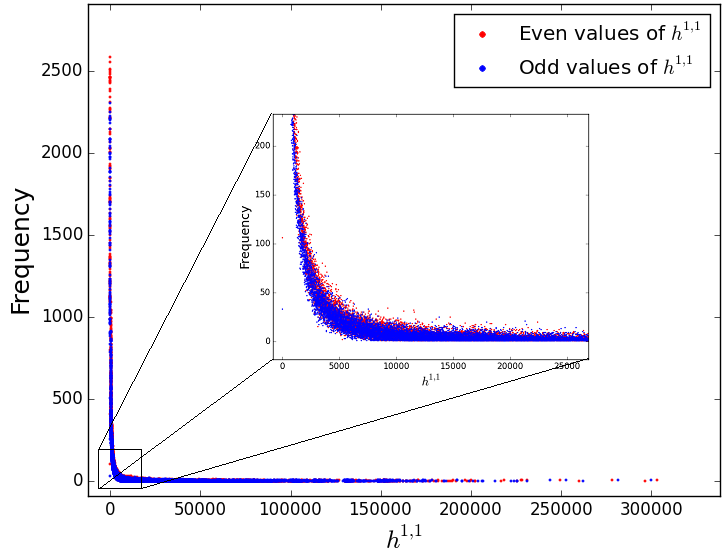}
        \captionN{ $h^{1,1}$ vs. frequency.}
        \label{Fig:F4h11}
    \end{subfigure}
	~
    \begin{subfigure}[h]{0.45\textwidth}
        \includegraphics[width=\textwidth]{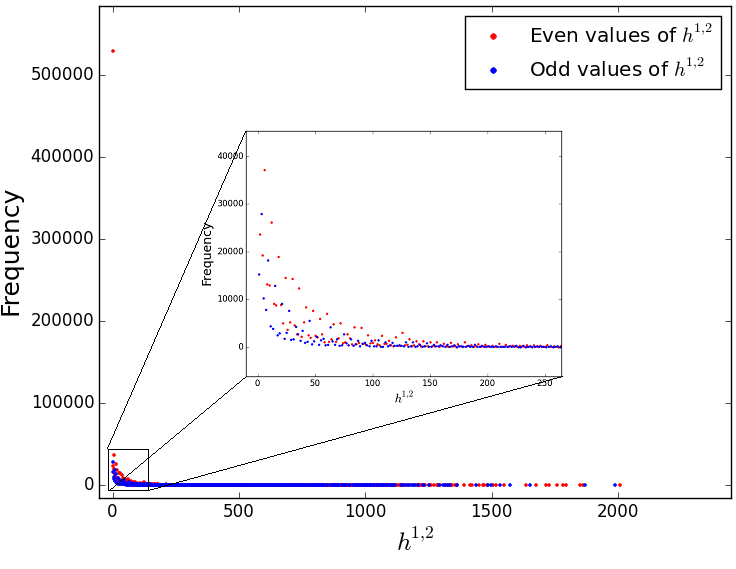}
        \captionN{ $h^{1,2}$ vs. frequency.}
        \label{Fig:F4h12}
    \end{subfigure}
    
 \end{figure}
\begin{figure}[H]
\ContinuedFloat 
  \centering 
  
    \begin{subfigure}[h]{0.45\textwidth}
        \includegraphics[width=\textwidth]{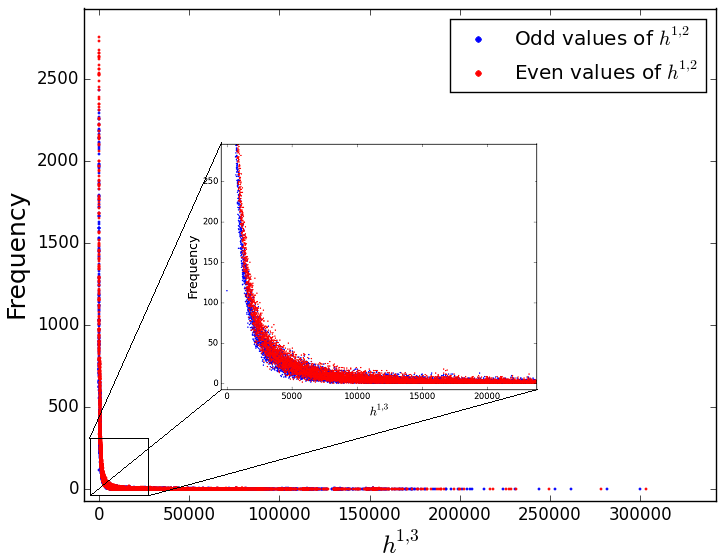}
        \captionN{ $h^{1,3}$ vs. frequency.}
        \label{Fig:F4h13}
    \end{subfigure}
    ~
       \begin{subfigure}[h]{0.45\textwidth}
        \includegraphics[width=\textwidth]{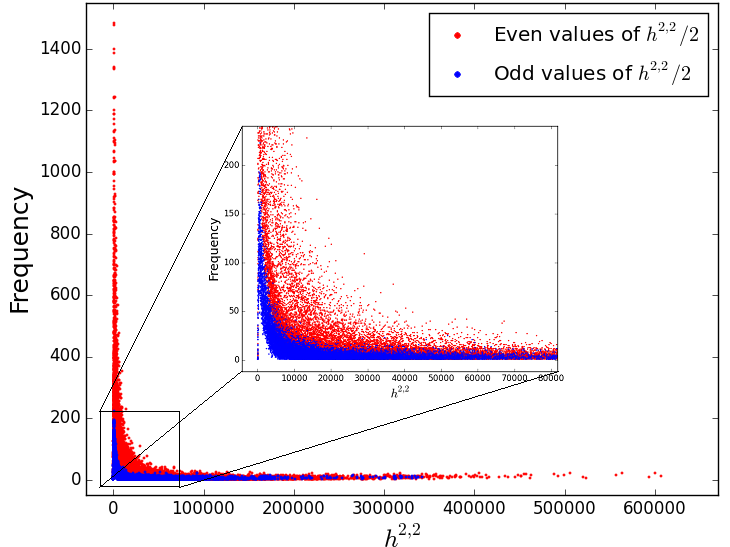}
        \captionN{ $h^{2,2}$ vs. frequency.}
        \label{Fig:F4h22}
    \end{subfigure}
       
    \captionN{The frequency for all the hodge $h^{i,j}$ numbers. Red points and blue are odd and even points respectively for the various Hodge numbers. The data points are very dense close to the origin making it difficult to properly illustrate the mixing of odd and even Hodge numbers. Only $h^{2,2}$ (c) has a clear separation between of an even values.}
    \label{Fig:AllFourHodge}
\end{figure}

\section{Conclusions and Outlook}\label{s:conc}
By examining the distribution of Hodge numbers of Calabi--Yau manifolds of complex dimension two, three and four, realized as hypersurfaces in toric varieties of one higher dimension as constructed by Kreuzer and Skarke based on the results of Batyrev and Borisov, we have found many hithertofore undiscovered patterns.
We summarize our key points as follows.
\begin{itemize}
\item
  For threefolds, there are $30108$ distinct pairs of Hodge numbers $(h^{1,1},h^{1,2})$ from $473800776$ reflexive polytopes, the frequency of both the half-Euler number $h^{1,1} - h^{1,2}$ and the sum $h^{1,1} + h^{1,2}$ are distributed according to whether the value is odd or even;
  \begin{itemize}
 \item The half-Euler number
    $h^{1,1} - h^{1,2}$ follows a modified pseudo-Voigt distribution
    \[
    f(x) = (1-\alpha)\frac{A'}{\sigma\sqrt{2\pi}}e^{\frac{-(x)^2}{2\sigma^2}} + \alpha\frac{A'}{\pi}\left[\frac{\sigma^2}{x^2+\sigma^2}\right] \ .
    \]
   where the modification is made in the amplitude $A$ of the distribution, such that
   \[ 
   A' = A_0 + b\cos(2\pi\cdot b) \ .
   \]
   There is fine periodic substructure in terms of curves indexed by an integer $r$. 
   Our model is accurate for low $r$-values ($r\in[36,110]$ and $r \in [37,99]$); using probability plots as test for goodness of fit, this modified pseudo-Voigt model is indeed the best one out of several standard candidates (cf.~Figure \ref{fig:ParamTableEvenOddHdiff} for all the $R^2$ and $p$ values).

   Among $A,\sigma,\alpha,b,a$, the parameters $\sigma, b, \alpha$ have a strong linear relationship with $r$:
\[
     \begin{array}{lll}
       & {\rm Even } \ \ r & {\rm Odd } \ \ r \\
       \sigma(r)= & 0.5097r - 12.7142 & 0.51379r - 13.2494 \\
       \alpha(r)= & 2\times 10^{-4}r  -0.0345 & 2.25\times 10^{-4}r  -0.0388,\\
       b(r)= & 3.7299\times 10^{-5}r + 0.6629 & 7.9101\times 10^{-5}r + 0.65956
     \end{array}
\]

   For a small subset of curves with a low $r$-value and an appropriate cut-off frequency, it is extraordinary that the model {\it exactly fits the data}. That is, it appears that the number of data points for each curve required, such that the model will result in a perfect fit is: 7 for even $r$-valued curves and $10$ for the odd valued $r$-curves, see Figure \ref{fig:PerfectFits}.

  \item The quantity
    $h^{1,1} + h^{1,2}$ follows a Planckian distribution
    \[
    f(x) = \frac{A}{x^n}\frac{1}{e^{b/(x-22)}-1}
    \]
    There is a substructure of curves, indexed by an integer $q$, each Planckian and with some periodic behavior.
    The curves $q_n$ appear clustered into groups of residue classes distinguished by $n$ mod 6, and the parameters $\log(A),n,b$ all have extremely strong relationships with the $q$ value.

    By substituting this relationship into the model, we have a function $f_k(x,q)$ that approximately describes the entire $h^{1,1} + h^{1,2}$ distribution up to a $q$ value of $69,100$:    
    \begin{equation}
      f_k(x,q) = \frac{e^{\sum_{i=0}^{4}A_{k,i}q^i}}{x^{\sum_{i=0}^{4}n_{k,i}q^i}}\frac{1}{\left(e^{\frac{\sum_{i=0}^{4}b_{k,i}q^i}{(x-22)}}-1\right)},
    \end{equation}
    with $k=0,1,\ldots 5$ and the coefficients given in \ref{Acoeffs},\ref{Bcoeffs},\ref{Ncoeffs}.

 \item The Euler number $\chi = 2(h^{1,1} - h^{1,2})$ follows the modified pseudo-Voigt distribution composed with a sinusoidal $A + A_0 + a\cos(2\pi b\cdot x)$ which is almost an exact fit, with the coefficients given by
    $(A_0,\sigma,\alpha,b,a) = 
   (1.9032\times 10^{9},
   75.8305889,
   0.00718459,
   0.58347826,
   8.7427\times 10^{7})$, at  $R^2$ = 0.99944 for even $\chi$ and\\   
   $(1.9032\times 10^{9},
   75.8305889,
   0.00718459,
   0.58347826,
   8.7427\times 10^{7})$ at $R^2$ = 0.99965 for odd $\chi$,
  
 The modified pseudo-Voigt distribution is remarkably accurate in predicting the overall and fine sub-structure of the Euler number distribution. 
  \end{itemize}
 
\item For K3 surfaces, we have looked at the distribution of the multiplicity with Picard number. We find that this distribution follows a standard pseudo-Voigt profile. Adding in the sinusoidal modification does not significantly increase the overall fit.
  The parameters are given by $(a,\mu,\sigma,\alpha) = (4517.45, 10.76, 2.97, -0.031)$ with $R^2 = 0.99908$.

\item For Calabi--Yau fourfolds, there is no exact mirror symmetry, due to incompleteness of available data. Nevertheless, by breaking up the data into three groups, we have
\begin{itemize}
\item Mirror symmetric partners with the same frequency: $27.35\%$\\
\item Mirror symmetric partners without the same frequency: $42.22\%$\\
\item Non mirror symmetric partners: $30.33\%$
\end{itemize}

By plotting the various $h^{i,j}$ vs frequency we see there is no distinction between even and odd data values for $h^{i,j}$, expect for $h^{2,2}/2$. This distinction is carried out further in the Euler number distribution where odd points are clustered on a band with much lower frequencies. The even values of $\chi/6$ appear to be distributed along to separate bands.
\end{itemize}

It is remarkable how well the pseudo-Voigt distribution, modified with a sinusoidal component, fits the distribution of topological numbers of toric Calabi--Yau manifolds, often giving an exact fit.
Of course, what we are studying at heart is the number of integer points inside
(\textit{cf.}~\eqref{hodgeDelta}) reflexive polytopes.
This is a highly non-trivial counting problem whose answer will ultimately give full analytic results for our distributions and we suspect that the answer should be some generalized pseudo-Voigt function.

Now, in addition of Calabi--Yau manifolds, stable vector bundles over various such manifolds in a variety of construction beyond Kreuzer--Skarke have also been studied algorithmically over the years in the context of heterotic compactification (\textit{cf.}~\textit{e.g.}, \cite{Anderson:2007nc,Gabella:2008id,Gao:2014nfa,Anderson:2012yf}).
One can see a somewhat pseudo-Voigt profile in these as well, even though there is no underlying polytope and the counting problem is dictated by certain Diophantine system.
It would be interesting to see why this shape is universal in such classifications.

\section*{Acknowledgements}
We are grateful to Cyril Matti for collaboration during the early stages of this project.
We thank Mark Dowdeswell for his input with regards to the goodness-of-fits for the various plots.
YHH is indebted to the Science and Technology Facilities Council, UK, for grant ST/J00037X/1, the Chinese Ministry of Education, for a Chang-Jiang Chair Professorship at NanKai University, and the city of Tian-Jin for a Qian-Ren Award.
YHH is also perpetually indebted to Merton College, Oxford for continuing to provide a quiet corner of Paradise for musing and contemplations.
VJ and LP are supported by the South African Research Chairs Initiative of the Department of Science and Technology and the National Research Foundation.

\appendix
\renewcommand\thefigure{\thesection.\arabic{figure}}    
\section{Appendix}
\setcounter{figure}{0}   

Here we include all additional plots to supplement the main body. This includes the relevant plots for the odd distributions --- since in the main text we only presented the plots for even distributions --- as well as the regression analysis statistics and parameter values for both distributions. 

\subsection{Supplementary plots for the $h^{1,1}-h^{1,2}$ distribution}
All even plot counterparts will be referenced in the figures. The plots appear in the same order as in the main body, with descriptions only if necessary.
\subsubsection{Plots for the odd distribution as counterparts to the even ones}
\begin{figure}[H]
	\begin{center}	
		\includegraphics[scale=0.4]{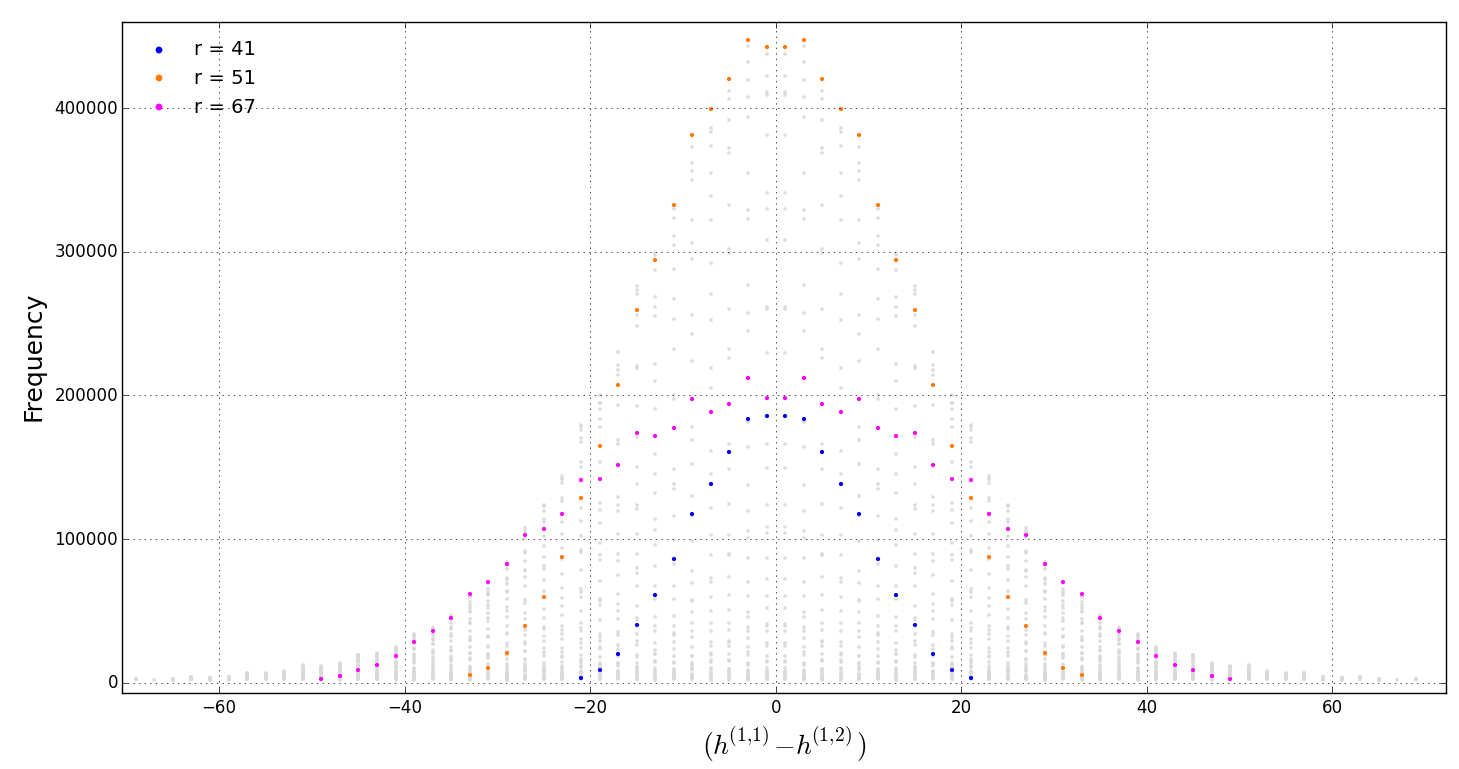}
	\end{center}
	\captionN{Three highlighted curves ($r = 41,51,67$) within the odd $h^{1,1} - h^{1,2}$ distribution. The transparent grey data dots is the rest of the distribution. Refer to Figure~\protect\ref{Fig:HdiffExample} for the even plot.}  	
	\label{Fig:HdiffExampleOdd}
\end{figure}

\begin{figure}[H]
    \centering
    \begin{subfigure}[h]{0.9\textwidth}
        \includegraphics[width=\textwidth]{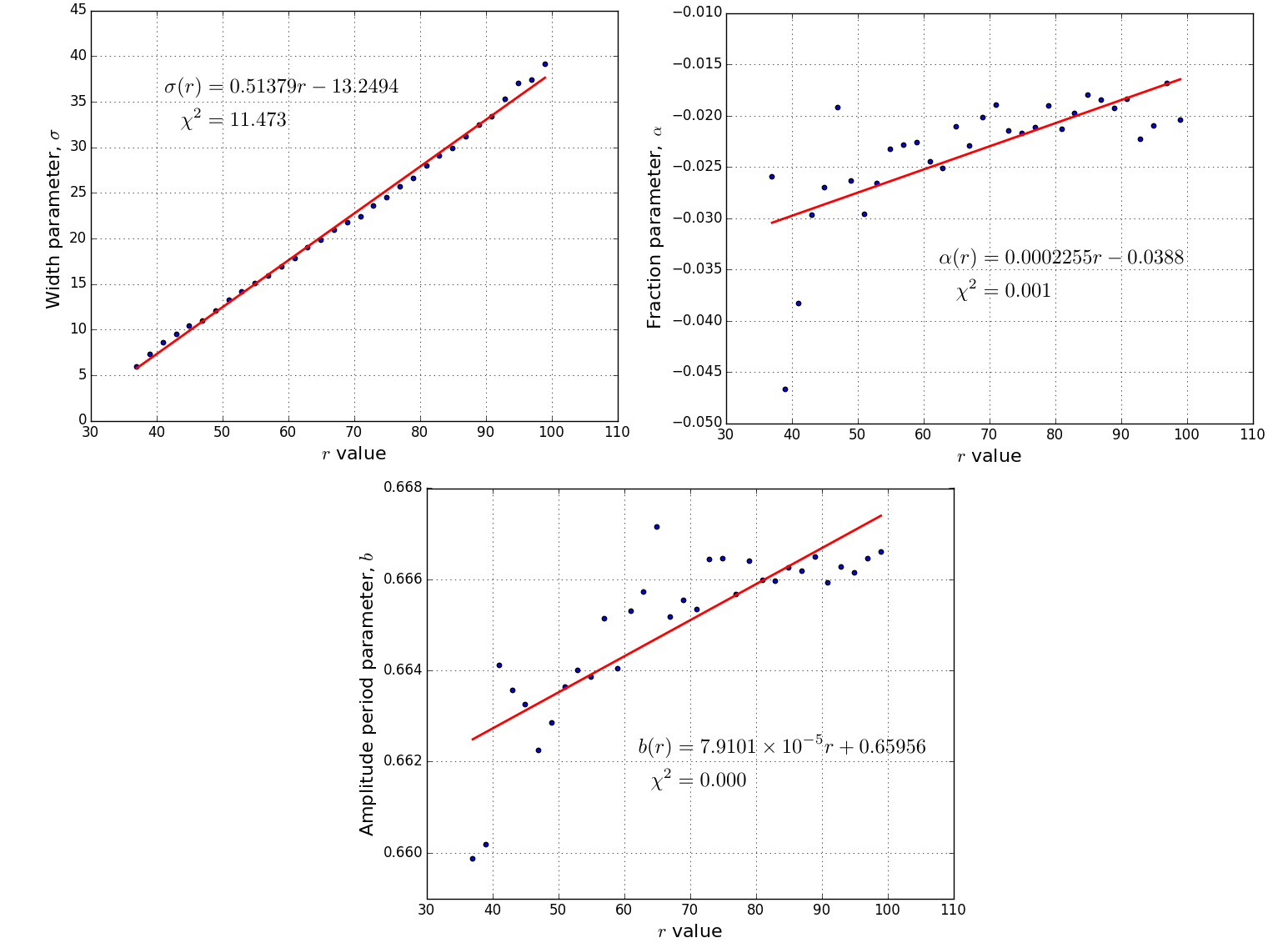}
        \captionN{The width parameter $\sigma$ has a linear relationship with $r$ such that $\sigma(r) = 0.51379r - 13.2494$. The amplitude period parameter,$b$, also has a linear relationship, however, since $r$ is at most order $3$ in magnitude, we can regard it approximately as a constant such that $b(r) = 0.65956 \sim 2/3$. The same goes for the fraction parameter,$\alpha$, we can regard it as a constant such that $\alpha(r) = -0.0388$. For even parameter fit statistics see Figure~\protect\ref{Fig:EvenHdiffParamAll}.}
        \label{fig:OddHdiffParam1}
    \end{subfigure} 
    \begin{subfigure}[h]{0.9\textwidth}
        \includegraphics[width=\textwidth]{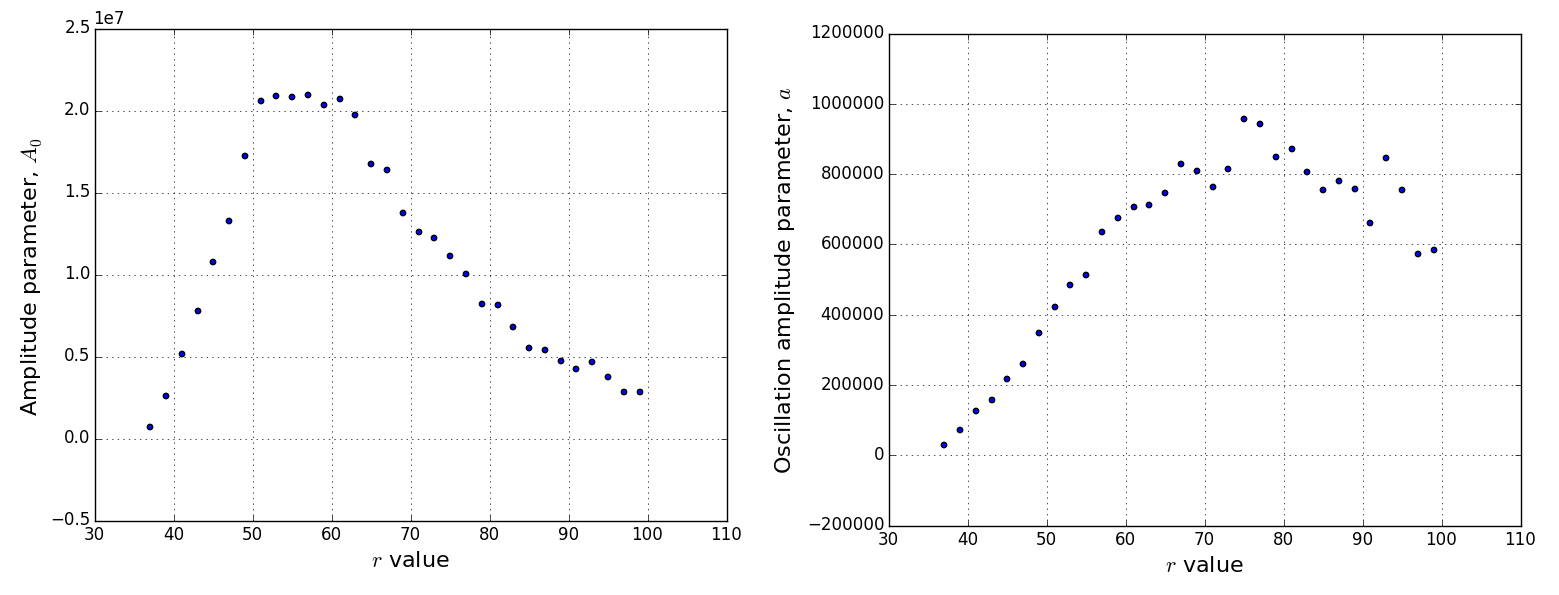}
        \captionN{Plots of $A_0$ vs $r$ (left) and  $a$ vs $r$ (right). Both exhibit a similar pattern, however it is difficult to find any nice relationships. For even parameter plots see Figure~\protect\ref{Fig:EvenHdiffParamAll}.}
        \label{fig:OddHdiffParam2}
    \end{subfigure}
    
    \captionN{The plots of the various parameters $A,\sigma,\alpha,b,a$ versus $r$ for odd values of $r$. }
    \label{fig:AllHdiffOddParam}
\end{figure}

\subsubsection{Comparative plots}\label{AppenModels}
Here we present a comparison of various models we used, by plotting them side by side with the relevant fit-statistics. We choose a single even curve, $r= 54$, and odd curve, $r=51$, to illustrate the difference between models.

\paragraph*{\textbf{Gaussian Model}}
\begin{align} 
f(x,A,\mu,\sigma) = \frac{A}{\sigma\sqrt{2\pi}}e^{-(x-\mu)^2/2\sigma^2}
\end{align}

\begin{figure}[H]
    \centering
    \begin{subfigure}[h]{0.8\textwidth}
        \includegraphics[width=\textwidth]{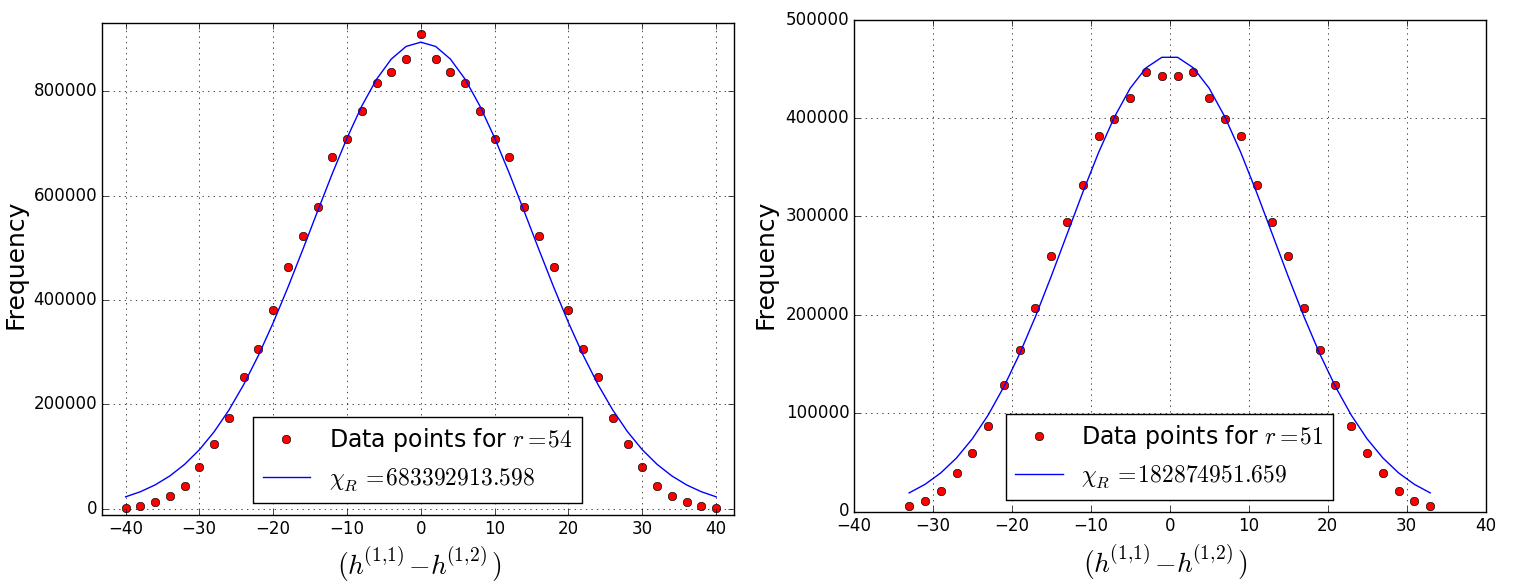}
        \captionN{Gaussian model.}
        \label{fig:GaussEO}
    \end{subfigure} 
\end{figure}

\paragraph*{\textbf{Lorentzian Model}}
\begin{align}
f(x,A,\mu,\sigma) = \frac{A}{\pi}\left[\frac{\sigma}{(x-\mu)^2+\sigma^2}\right]
\end{align}

\begin{figure}[H]
  \ContinuedFloat 
  \centering     
    \begin{subfigure}[h]{0.8\textwidth}
        \includegraphics[width=\textwidth]{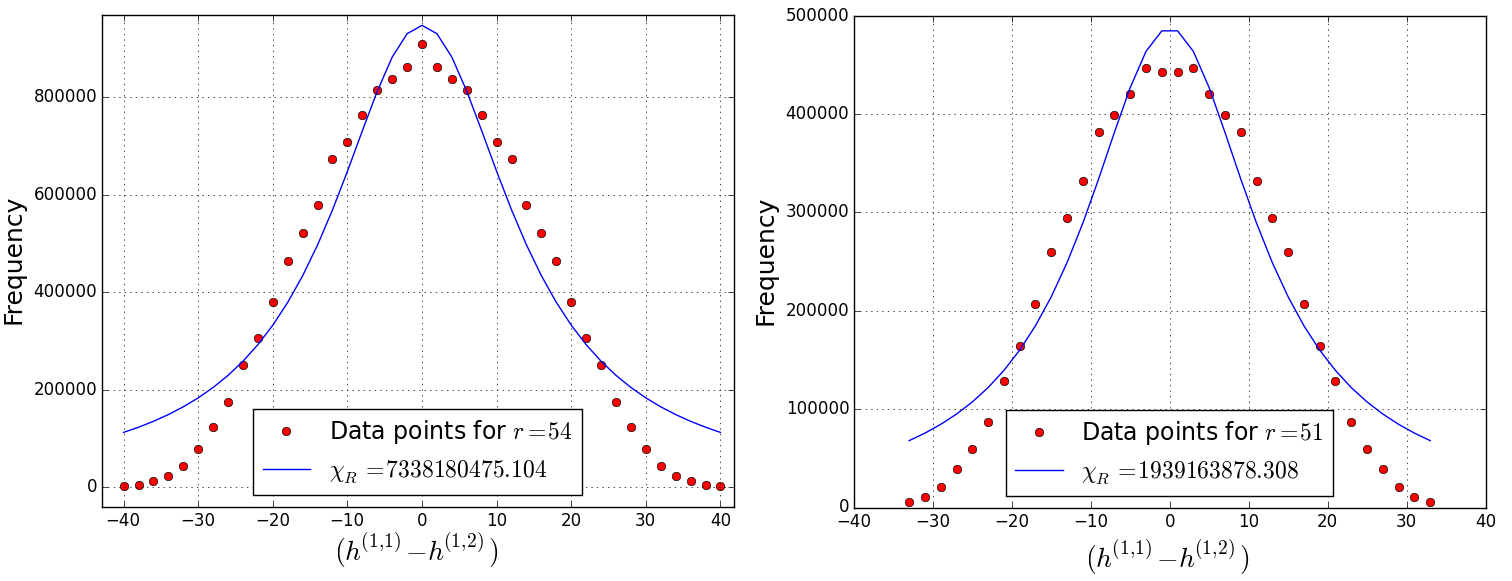}
        \captionN{ Lorentzian (Cauchy) model.}
        \label{fig:LorentzEO}
    \end{subfigure}
\end{figure}
    
\paragraph*{\textbf{Pearson7 Model}}

\begin{align}
f(x,A,\mu,\sigma,m) = \frac{A}{\sigma\beta(m-\frac{1}{2},\frac{1}{2})}\left[1+\frac{(x-\mu)^2}{\sigma^2}\right]^{-m},
\end{align}
where $\beta$ is the Beta function.
    
 \begin{figure}[H]
  \ContinuedFloat 
  \centering  
    \begin{subfigure}[h]{0.8\textwidth}
        \includegraphics[width=\textwidth]{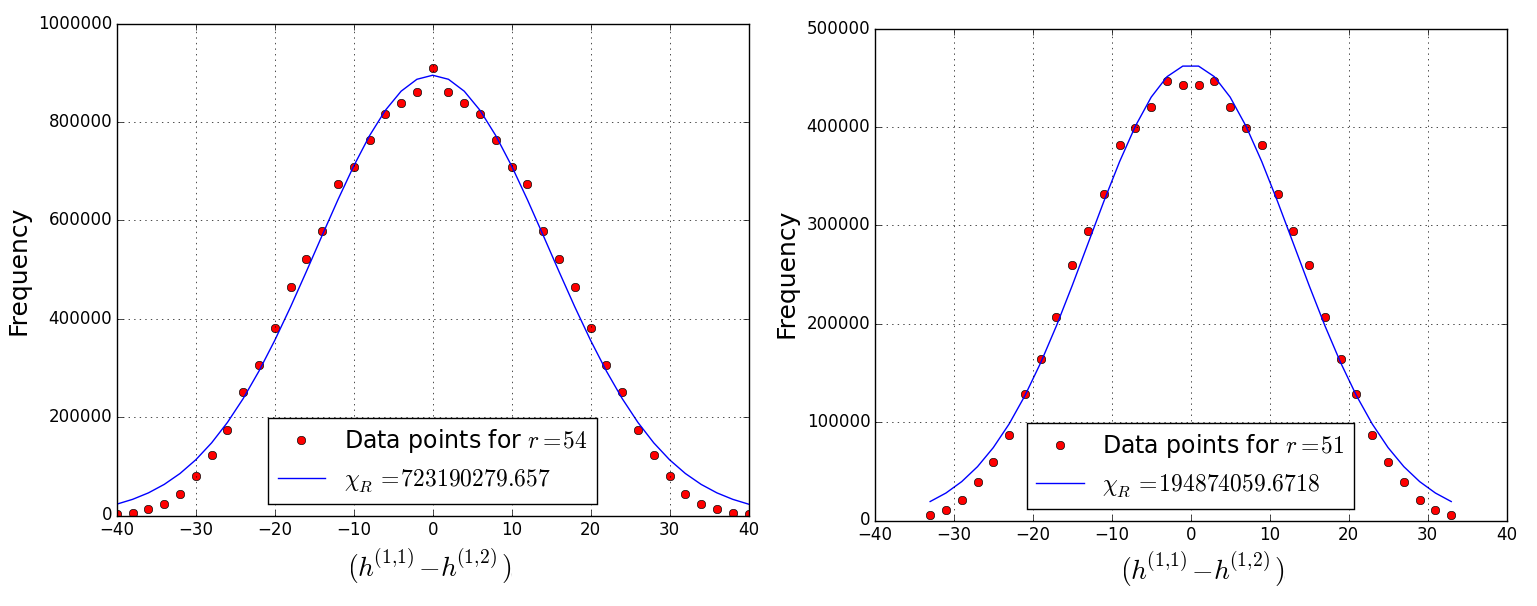}
        \captionN{Pearson7 model.}
        \label{fig:Pearson7EO}
    \end{subfigure}
\end{figure}
    
\paragraph*{\textbf{Breit-Wigner Model}}\mbox{}\\
\\
This model is based on the Breit-Wigner function.
\begin{align}
f(x,A,\mu,\sigma,t) = \frac{A(t\sigma/2 + x - \mu)^2}{(\sigma/2)^2+(x-\mu)^2}
\end{align}     

 \begin{figure}[H]
  \ContinuedFloat 
  \centering 

    \begin{subfigure}[h]{0.8\textwidth}
        \includegraphics[width=\textwidth]{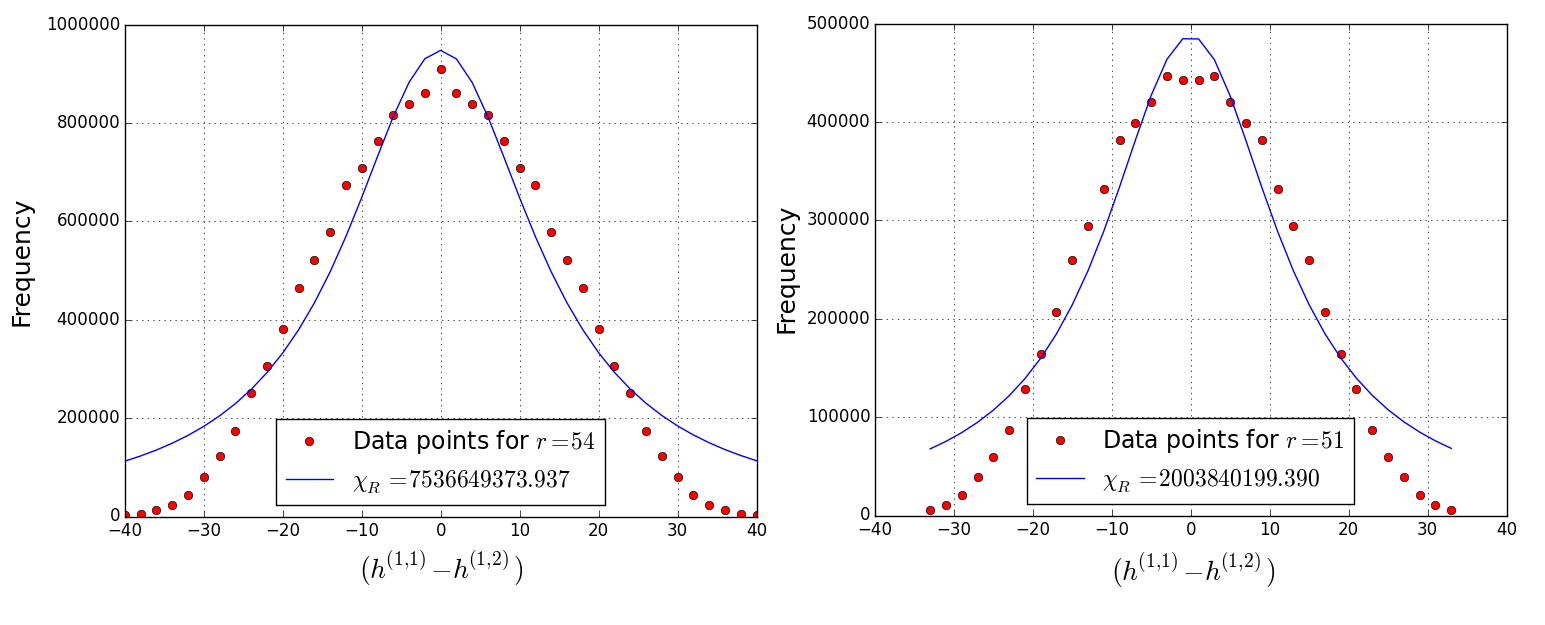}
        \captionN{Breit--Wigner model.}
        \label{fig:BreitEO}
    \end{subfigure}
\end{figure}
   
\paragraph*{\textbf{Voigt Model}}

\begin{equation}
f(x,A,\mu,\sigma,\gamma) = \frac{a\text{Re}[(z)]}{\sigma\sqrt{2\pi}}
\end{equation}
where
\begin{equation}
z =  \frac{x-\mu + i\gamma}{\sigma\sqrt{2}}\ , \quad
w(z) = e^{-z^2}\text{erfc}(-iz)
\end{equation}
The Voigt model is a convolution of the Gaussian and Lorentzian models. 
  \begin{figure}[H]
  \ContinuedFloat 
  \centering 
  
    \begin{subfigure}[h]{0.8\textwidth}
        \includegraphics[width=\textwidth]{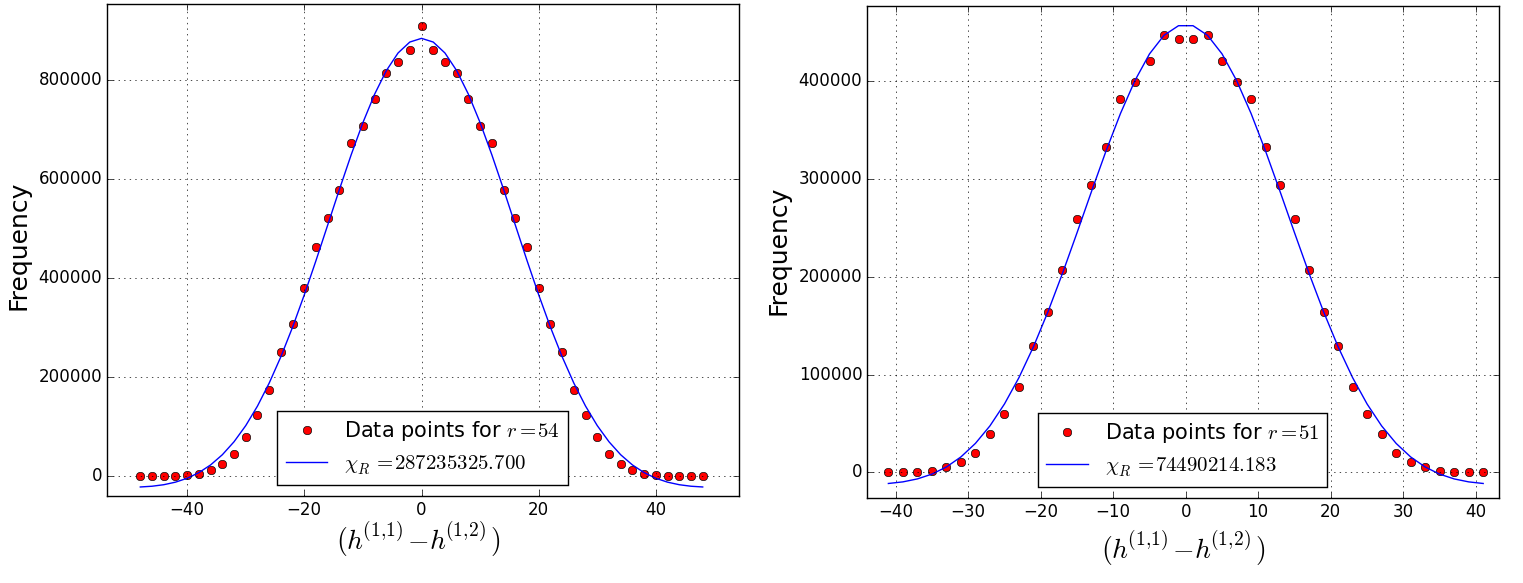}
        \captionN{Voigt model.}
        \label{fig:VoigtEO}
    \end{subfigure}
\end{figure}

\paragraph*{\textbf{pseudo-Voigt Model}}

\begin{equation}
f(x,A,\mu,\sigma,\alpha) = (1-\alpha)\frac{A}{\sigma\sqrt{2\pi}}e^{\frac{-(x-\mu)^2}{2\sigma^2}} + \alpha\frac{A}{\pi}\left[\frac{\sigma^2}{(x-\mu)^2}+\sigma^2\right]
\end{equation}   
 
  \begin{figure}[H]
  \ContinuedFloat 
  \centering 
  
     \begin{subfigure}[h]{0.8\textwidth}
        \includegraphics[width=\textwidth]{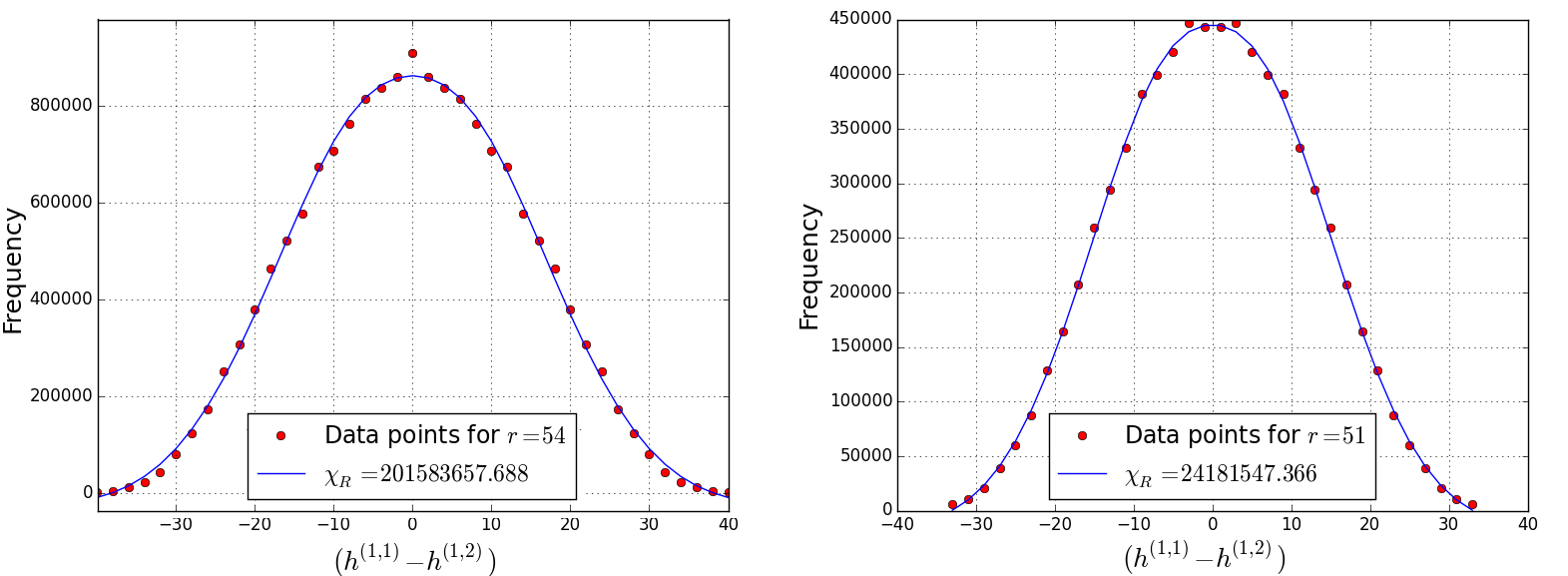}
        \captionN{pseudo-Voigt model.}
        \label{fig:PVEO}
    \end{subfigure}
\end{figure}  
 
We present the standardized and shifted probability plots for the above comparisons: 
 
  \begin{figure}[H]
  \ContinuedFloat 
  \centering 
  
     \begin{subfigure}[h]{0.9\textwidth}
        \includegraphics[width=\textwidth]{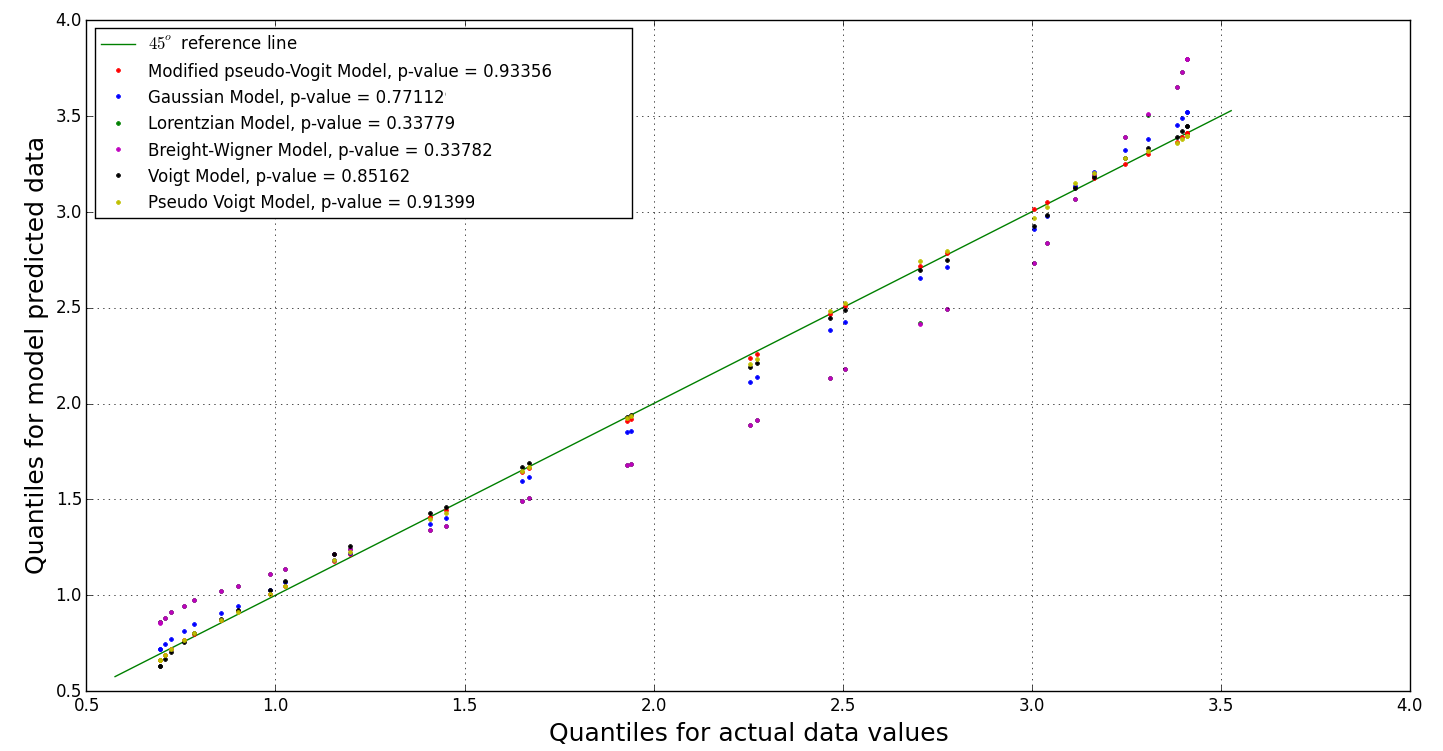}
        \captionN{The probability plot for $r = 51$. }
        \label{fig:Allqq51}
    \end{subfigure}
 \end{figure}    
 
   \begin{figure}[H]
  \ContinuedFloat 
  \centering 
  
     \begin{subfigure}[h]{0.9\textwidth}
        \includegraphics[width=\textwidth]{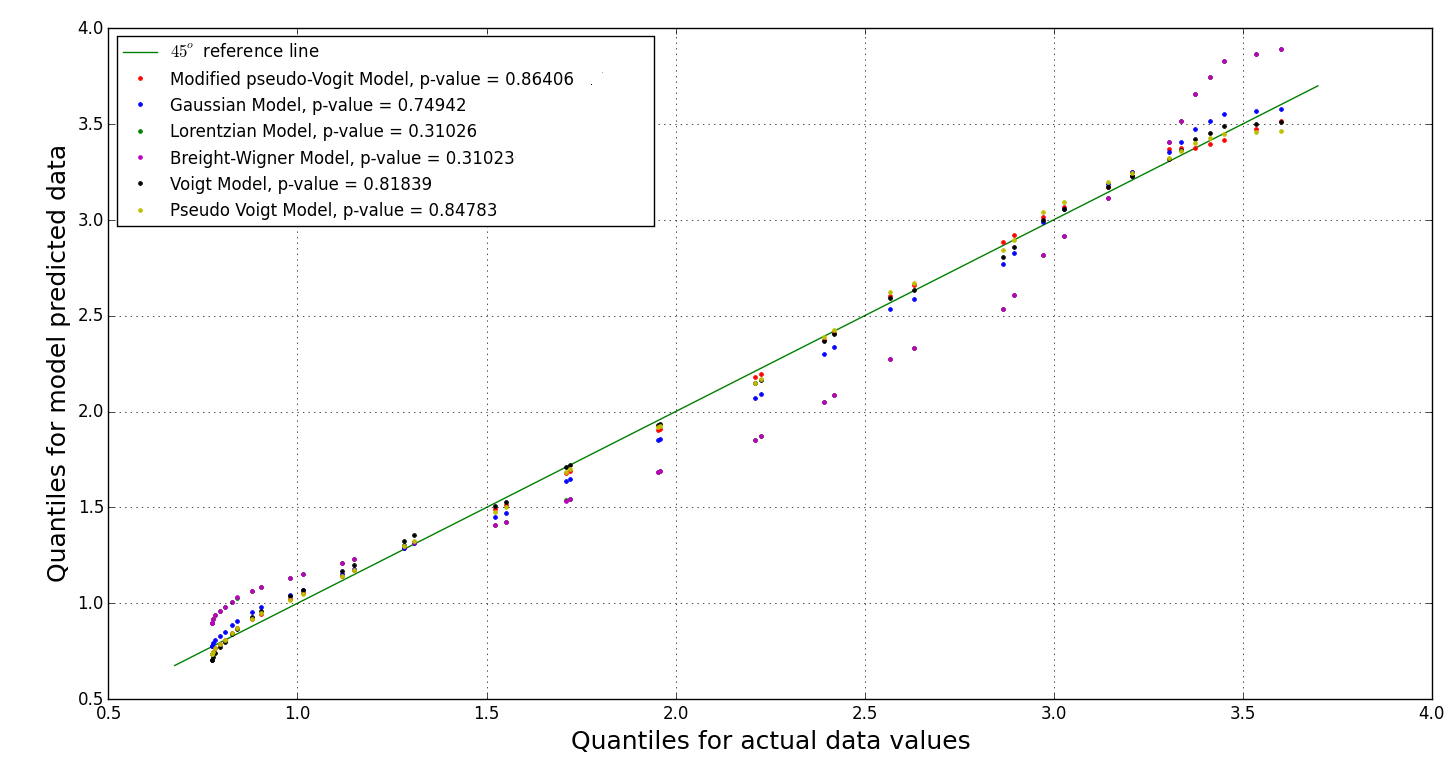}
        \captionN{The probability plot for $r = 54$. }
        \label{fig:Allqq54}
    \end{subfigure}

    \captionN{For all models, the left hand graph is for $r=54$ and the right is for $r=51$. The probability plot presents all the models together. All the above mentioned modeled are included to compare their resemblance with the actual data. The larger the $p$ value the better the line $y=x$ fits the data, implying the better the model is at describing the data. }
    \label{fig:AppenAllModels}
\end{figure}

\subsubsection{A first approximation to the data}\label{Appenfirst}

The overall behavior of the data across each curve is modeled extremely well using the pseudo-Voigt model. Here we present a few plots illustrating a first approximation to the data. A second approximation can be made by introducing an oscillating amplitude as described in Section~\ref{Hdiff}

   \begin{figure}[H]
  \centering 
  
     \begin{subfigure}[h]{0.9\textwidth}
        \includegraphics[width=\textwidth]{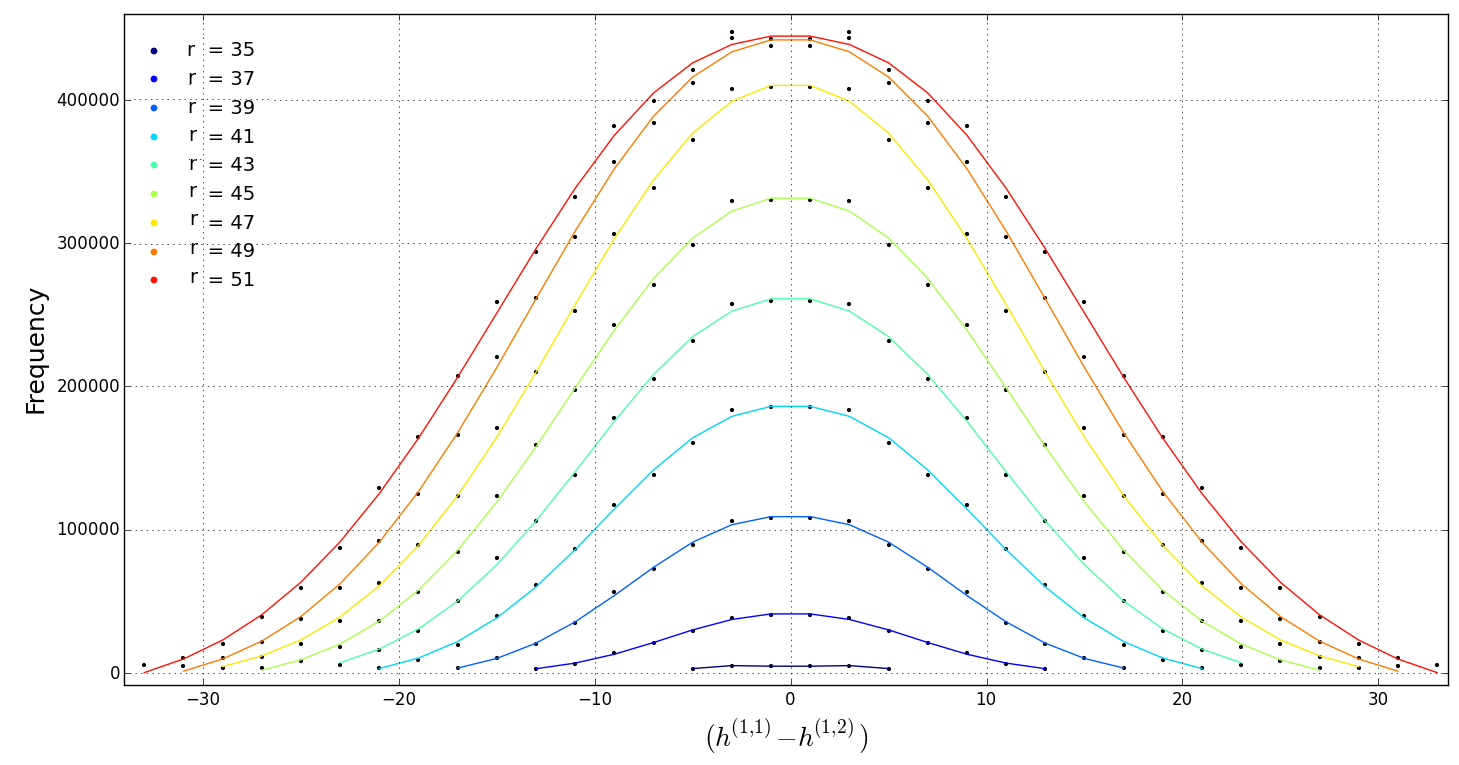}
        \captionN{Regression lines for few select even $r$ values, with $r \in [35,51]$.}
        \label{fig:AllPVOdd1}
    \end{subfigure}
    
     \begin{subfigure}[h]{0.9\textwidth}
        \includegraphics[width=\textwidth]{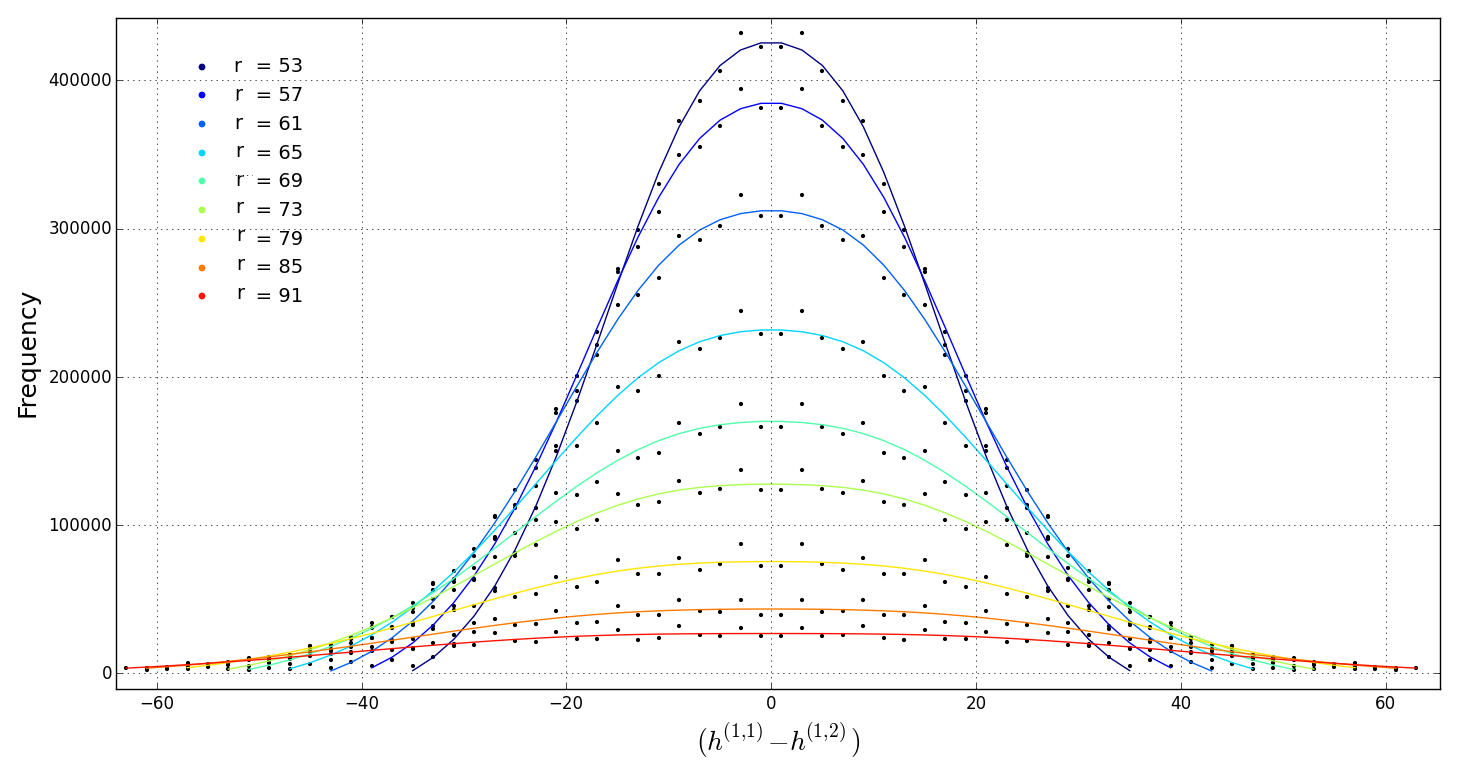}
        \captionN{Regression lines for few select even $r$ values, with $r>51$.}
        \label{fig:AllPVOdd1}
    \end{subfigure}          
     
    \captionN{Best fit curve based on the pseudo-Voigt model for the same sets of curves as seen in Figure~\protect\ref{Fig:OddHdiffs}.}
    \label{Fig:ApOddHdiffs}
\end{figure}

\begin{figure}[H]
    \centering
    \begin{subfigure}[h]{0.9\textwidth}
        \includegraphics[width=\textwidth]{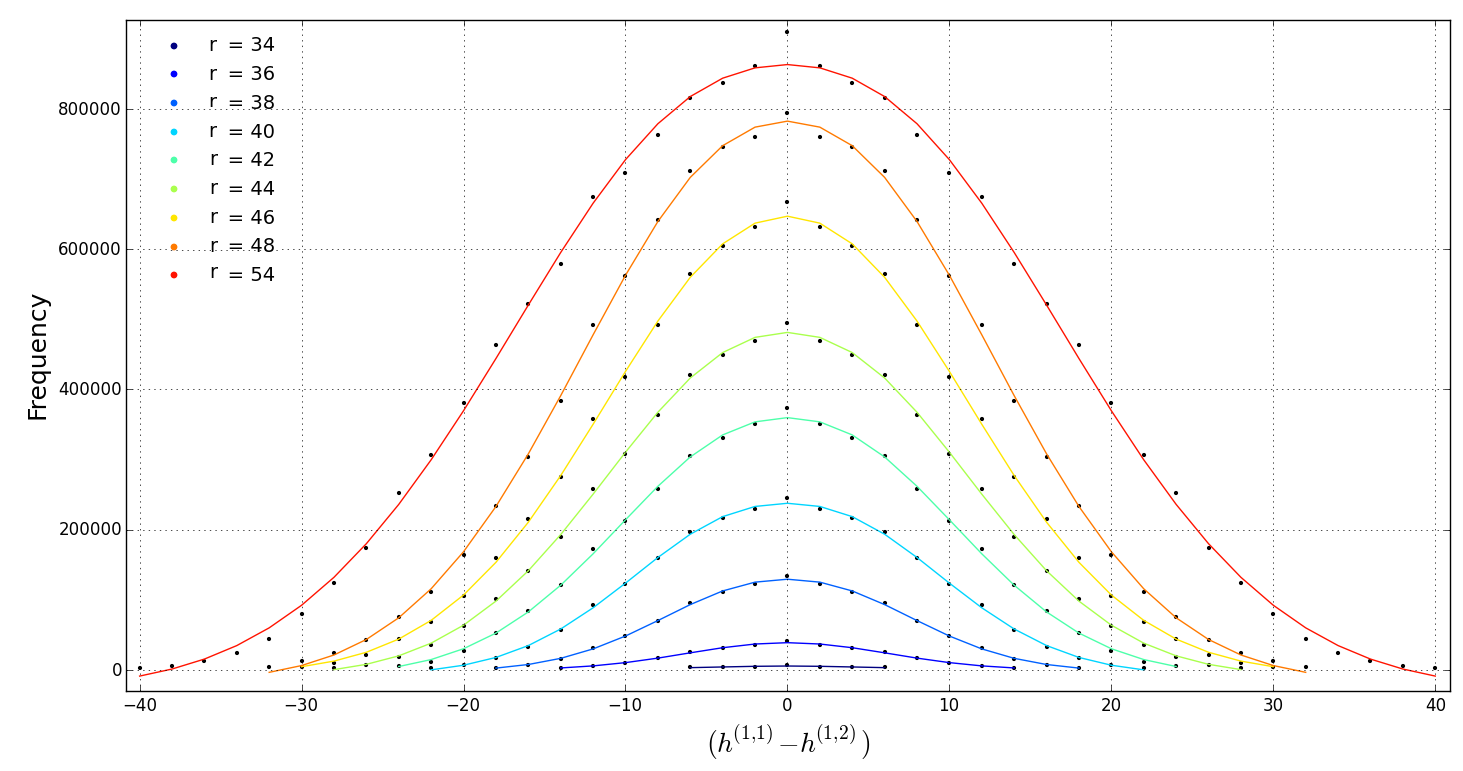}
        \caption{Regression lines for few select even $r$ values, with $r\leq 54$.}
        \label{fig:AllPVEven1}
    \end{subfigure}

    \begin{subfigure}[h]{0.9\textwidth}
        \includegraphics[width=\textwidth]{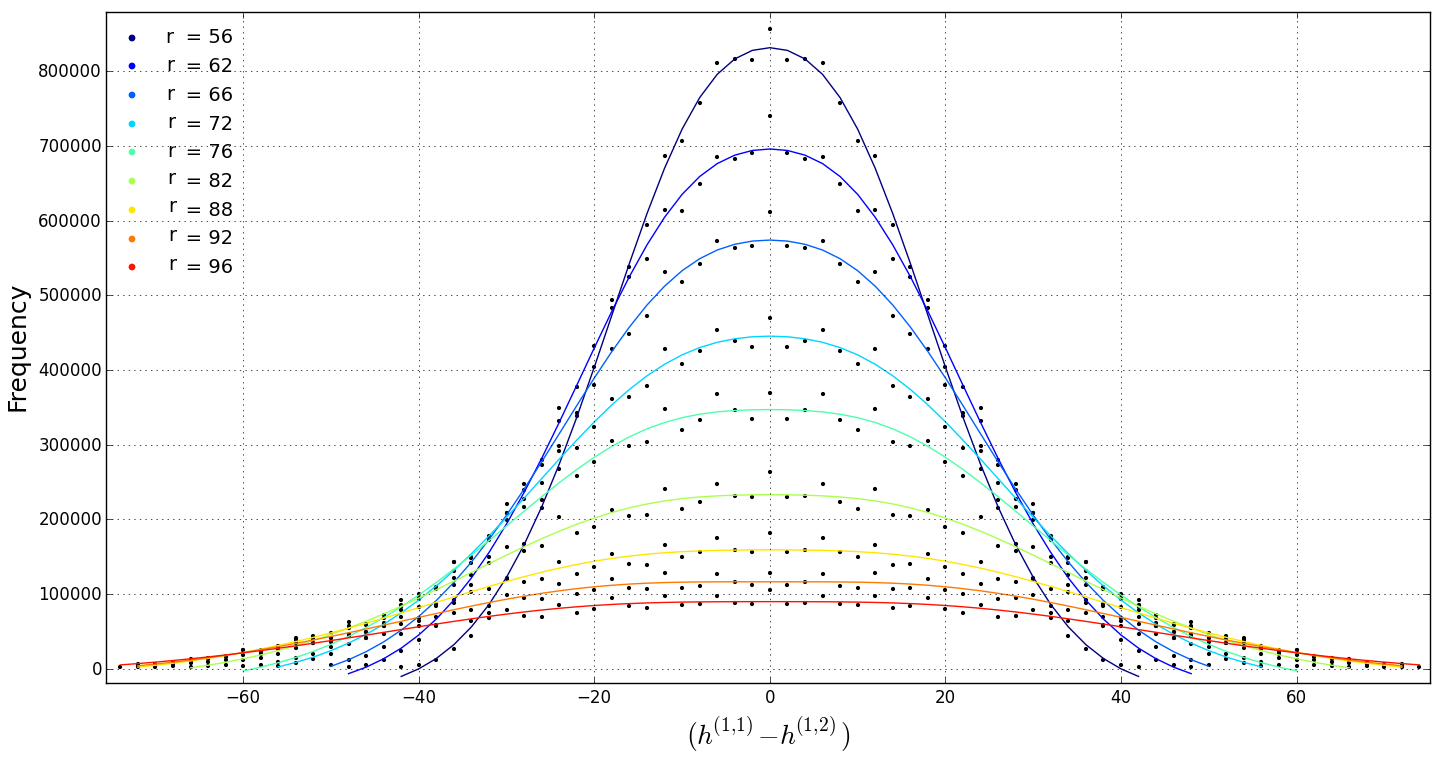}
        \caption{Regression lines for few select even $r$ values, with $r>54$.}
        \label{fig:AllPVEven2}
    \end{subfigure}
    
    \caption{Best fit curve based on the pseudo-Voigt model for the same sets of curves as seen in Figure~\protect\ref{Fig:EvenHdiffs}. }
    \label{Fig:ApEvenHdiffs}
\end{figure}

\subsubsection{Table of parameter values and statistics}\label{ADiffPValues}
Here we present the parameter values as well as the reduced $\chi$ value, $\chi_R$, in a tabular format for all even $r$ curves --- $r \in [34,120]$ --- and for all odd $r$ curves --- $r\in[35,99]$.

\begin{figure}[H]
    \centering
    \begin{subfigure}[h]{0.9\textwidth}
        \includegraphics[width=\textwidth]{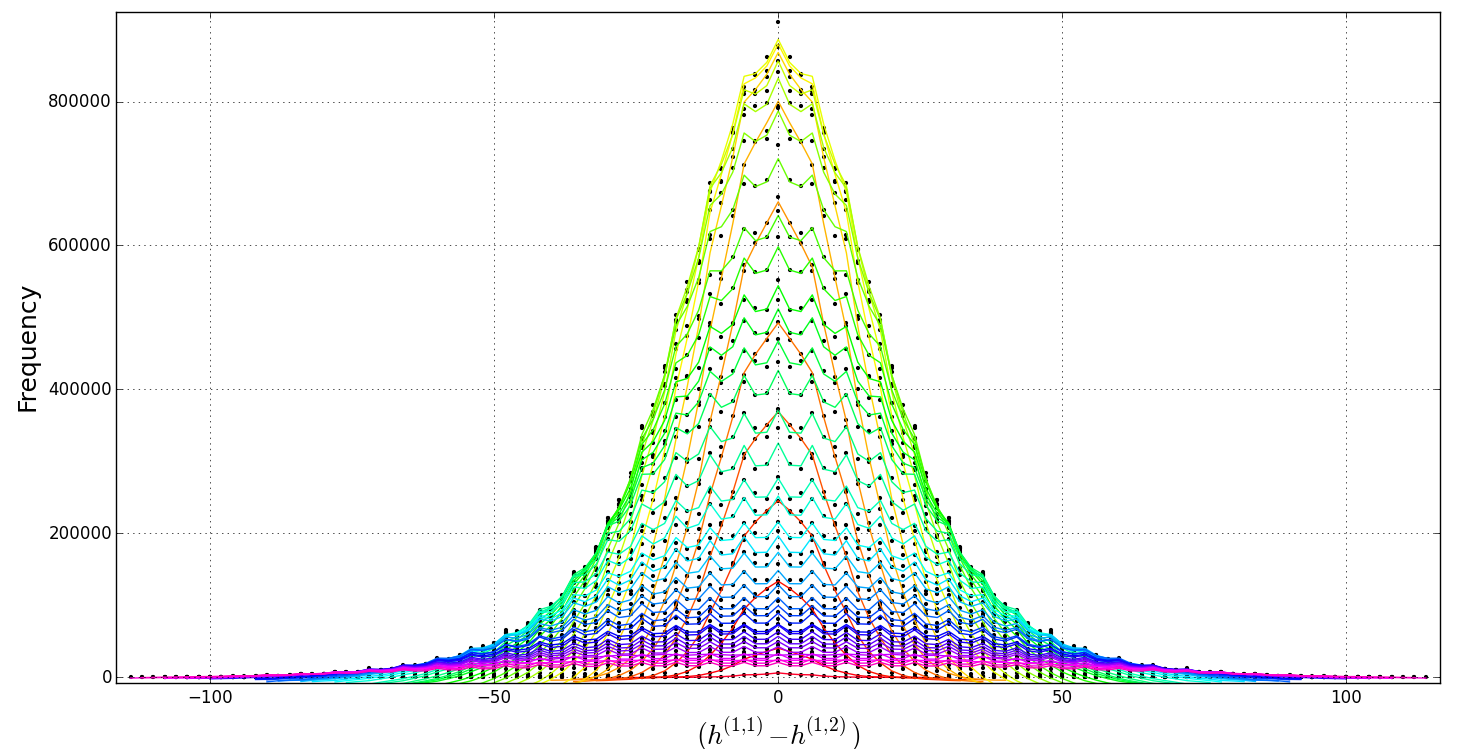}
        \caption{Every fitted even curve from $r = 34$ until $r=120$.}
        \label{fig:AllPVEven2}
    \end{subfigure}

    \begin{subfigure}[h]{0.9\textwidth}
        \includegraphics[width=\textwidth]{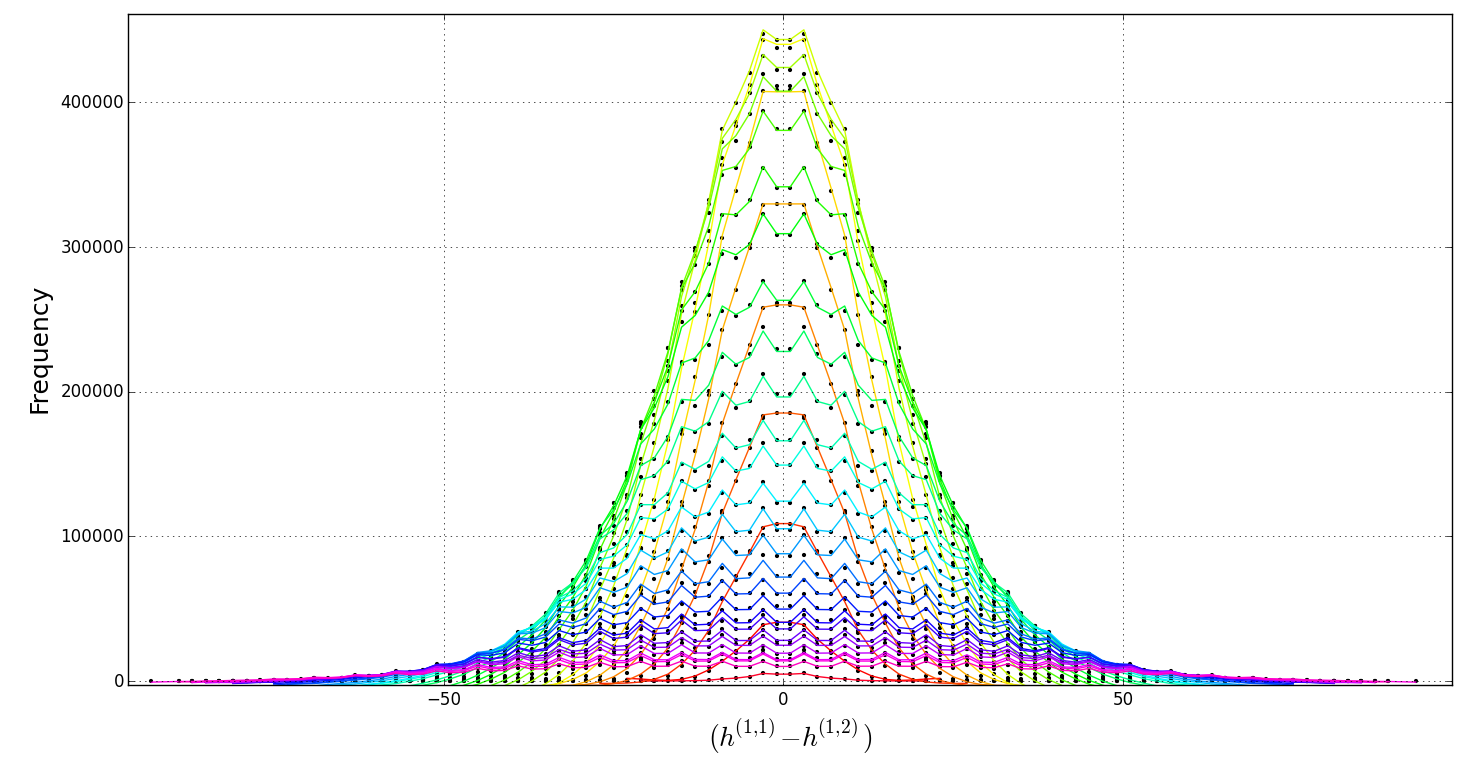}
        \caption{Every fitted even odd from $r = 35$ until $r=99$.}
        \label{fig:AllPVOdd2}
    \end{subfigure}
    
    \caption{This is what the entire distribution looks like using our modified pseudo-Voigt model. See Figure~\protect\ref{fig:ParamTableEvenOddHdiff} for the fitted coefficients as well as the fits for every curve given by the probability plots.}
    \label{fig:AllOddEvenMPV}
\end{figure}

\begin{figure}[H]
	\begin{center}	
		\includegraphics[scale=0.45]{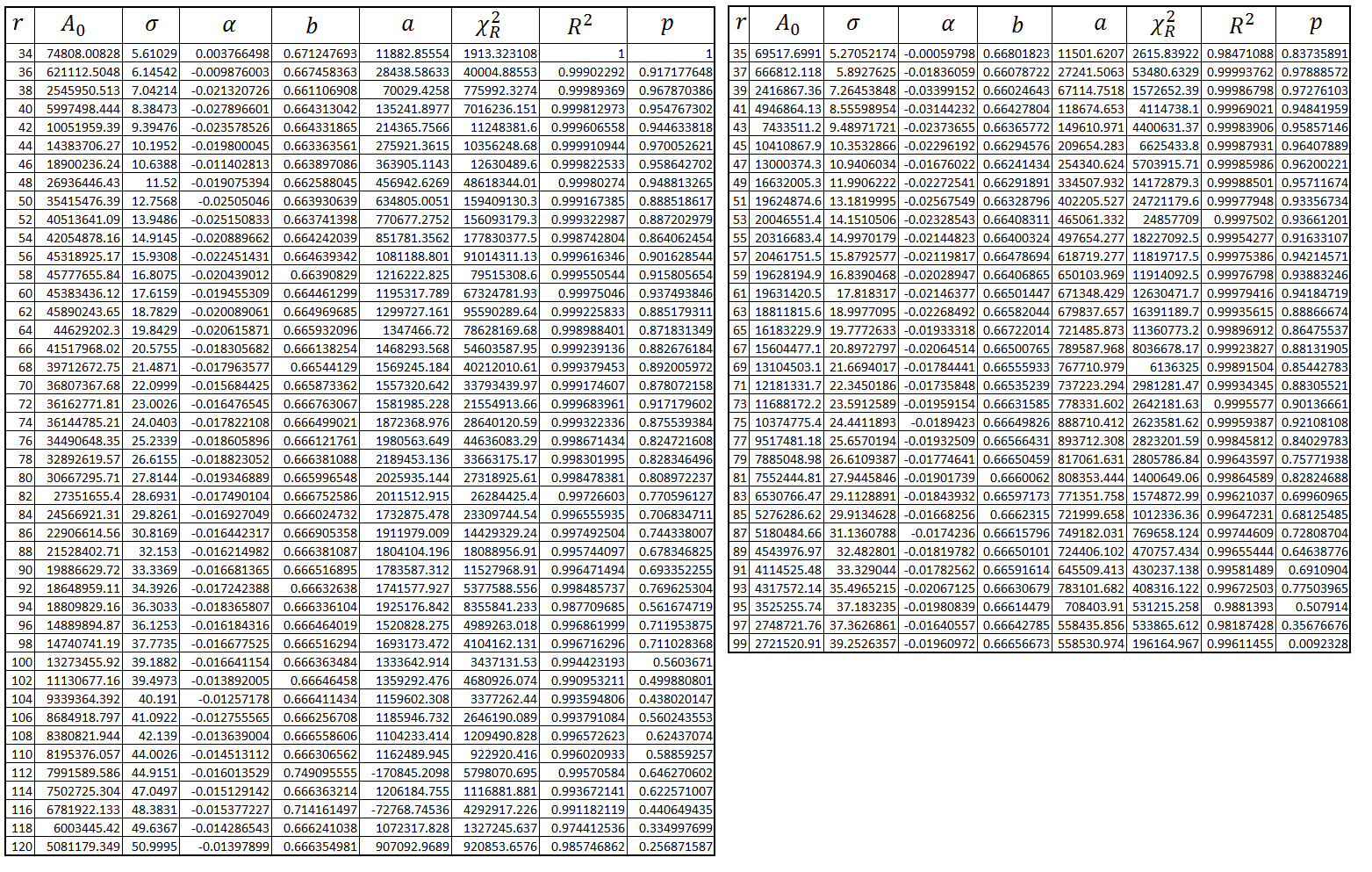}
	\end{center}
	\captionN{ Left : list of best fit coefficients for all even curves $r \in [34,120]$. Right: List of best fit coefficients for all odd curves $r\in[35,99]$. In both tables, the last two columns represent the $R^2$ and $p$ values for the probability plot for each curve. The $p$-values were obtained by first performing a Z-Standardization on the data.}
	\label{fig:ParamTableEvenOddHdiff}
\end{figure}

\begin{figure}[H]
	\begin{center}	
		\includegraphics[scale=0.6]{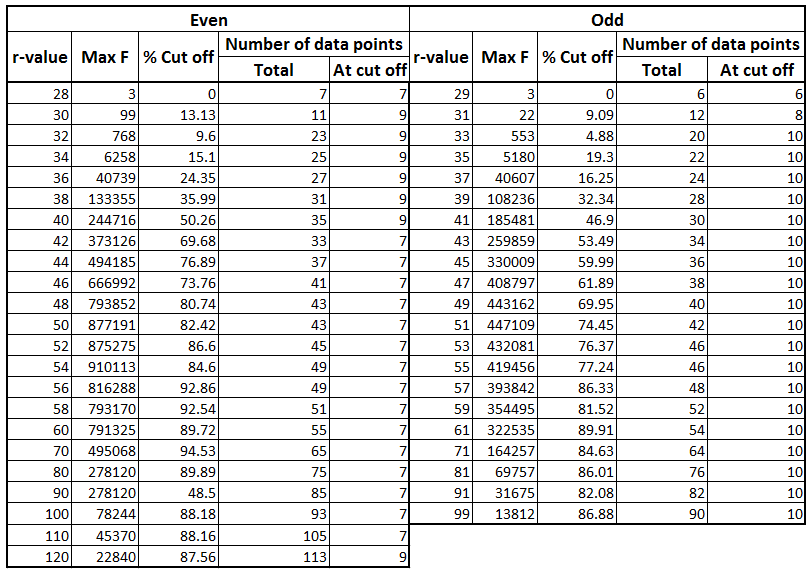}
	\end{center}
	\captionN{A list showing the number of data points left after increasing the cut off frequency to achieve a perfect fit. Conversely, one may state is as, the number of data points for each curve required such that the model will result in a perfect fit.}
	\label{fig:PerfectFits}
\end{figure}

\subsection{Supplementary plots for the $h^{1,1}+h^{1,2}$ distribution}

\subsubsection{Plots for the odd distribution as counterparts to the even ones}
All even plot counterparts will be referenced in the figures. The plots appear in the same order as in the main body, with descriptions only if necessary.

\begin{figure}[H]
	\begin{center}	
		\includegraphics[scale=0.40]{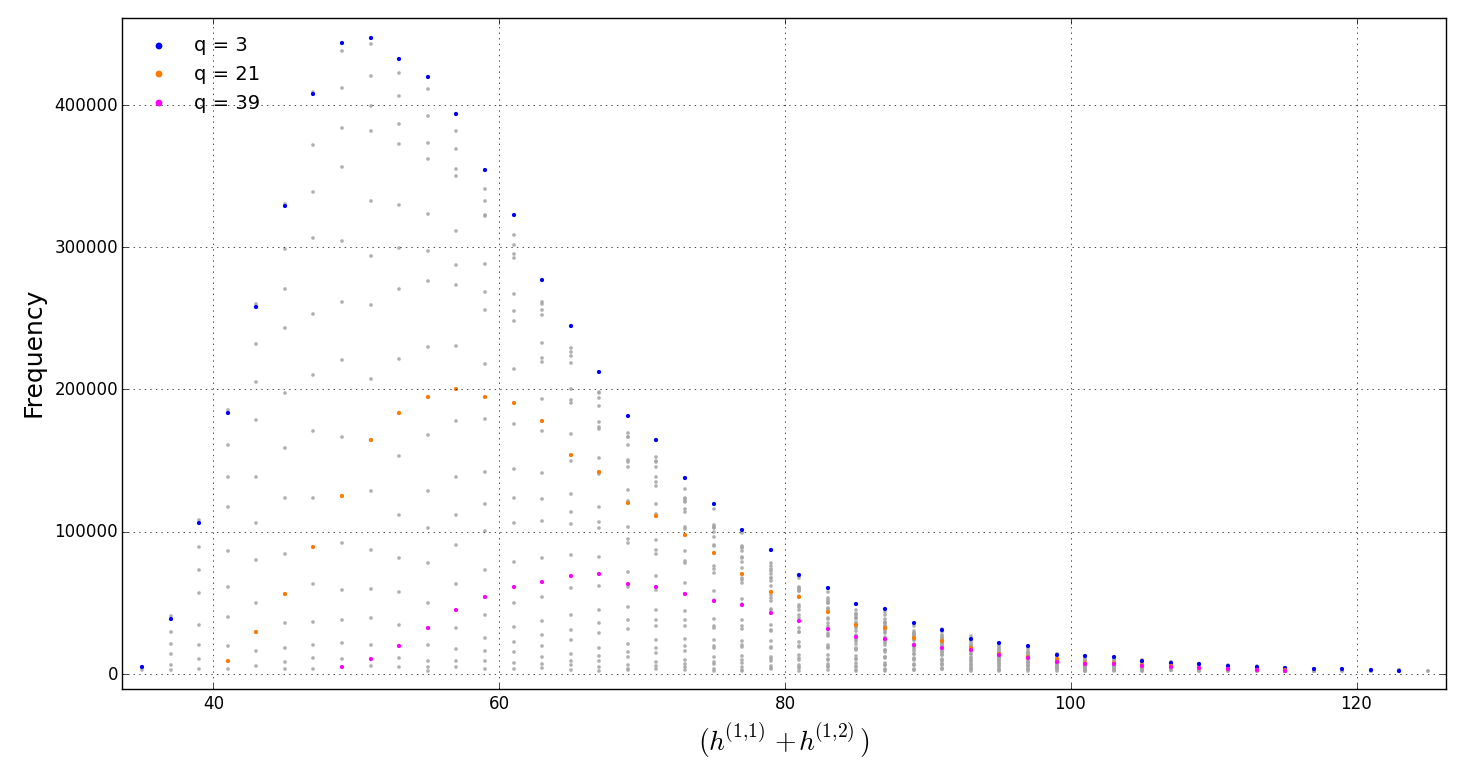}
	\end{center}
	\captionN{Three highlighted curves ($q = 3,19,31$) within the odd $h^{1,1} + h^{1,2}$ distribution. The transparent grey data dots are all the data plots for the distribution. Refer to Figure~\protect\ref{fig:CurveExample} for the even plot.}
	\label{fig:ExampleOdd1}
\end{figure}

\begin{figure}[H]
    \centering
    \begin{subfigure}[h]{0.9\textwidth}
        \includegraphics[width=\textwidth]{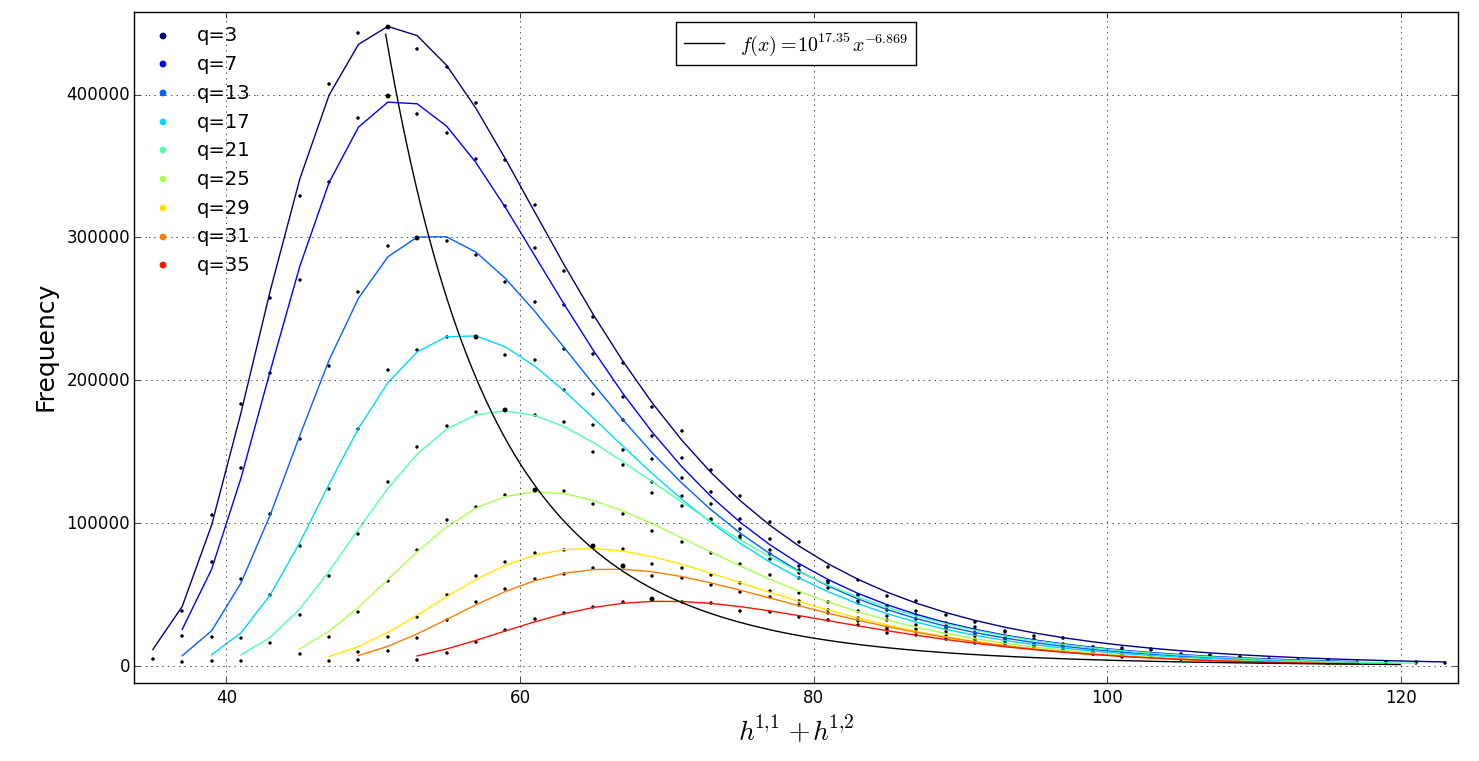}
        \captionN{Lines of best fit from a regression analysis for a few select curves. The black data points represent the maximum frequency for that particular $q-curve$. the black line is a line of best fit to describe the points of maximum frequency --- this is analogous to a blackbody spectrum. See Figure~\protect\ref{Fig:CurvesEven1} for the curves within the even distribution. }
        \label{fig:SeveralOddActual}
    \end{subfigure}
\end{figure}
\begin{figure}[H]
\ContinuedFloat 
  \centering 
    \begin{subfigure}[h]{0.9\textwidth}
        \includegraphics[width=\textwidth]{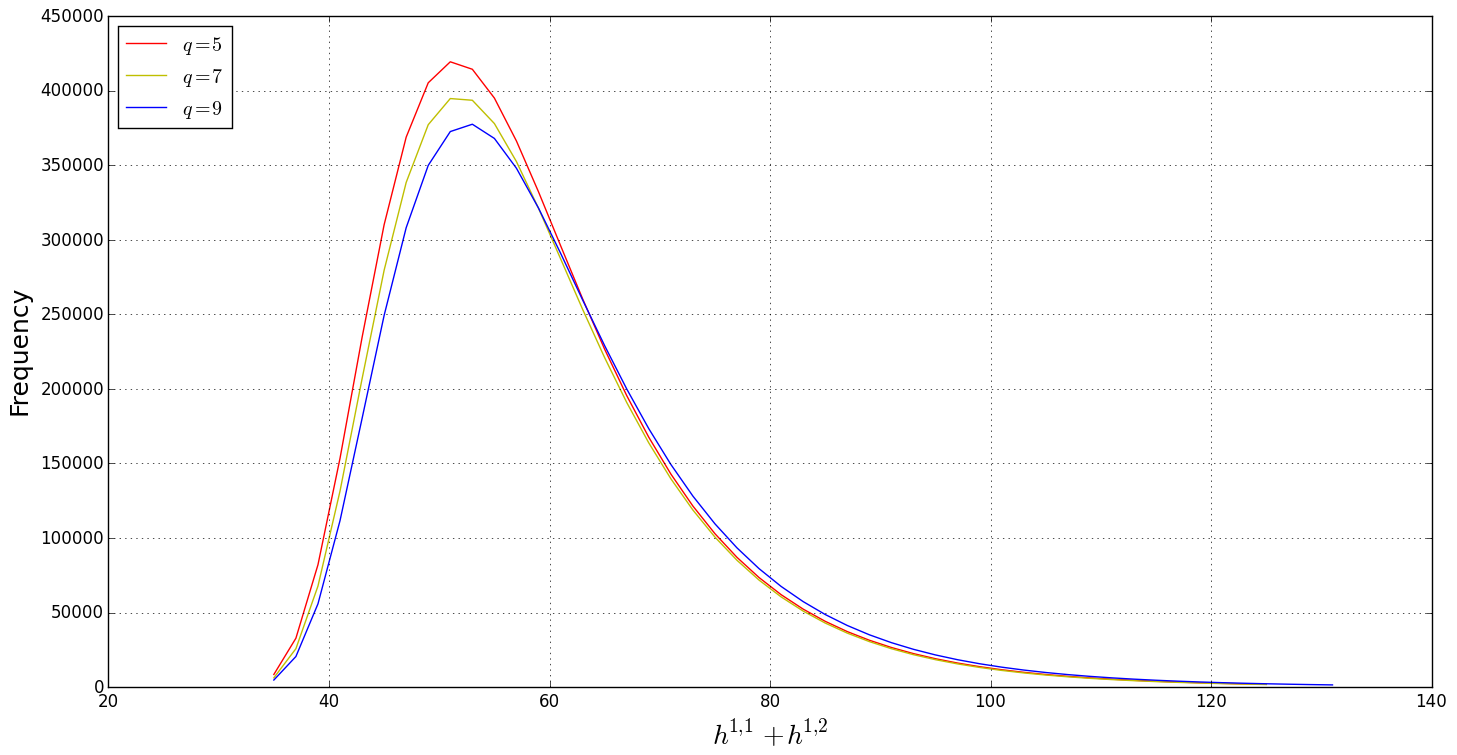}
        \captionN{The curves segregate into three classes determined by the value of the even integer modulo $6$. A similar pattern occurs in the even distribution; see Figure~\protect\ref{Fig:Overlap1}.}
        \label{fig:OverLapOdd}
    \end{subfigure}
    
    \captionN{In the attempt to describe the data analogously to a blackbody distribution (a), we discover some subtle structure, (b). These are the odd counterparts to Figure~\protect\ref{Fig:BlackEven}.}
    \label{Fig:BlackOdd}
\end{figure}

\begin{figure}[H]
    \centering
    \begin{subfigure}[h]{0.9\textwidth}
        \includegraphics[width=\textwidth]{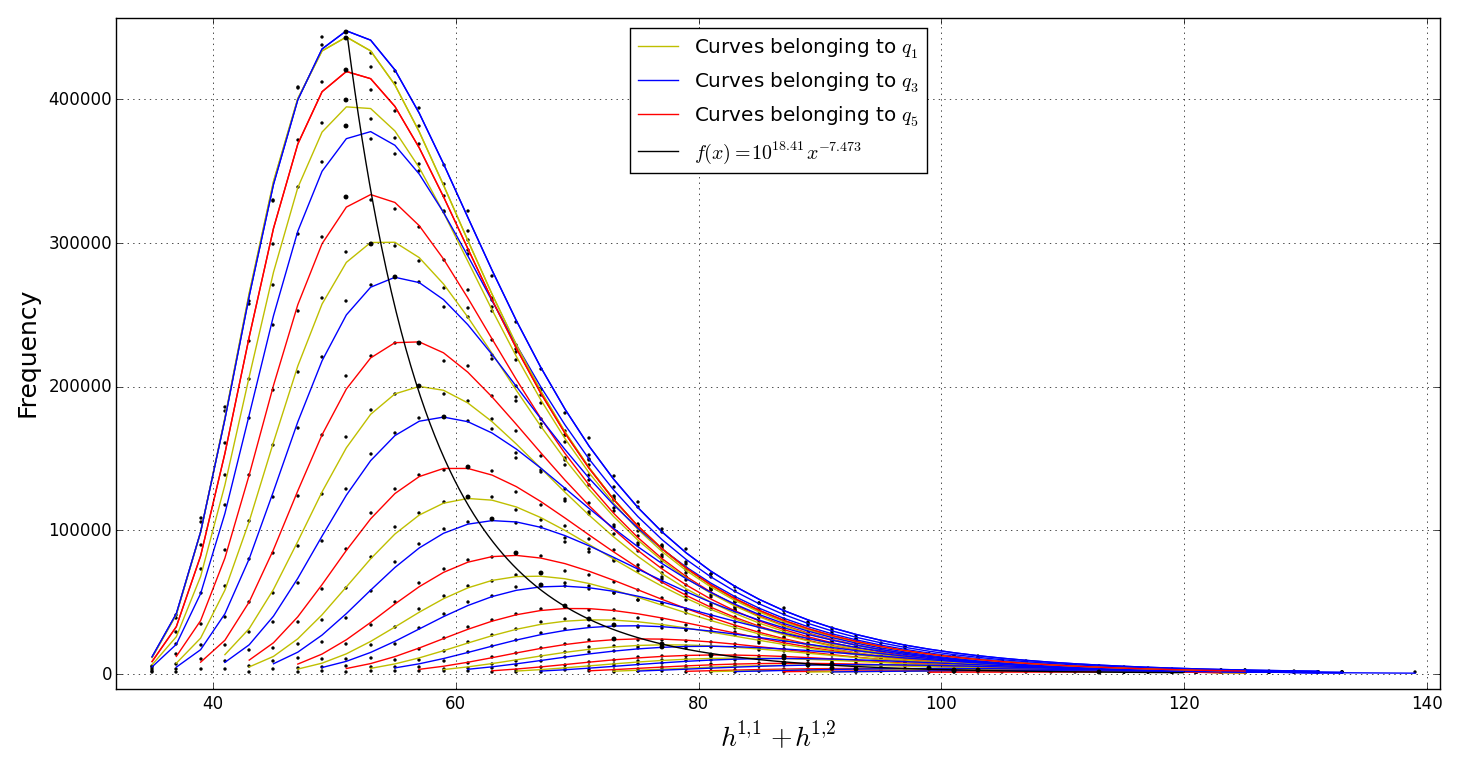}
        \captionN{All the curves color coded according to what residue class their curves $q_n$ belongs to. }
        \label{Fig:OddHdiff1}
    \end{subfigure}
    
\end{figure}
\begin{figure}[H]
\ContinuedFloat 
  \centering 
  
    \begin{subfigure}[h]{0.4\textwidth}
        \includegraphics[width=\textwidth]{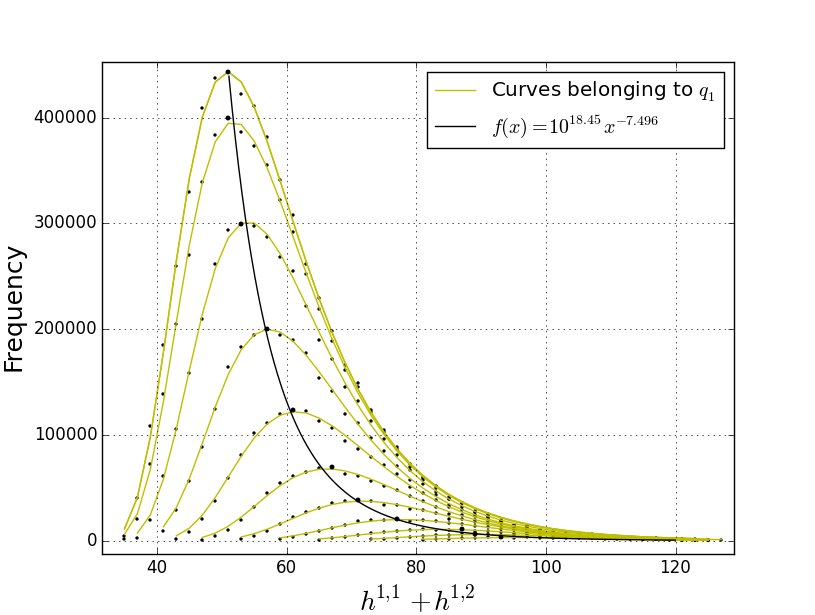}
        \captionN{Family of curves all belonging to $q_1$.}
        \label{Fig:Curvesq_1}
    \end{subfigure}
    ~
     \begin{subfigure}[h]{0.4\textwidth}
        \includegraphics[width=\textwidth]{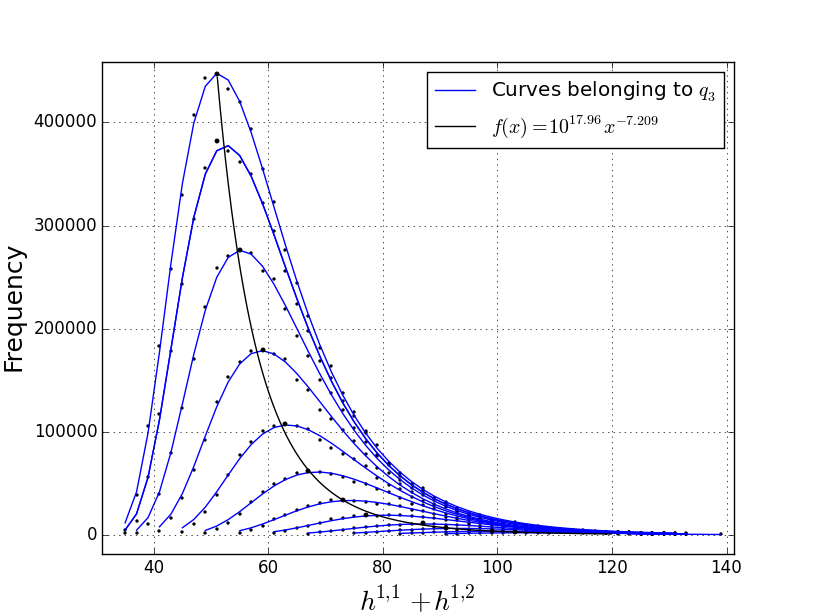}
        \captionN{Family of curves all belonging to $q_3$.}
        \label{Fig:Curvesq_3}
    \end{subfigure}

     \begin{subfigure}[h]{0.4\textwidth}
        \includegraphics[width=\textwidth]{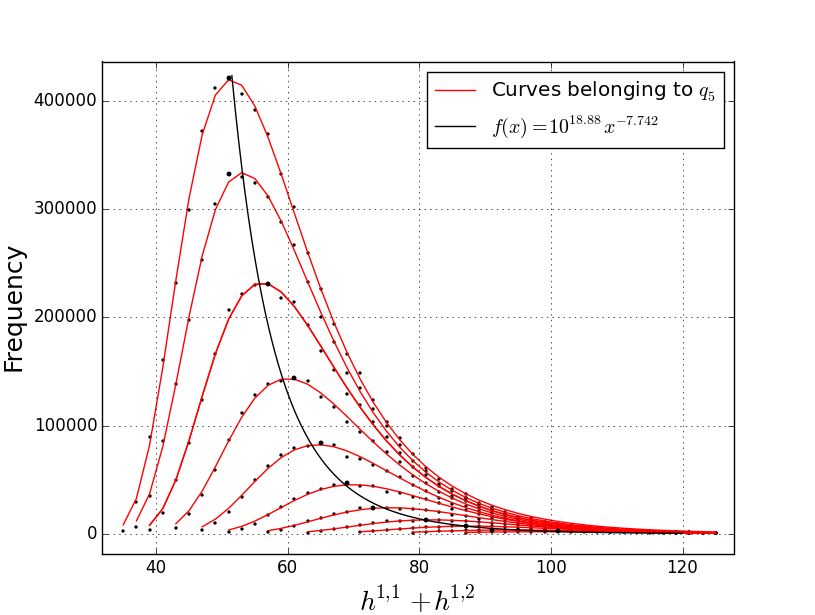}
        \captionN{Family of curves all belonging to $q_5$.}
        \label{Fig:Curvesq_5}
    \end{subfigure}

    \caption{We illustrate the added structure for odd $h^{1,1}+h^{1,2}$ data, by displaying how the regression curves can be divided into residue classes. For the list of even curves, refer to Figure~\protect\ref{Fig:AllEvenResidue}.}
    \label{Fig:AllOddResidue}
\end{figure}

\begin{figure}[H]
    \centering
    \begin{subfigure}[h]{\textwidth}
        \includegraphics[width=\textwidth]{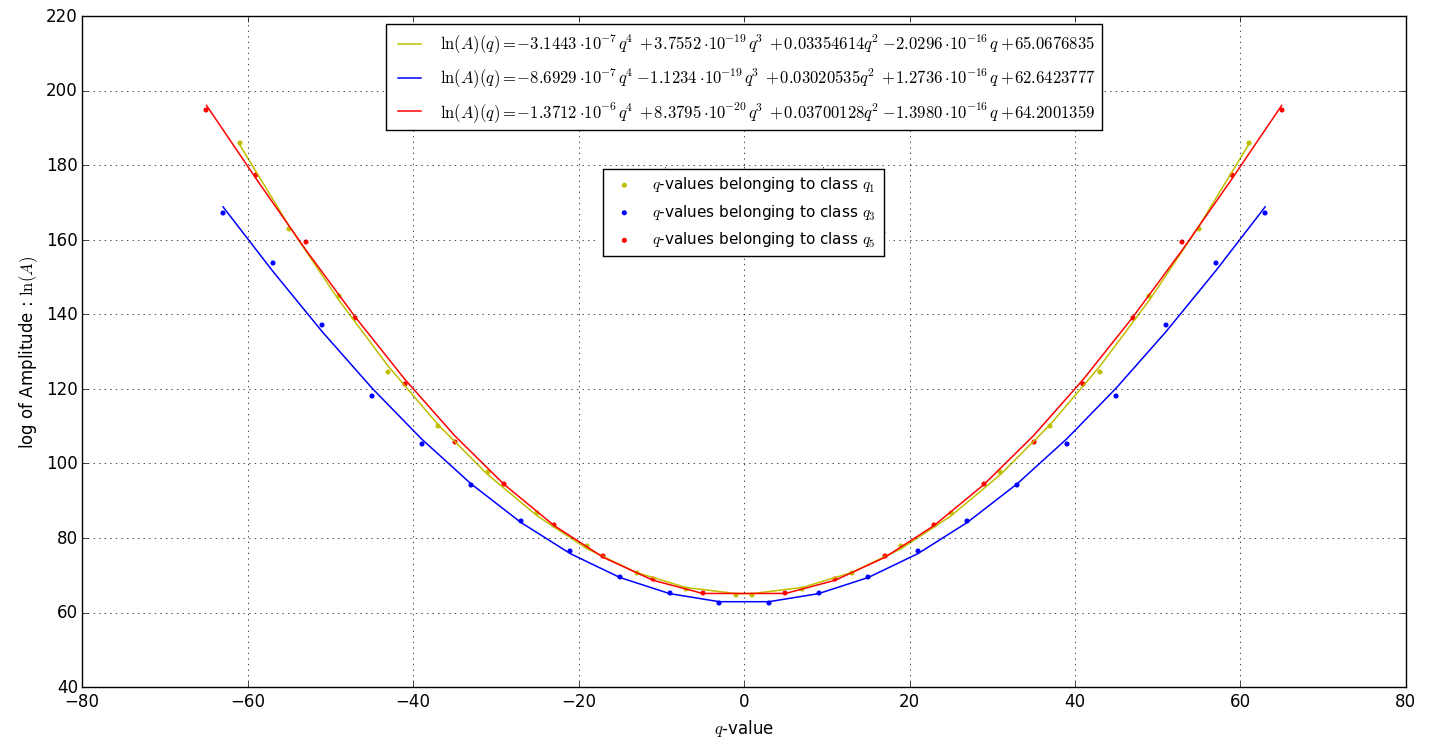}
        \captionN{Plotting the $q$- value parameter vs the $\log(A)$ parameter.}
        \label{fig:OParamLogA}
    \end{subfigure}

    \begin{subfigure}[h]{\textwidth}
        \includegraphics[width=\textwidth]{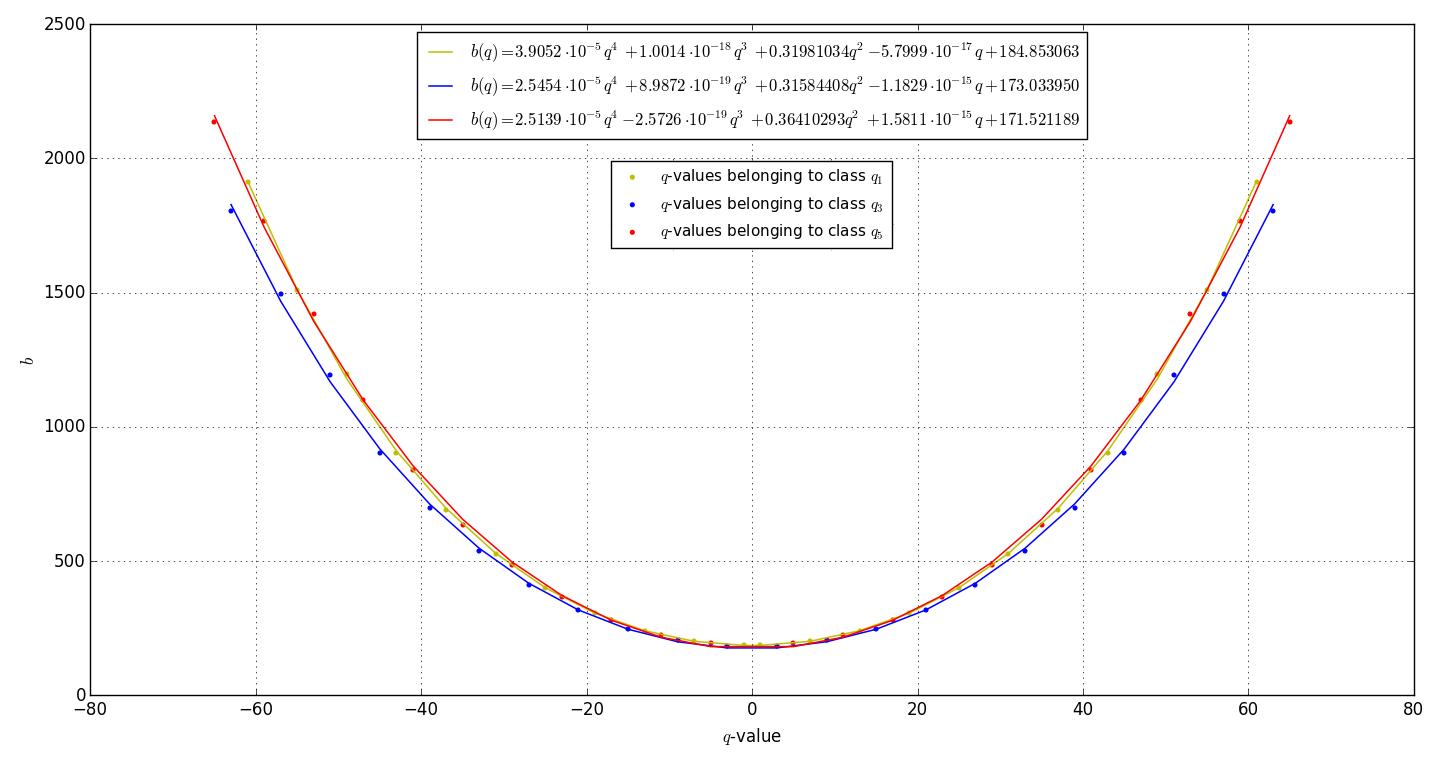}
        \captionN{Plotting the $q$- value parameter vs the $b$ parameter.}
        \label{fig:OParamB}
    \end{subfigure}
\end{figure}

\begin{figure}[H]
 \centering
 \ContinuedFloat
    \begin{subfigure}[t]{\textwidth}
        \includegraphics[width=\textwidth]{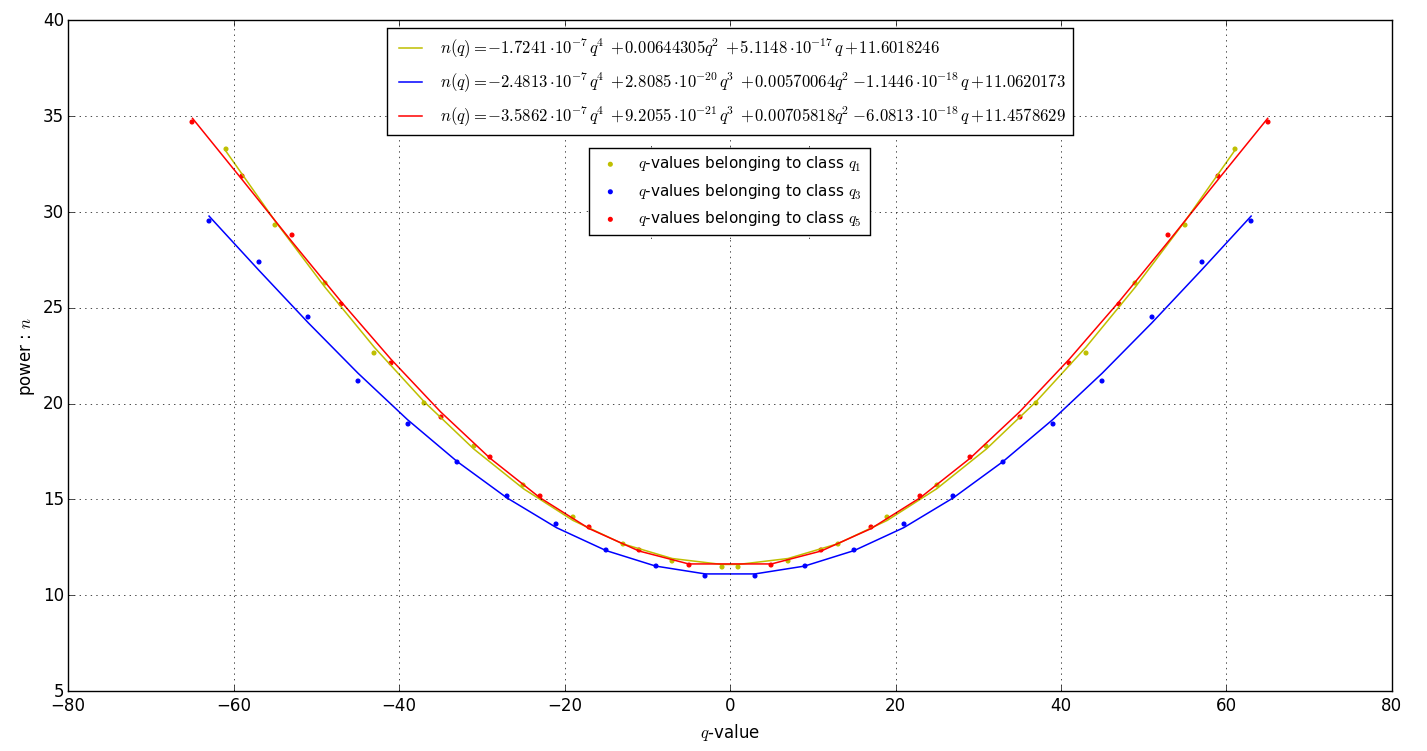}
        \captionN{Plotting the $q$- value parameter vs the power $n$ parameter.}
        \label{fig:OParamN}
    \end{subfigure}
    \captionN{The parameter plots are color coded according to what residue class their $q$ value belong to. For the relationships in the even distribution, see Figure~\protect\ref{fig:AllEvenHsumParam}.}
    \label{fig:AllOddHsumParam}
\end{figure}

\begin{figure}[H]
	\begin{center}	
		\includegraphics[scale=0.42]{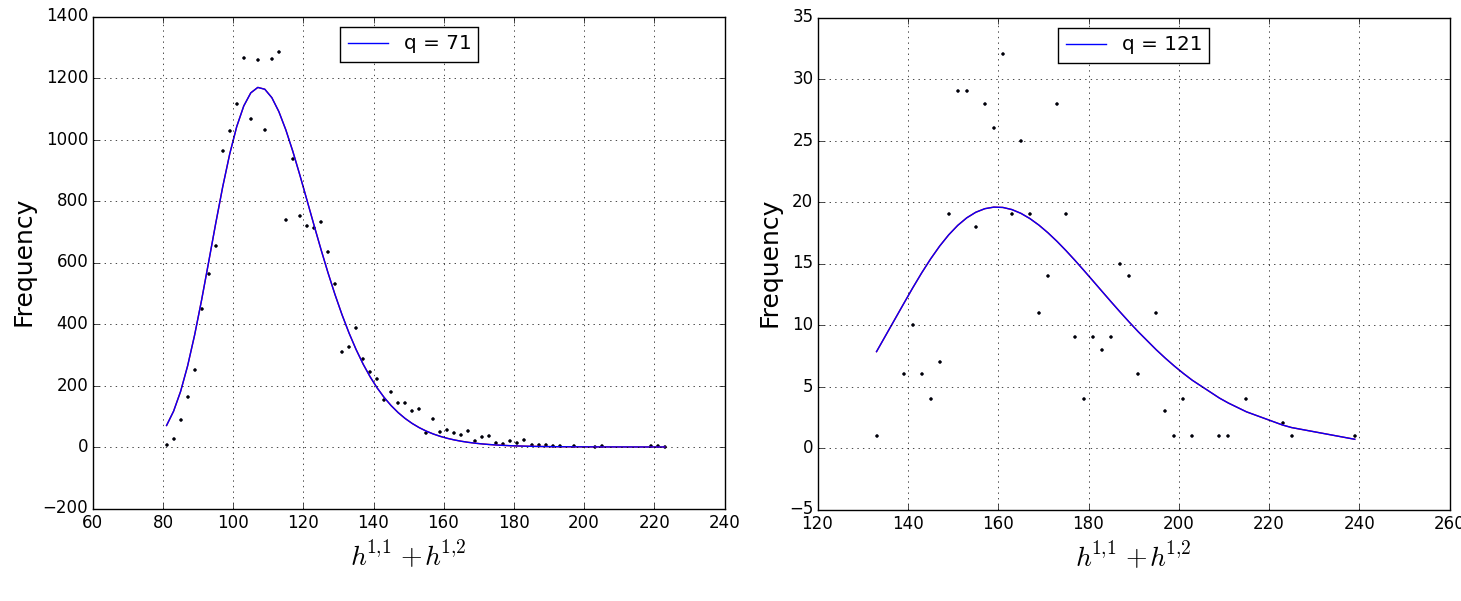}
	\end{center}
	\captionN{Left figure is the fitted model(blue line) for a $q$ value of 71 and right has a $q$ value of 121. As the $q$-value increases, the scattering of the data points within $h^{1,1} + h^{1,2}$ increases to the point where the model works no longer. For an example of how the model begins to break down at large $q$, see Figure~\protect\ref{fig:Highq1}.}
	\label{fig:ExampleLargeOdd}
\end{figure}

\subsubsection{Table of parameter values, coefficient values and statistics}\label{AHsumTables}
\begin{figure}[H]
	\begin{center}	
		\includegraphics[scale=0.57]{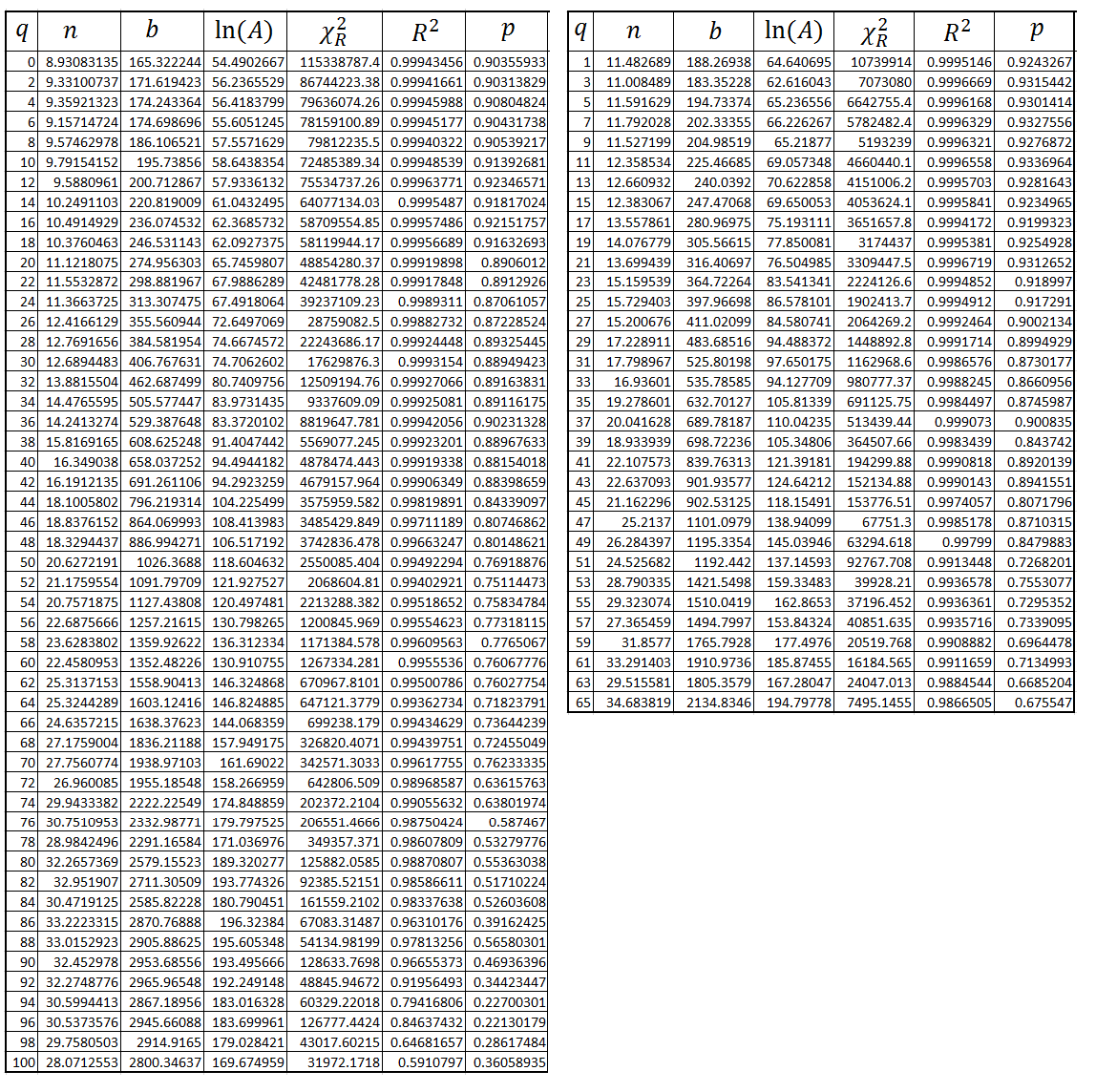}
	\end{center}
	\captionN{Left : list of best fit coefficients for all even curves $q \in [0,100]$. Right: List of best fit coefficients for all odd curves $q\in[1,65]$.}
	\label{fig:ParamOddEvenValues}
\end{figure}
\paragraph*{\textbf{Coefficient values for the description of the entire $h^{1,1}+h^{1,2}$ distribution}}
\begin{equation}\label{Acoeffs}
A_{k,i}= \left( \begin{array}{ccccc}
	54.2664195		& 	 2.9066\times 10^{-16} 	& 	0.02414823 & 	-5.4137\times 10^{-20} 		&		-7.2635\times 10^{-7} \\
  	65.0676835		&  	-2.0296\times 10^{-16} 	& 	0.03354614 & 	 3.7552\times 10^{-19}		& 		-3.1443\times 10^{-7} \\
   	54.8909275		&	-2.0323\times 10^{-16} 	& 	0.02753302 & 	-2.7091\times 10^{-20} 		&		-9.1972\times 10^{-7} \\
   	62.6423777		&	 1.2736\times 10^{-16} 	& 	0.03020535 & 	-1.1234\times 10^{-19}		&  		-8.6929\times 10^{-7} \\
   	54.5840853		&	 2.9011\times 10^{-16}	&	0.02748121 &	-9.4235\times 10^{-20}		&		-9.3840\times 10^{-7} \\
    64.2001359		&	-1.3980\times 10^{-16}	&	0.03700128 &	 8.3795\times 10^{-20}		&		-1.3712\times 10^{-7}   	
   \end{array} \right)
\end{equation}
\begin{equation}\label{Bcoeffs}
b_{k,i}= \left( \begin{array}{ccccc}
	132.357878		& 	 3.3411\times 10^{-15} 	& 	0.32753297 & 	-8.6619\times 10^{-19} 		&		 4.5825\times 10^{-6} \\
  	184.853063		&  	-5.7999\times 10^{-17} 	& 	0.31981034 & 	 1.0014\times 10^{-18}		& 		 3.9052\times 10^{-5} \\
   	117.228782		&	-1.2791\times 10^{-15} 	& 	0.36989364 & 	-8.5325\times 10^{-20} 		&		 2.9743\times 10^{-6} \\
   	173.033950		&	-1.1829\times 10^{-15} 	& 	0.31584408 & 	 8.9872\times 10^{-19} 		&		 2.5454\times 10^{-5} \\
   	105.298297		&	 5.7916\times 10^{-15}	&	0.37843953 &	-1.5078\times 10^{-18}		&		 1.3974\times 10^{-6} \\
    171.521189		&	 1.5811\times 10^{-15}	&	0.36410293 &	-2.5726\times 10^{-19}		&		 2.5139\times 10^{-5}   	
   \end{array} \right)
\end{equation} 
 
\begin{equation} \label{Ncoeffs}
n_{k,i}= \left( \begin{array}{ccccc}
	8.98205242		& 	 2.9066\times 10^{-17} 	& 	0.00434183 & 	-6.7671\times 10^{-21} 		&		-1.5512\times 10^{-7} \\
  	11.6018246		&  	 5.1148\times 10^{-17} 	& 	0.00644305 & 	 0							& 		-1.7241\times 10^{-7} \\
   	9.19515076		&	 4.3161\times 10^{-17} 	& 	0.00496066 & 	-1.3763\times 10^{-20} 		&		-1.9163\times 10^{-7} \\
   	11.0620173		&	-1.1446\times 10^{-18} 	& 	0.00570064 & 	 2.8085\times 10^{-20}		&  		-2.4813\times 10^{-7} \\
   	9.15798913		&	 5.0109\times 10^{-17}	&	0.00493009 &	-2.3559\times 10^{-20}		&		-1.9210\times 10^{-7} \\
    11.4578629		&	-6.0813\times 10^{-18}	&	0.00705818 &	 9.2055\times 10^{-21}		&		-3.5862\times 10^{-7}   	
   \end{array} \right)
\end{equation}

\subsection{Supplementary plots for the fourfold data.}
When looking for mirror symmetry in the fourfold data, we only observed partial mirror symmetry. Below is the full break down of the data set.
\begin{figure}[H]
	\begin{center}	
		\includegraphics[scale=0.43]{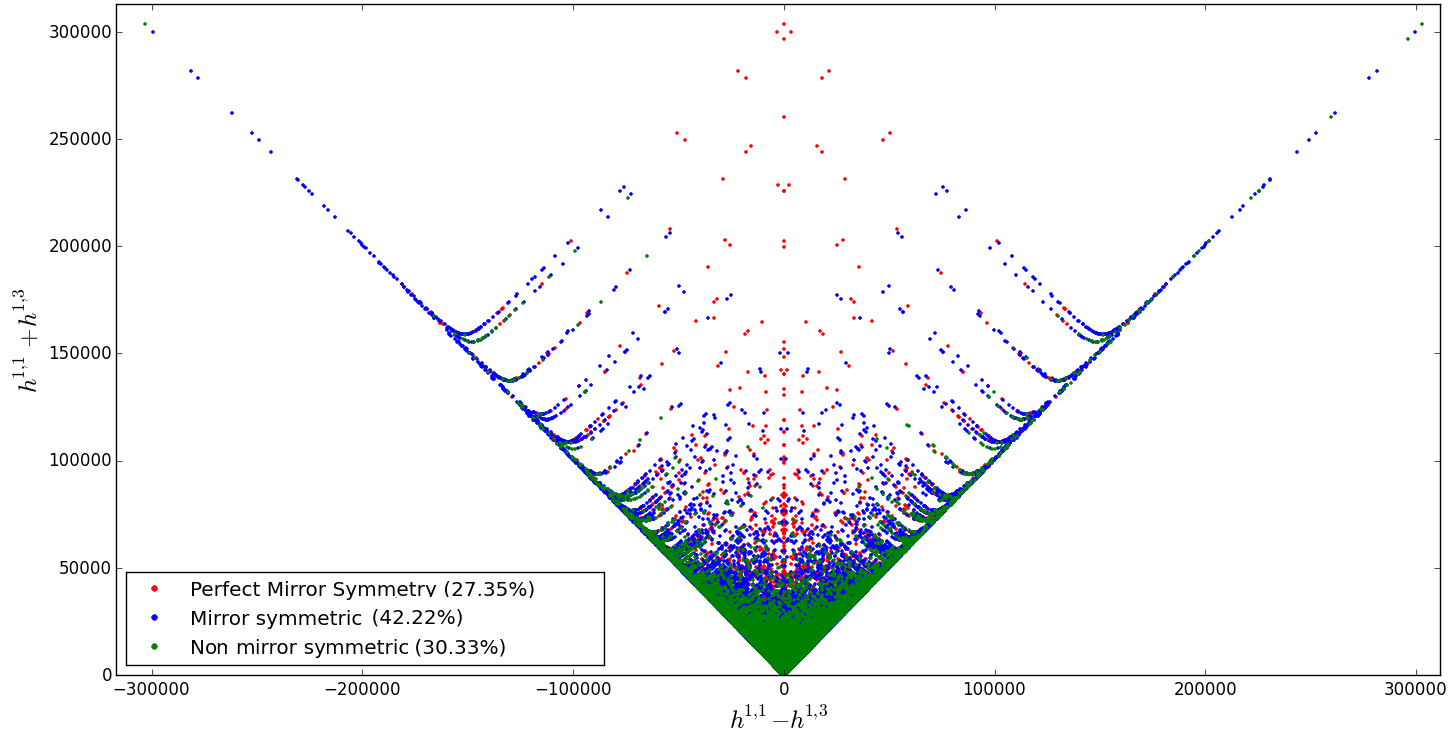}
	\end{center}
	\caption{Mirror symmetry is incomplete in the fourfold data set.}
	\label{Fig:AAllMirror1}
\end{figure}

\end{document}